\documentclass[dvips,a4paper,12pt,twoside]{book}
\usepackage[dvips]{graphicx,color}
\usepackage[dvips,pagebackref]{hyperref}
\hypersetup{
  pdftitle={Determination of the total width of the eta' meson},
  pdfauthor={Eryk Miros{\l}aw Czerwi{\'n}ski},
  pdfkeywords={eta prime, meson, width},
  unicode=false,
  pdfstartview={Fit},
  pdffitwindow=false,
  colorlinks=true,
  linkcolor=black,
  citecolor=black,
  breaklinks=true,
}
\usepackage[english]{babel}
\usepackage[numbers,sort&compress]{natbib}
\usepackage{hypernat}
\usepackage[T1]{fontenc}
\usepackage[utf8]{inputenc}
\usepackage{times}
\usepackage{geometry}
\usepackage{fancyhdr}
\usepackage{amsfonts}
\usepackage{amsmath}
\usepackage{mathrsfs}
\usepackage{wasysym}
\newcommand{\ep}{$\eta'$}
\newcommand{\e}{$\eta$}
\newcommand{\epw}{$\Gamma_{\eta'}$}
\newcommand{\pp}{$pp\to pp$}
\newcommand{\ppep}{$pp\to pp\eta'$}
\newcommand{\ppx}{$pp\to ppX$}
\newcommand{\cc}{COSY--11}
\fancypagestyle{plain}{%
   \fancyhead{} 
   \fancyhf{}%
}
\begin{document}

\pdfbookmark[-1]{Title pages}{title_en}
\cleardoublepage
\pdfbookmark[0]{Title page}{title_en}
\begin{titlepage}
\begin{center}
\vspace{0.7cm}
\begin{Huge}
JAGIELLONIAN UNIVERSITY\\
INSTITUTE of PHYSICS\\
\end{Huge}
\begin{figure}[!h]
\begin{center}
\includegraphics[width=0.17\textwidth]{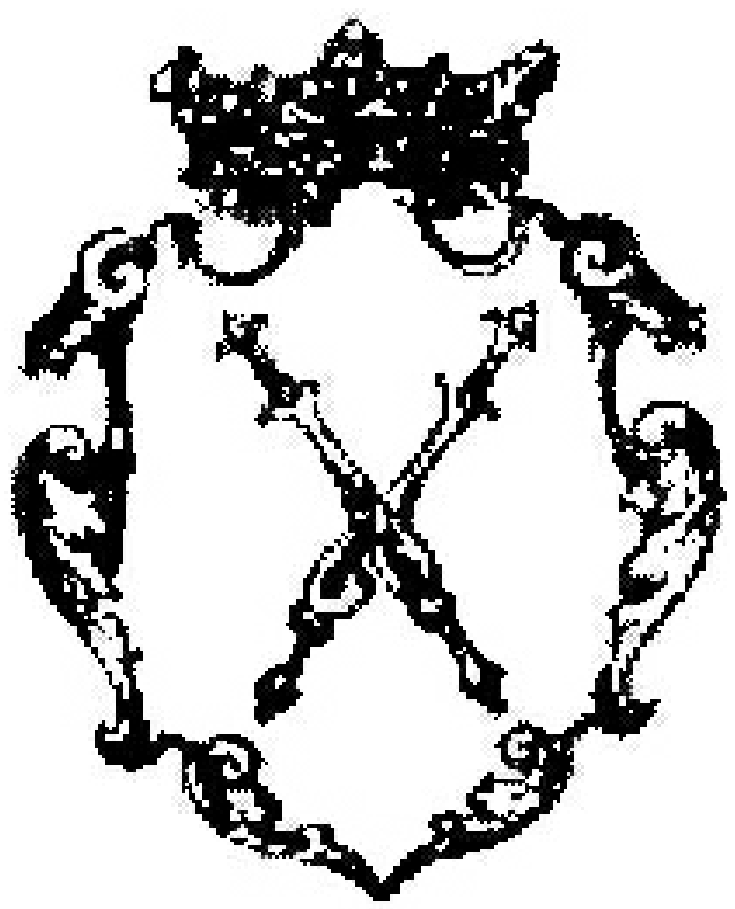}
\end{center}
\end{figure}
\vspace{1cm}
\begin{LARGE}
Determination of the total width of the $\eta'$ meson\\
\end{LARGE}
\vspace{1.5cm}
\begin{Large}
Eryk Miros{\l}aw Czerwi{\'n}ski
\end{Large}
\vspace{6.0cm}
\begin{Large}
\\PhD thesis prepared in
the Department of Nuclear Physics of the Jagiellonian University
and in the Institute\\of Nuclear Physics of the Research Center J{\"u}lich
\\under supervision of Prof. Paweł Moskal\\
\end{Large}
\begin{large}
\vspace{1.5cm}
Cracow 2009
\end{large}
\end{center}
\end{titlepage}

\thispagestyle{plain}

\cleardoublepage
\pdfbookmark[0]{Strona tytu{\l}owa}{title_pl}
\begin{titlepage}
\begin{center}
\vspace{0.7cm}
\begin{Huge}
UNIWERSYTET JAGIELLOŃSKI\\
INSTYTUT FIZYKI\\
\end{Huge}
\begin{figure}[!h]
\begin{center}
\includegraphics[width=0.17\textwidth]{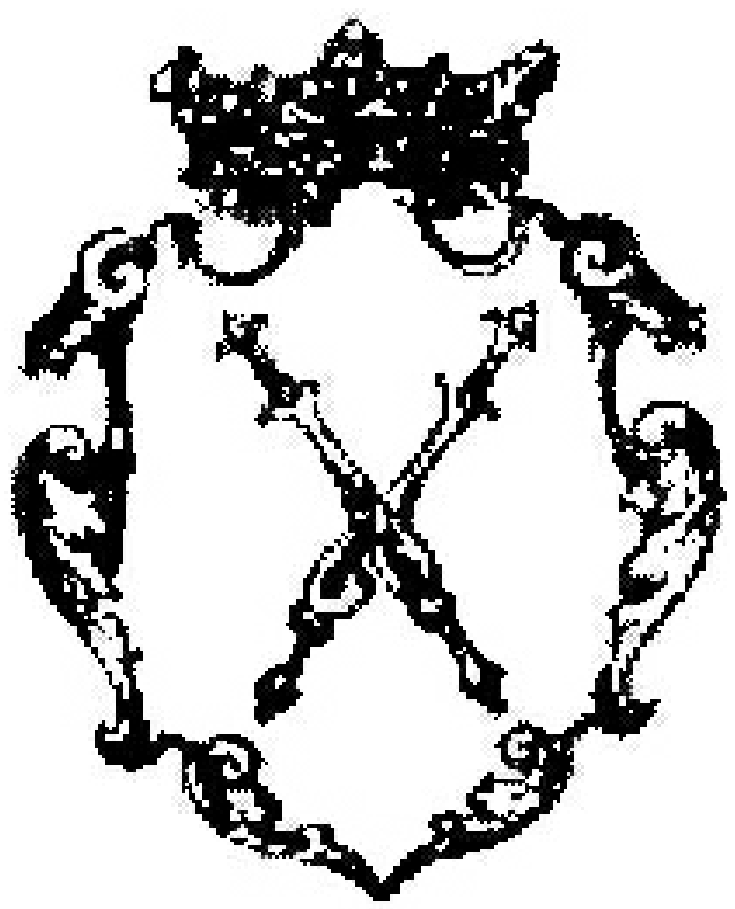}
\end{center}
\end{figure}
\vspace{1cm}
\begin{LARGE}
Wyznaczenie szeroko{\'s}ci ca{\l}kowitej mezonu $\eta'$
\end{LARGE}
\vspace{1.5cm}
\begin{Large}
Eryk Miros{\l}aw Czerwi{\'n}ski
\end{Large}
\vspace{6.0cm}
\begin{Large}
\\Praca doktorska
wykonana w Zak{\l}adzie Fizyki J\k{a}drowej
Uniwersytetu Jagiellońskiego
oraz w Instytucie Fizyki J\k{a}drowej
w Centrum Badawczym w J{\"u}lich
pod kierunkiem dr. hab. Paw{\l}a Moskala, prof. UJ\\
\end{Large}
\vspace{1.5cm}
\begin{large}
Krak{\'o}w 2009
\end{large}
\end{center}
\end{titlepage}

\thispagestyle{plain}

\cleardoublepage
\thispagestyle{plain}
\hspace{1cm}\\
\vfill
\begin{flushright}
\emph{ The reasonable man adapts himself to the world;\\
the unreasonable one persists in trying to adapt the world to himself.\\
Therefore all progress depends on the unreasonable man.}
\\George Bernard Shaw
\end{flushright}

\thispagestyle{plain}

\cleardoublepage
\pdfbookmark[-1]{Abstracts}{abstract_en}
\pdfbookmark[0]{Abstract}{abstract_en}
\begin{titlepage}
\begin{center}
\bf{Abstract}
\end{center}
The aim of this work was to determine the total width of the \ep\ meson.
The investigated meson was produced via the \ppep\ reaction
in the collisions of beam protons from COSY synchrotron 
with protons from a hydrogen cluster target.
The \cc\ detector was used for the
measurement of the four-momentum vectors of outgoing protons.
The mass of unregistered meson was determined via the missing mass technique, while
the total width was directly derived from the mass distributions
established at five different beam momenta.
Parallel monitoring of the crucial parameters (e.g. size and position of the target stream)
and the measurement close-to-threshold permitted to obtain
mass resolution of FWHM~=~0.33~MeV/c$^2$.

Based on the sample of more than 2300 reconstructed \ppep\ events
the determined total width of the \ep\ meson  amounts to
$\Gamma_{\eta'}=0.226\pm0.017(\textrm{stat.})\pm0.014(\textrm{syst.})$~MeV,
which is the most precise measurement until now.
\end{titlepage}

\thispagestyle{plain}

\cleardoublepage
\pdfbookmark[0]{Streszczenie}{abstract_pl}
\begin{titlepage}
\begin{center}
\bf{Streszczenie}
\end{center}
Celem tej pracy by{\l}o wyznaczenie szeroko\'sci ca{\l}kowitej mezonu \ep.
Badany mezon by{\l} produkowany w reakcji \ppep\ w zderzeniach proton\'ow wi{\k{a}}zki
synchrotronu COSY oraz proton\'ow z wodorowej tarczy klastrowej.
Do pomiaru czterop\k{e}d\'ow wylatuj\k{a}cych proton\'ow u\.zyty zosta{\l} detektor \cc.
Masa nierejestrowanego mezonu by{\l}a wyznaczona dzi\k{e}ki metodzie masy brakuj\k{a}cej,
podczas gdy ca{\l}kowita szeroko\'s\'c zosta{\l}a otrzymana bezpo\'srednio z widm
masy brakuj\k{a}cej uzyskanych dla pi\k{e}ciu r\'o\-\.znych p\k{e}d\'ow wi\k{a}zki.
R\'ownoczesne monitorowanie kluczowych parametr\'ow (np. takich jak ro\-zmiar i pozycja strumienia tarczy)
oraz wykonanie pomiaru
w pobli\.zu progu kinematycznego na produkcj\k{e} mezonu \ep\
pozwoli{\l}o otrzyma\'c dok{\l}adno\'s\'c wyznaczenia masy r\'ow\-n\k{a} FWHM~= 0.33~MeV/c$^2$.

W oparciu o ponad 2300 zrekonstruowanych zdarze\'n \ppep\
wyznaczona szeroko\'s\'c ca{\l}kowita mezonu \ep\ wynosi
$\Gamma_{\eta'}=0.226\pm0.017(\textrm{stat.})\pm0.014(\textrm{syst.})$~MeV,
co jest najdok{\l}adniejszym dotychczas wynikiem pomiaru tej wielko\'sci.
\end{titlepage}

\thispagestyle{plain}

\frontmatter
\cleardoublepage
\pdfbookmark[-1]{Contents}{tableofcontents}
\tableofcontents
\mainmatter
\chapter{Introduction}
Enlarging the knowledge about nature can be realised either
by taking into account a larger field for investigations or by focusing on
the improvement of the quality of the already existing information
e.g. by improving significantly the precision of measurements.
This work is an example for the second method in order to deepen the
understanding of properties of hadronic matter, precisely,  the value of
the total width of the \ep\ meson (\epw).

Although the value of
\epw\ is known since 30 years~\cite{Binnie}, there are only two measurements
so far~\cite{Binnie,Wurzinger} with results which are admittedly in agreement within
the limits of the achieved accuracy, but the reported \mbox{$\sim$30--50\%} errors cause the average of these values 
not to be recommended by the Particle Data Group~(PDG)~\cite{pdg}.
Instead, the value resulting from a fit to 51 measurements
of partial widths, integrated cross sections, and
branching ratios is quoted by PDG~\cite{pdg}. However, both values
(the measured average and the fit result) are not consistent and, additionally,
the value recommended by PDG may cause
some difficulties when interpreting experimental data
due to the strong correlation between $\Gamma_{\eta'}^{\textrm{PDG}}$
and $\Gamma(\eta'\to\gamma\gamma)$.
This is the case e.g. in the investigations
aiming for the determination
of the gluonium contribution to the \ep\ meson wave function~\cite{Biagio}.

Though there is no
theoretical prediction about \epw, there is strong interest
in the precise determination of \epw\ to translate
branching ratios (BR) into partial widths, especially for the \ep\ meson decay channels to
$\pi^{+}\pi^{-}\eta$, $\rho\gamma$, and $\pi^{0}\pi^{0}\eta$ as
inputs for the phenomenological description of Quantum Chromo-Dy\-na\-mics in the non-perturbative regime~\cite{Borasoy}.

It is also worth to note that an improvement of the experimental
resolution by an order of magnitude in comparison to
 previous experiments~\cite{Binnie,Wurzinger} could resolve fine structures in the \ep\ signal,
which cannot be excluded \emph{a priori}.

The above-mentioned examples visualise that
a precise determination of \epw\ will provide important information for
a better understanding of meson physics at low energies and, in particular, for structure and
decay processes of the \ep\ meson.
Therefore, a more accurate than so far determination of \epw\ constitutes the main motivation for this thesis.
A more detailed motivation for such studies is presented in Chapter~\ref{motivation}.

This work focuses on a measurement of \epw\ performed in 2006
at the cooler synchrotron COSY with the \cc\ detector setup, where
the \ep\ mesons were produced in collisions of protons from the circulating beam with protons
from the cluster target
stream~\cite{proposal}.
The measurement was carried out at five beam momenta very close to the \ep\ production threshold.
The identification of the \ppep\ reaction is based on the
reconstruction of the four-momentum vectors of the outgoing protons and
on the calculation of the \ep\ meson four-momentum vector from energy and momentum conservation.
The total width of the \ep\ meson is directly determined from the missing mass spectra.
The mass resolution of the \cc\ detector was improved to such limits that \epw\ could have been
obtained directly from the mass distribution established with a precision comparable
to the width itself. Applied improvements are:
(i)~measurement very close to the kinematic threshold
to decrease the uncertainties of the missing mass determination,
since at threshold the value of $\partial(mm)\slash\partial\boldsymbol{p}$ approaches zero
(\emph{mm} $\equiv$ missing mass, {\bf \emph{p}} $\equiv$ momentum of the outgoing protons),
(ii)~higher voltage at the drift chambers to improve the spatial resolution for
track reconstruction,
(iii)~reduced width of the cluster target stream to decrease the effective momentum spread of the beam
due to the dispersion and to improve the momentum reconstruction, and finally (iv)~measurements at five different
beam momenta to reduce the systematic uncertainties.

The principle of the measurement together with the description of
the experimental setup is given in Chapter~\ref{idea}.
Information about the calibration of the detectors used for the registration
of the protons and checks of the experimental conditions can be found in Chapter~\ref{first}.
Further, in Chapter~\ref{identyfication}
the identification of the \ppep\ reaction and the extraction of the background-free missing mass spectra is presented.
The value of \epw\ was obtained
via a comparison of the experimental missing mass distributions to the Monte Carlo generated spectra
including the value of \epw\ as a free parameter, as it is presented in Chapter~\ref{determination}
together with estimations of the statistical and systematic uncertainties.
Finally, the discussion of the achieved result and conclusions
are presented in the last chapter.

\chapter{Motivation for the determination of the total width of the
\texorpdfstring{\ep}{eta prime} meson}
\label{motivation}
The total width of an unstable particle may be defined as a full width at half maximum (FWHM)
of its mass distribution.

The first information about the observation of a meson with the mass 958~MeV/c$^2$ 
came out in
May 1964~\cite{Kalbfleisch,Goldberg}\footnote{$X_0$ was an other name
for the \ep\ meson at that time.} together with an upper limit for the total width \epw<12MeV.
Afterwards several investigations about the properties of the \ep\ meson were
performed (see~e.g.~\cite{Dauber,Duane,London}). Soon it became clear, that physics connected
with the \ep\ meson has many interesting puzzles.

One of the still unsolved problems are the values and nature of decay constants. The predictions
made on the quark flavour basis are done under the assumption,
that the decay constants in that basis follow the pattern of particle state mixing~\cite{Feldmann,Kroll}.
The quark-flavor mixing scheme can also be used for calculations of pseudoscalar transition form
factors~\cite{Huang} and the degree of nonet symmetry and SU(3) breaking~\cite{Benayoun}.
Such studies can be done via measurements or calculations of inter alia $\Gamma(\eta'\to\gamma\gamma)$
and $\Gamma(\eta'\to\rho\gamma)$.
However, the pseudoscalar mixing angle depends on the still unknown
and vigorously investigated
gluonium content of the \e\ and \ep\ wave functions~\cite{Bass,Bass2,Moskal,Moskal2,
Aloisio,Escribano,Ambrosino,Biagio,Fritzsch,JKlaja}. 
There are indications about large contributions
of glue in both \e\ and \ep\ mesons~\cite{Bass,Bass2}
although at the same time there are phenomenological analyses showing no evidence
of a gluonium admixture in these mesons 
~\cite{Escribano}.
On the quark flavour basis the physical states \e\ and \ep\ are assumed to be a linear combination
of the states $|\eta_q\rangle\equiv 1/\sqrt{2}|u\bar u+d\bar d\rangle\ ,
|\eta_s\rangle\equiv |s\bar s\rangle\ , \textrm{and}\ |G\rangle\equiv |\mbox{gluonium}\rangle\ $~\cite{Escribano}:
\begin{equation}
\label{gluoniumeq}
|\eta\rangle=X_\eta |\eta_q\rangle+Y_\eta |\eta_s\rangle+Z_\eta |G\rangle\ ,\qquad
|\eta^\prime\rangle=
X_{\eta^\prime}|\eta_q\rangle+Y_{\eta^\prime}|\eta_s\rangle+Z_{\eta^\prime}|G\rangle\ ,
\end{equation}
where $X^{2}+Y^{2}+Z^{2}=1$, and a possible gluonium component corresponds to $Z^{2}>0$.
Experimental results indicate values which differ from zero:
$Z^{2}_{\eta'}=0.06^{+0.09}_{-0.06}$~\cite{Aloisio}, 
$Z^{2}_{\eta'}=0.14\pm0.04$~\cite{Ambrosino},
$Z^{2}_{\eta'}=0.11\pm0.04$~\cite{Biagio}.
Here the values of $\Gamma(\eta'\to\gamma\gamma)$
and $\Gamma(\eta'\to\rho\gamma)$ are important as constraints for $X^{2}_{\eta'}$ and $Y^{2}_{\eta'}$.
However, the value of \epw\ recommended by the PDG ($\Gamma_{\eta'}^{\textrm{PDG}}$) is strongly correlated
with $\Gamma(\eta'\to\gamma\gamma)$ as the most precise determined quantity contributing to the fit procedure~\cite{pdg},
what causes problems when both $\Gamma_{\eta'}^{\textrm{PDG}}$ and $\Gamma(\eta'\to\gamma\gamma)$ 
are needed for the interpretation of the results~\cite{Biagio}.
A direct measurement of \epw\ would allow to determine partial widths independently of $\Gamma(\eta'\to\gamma\gamma)$.

The precise determination of \epw\ will also allow to establish more precisely partial
widths, useful in many other interesting investigations. For example,
the partial widths of $\eta'\to\pi^{+}\pi^{-}\pi^{0}$ and $\eta'\to\pi^{+}\pi^{-}\eta$
are interesting as a tool for investigations of the quark mass difference $m_{d}-m_{u}$~\cite{Borasoy2,Zielinski,Kupsc},
which induces isospin breaking in Quantum Chromo-Dynamics (QCD)~\cite{Borasoy,Borasoy2,Beisert}.
The box anomaly of QCD, which breaks the symmetry under certain chiral transformations, together
with the axial U(1) anomaly, preventing the particle from being a Goldstone boson in the
limit of vanishing light quark masses, can be explored via anomalous decays of \ep\ into
$\pi^{+}\pi^{-}l^{+}l^{-}$ (with $l=e,\ \mu$) in a chiral unitary approach~\cite{Nissler}.
In all above considerations values of partial widths of \ep\ decays are necessary as
input values or as the cross checks for the assumptions.

From the experimental point of view the partial width can be determined either
by an extraction of the corresponding branching ratio,
or by a measurement of the ratio of the corresponding branching ratio
and the branching ratio of another decay channel. In the first method
the value of the total width has to be known,
whereas in the second approach the partial width of the second decay channel is required,
which refers again to the first method and the determination of the total width
(or to the decay into two photons)\footnote{Only
$\Gamma(X\to\gamma\gamma)$ can be derived separately due to the calculated dependence
between the production cross section of \emph{X} in two photons collisions and partial width~\cite{Poppe,Brodsky}.}.

Determinations of the total width of the \ep\ meson
via production processes
were performed in '79~\cite{Binnie} and
'94~\cite{Wurzinger}\footnote{In fact, there is a third measurement of \epw\ from
2004 obtained as a by-product during $J/\psi$ decay studies~\cite{Bai},
however, it is not used by the Particle Data Group.}
,
however, the achieved accuracy on the \mbox{$\sim$30\%} and \mbox{$\sim$50\%} level, respectively,
is not sufficient for studies discussed above.
The average value
of the two measurements is
$\Gamma_{\eta'}^{\textrm{\ average}}=(0.30\pm0.09)$~MeV~\cite{pdg}.

The indirect determination of \epw\ ($\Gamma_{\eta'}^{\textrm{PDG}}=(0.205\pm0.015)$~MeV)
based on partial widths and branching ratios,
recommended by PDG~\cite{pdg},
provides a satisfactory result due to the high number of accurate measurements
of branching ratios and of the probability of the \ep\ meson formation
in two photons collisions.
It is based on the fit of partial widths,
two combinations of particle widths obtained from integrated cross sections and on
16 branching ratios. Altogether PDG uses 51 measurements for the fit~\cite{pdg}.
The partial width of the
\ep\ meson decay into two photons is crucial in such approach and can be derived from the
following equation
(for details see e.g.~\cite{Acciarri,Karch,Williams}):
\begin{equation}
N_{\eta'}=\Gamma_{\gamma\gamma}\tilde{\sigma}(\gamma^{*}\gamma^{*}\to\eta')\mathrm{BR}(\eta'\to X)\mathscr{L}_{ee}\epsilon\ ,
\end{equation}
where $N_{\eta'}$ corresponds to the number of the \ep\ 
mesons observed in the reaction chain $e^{+}e^{-}\to e^{+}e^{-}\gamma^{*}\gamma^{*}\to e^{+}e^{-}\eta'\to e^{+}e^{-}X$,
$\Gamma_{\gamma\gamma}$ denotes the partial width of the \ep\
meson decay into two photons, $\mathrm{BR}(\eta'\to X)$ denotes the branching ratio for a measured decay channel,
$\mathscr{L}_{ee}$ is the integrated luminosity, and $\epsilon$ is the overall efficiency for the
registration of the $e^{+}e^{-}\to e^{+}e^{-}\gamma^{*}\gamma^{*}\to e^{+}e^{-}\eta'\to e^{+}e^{-}X$ reaction. However,
one needs to keep in mind, that the estimation
of the cross section ($\tilde{\sigma}(\gamma^{*}\gamma^{*}\to\eta')$) depends on the
form factor, which must be derived from theory~\cite{Lepage,Brodsky,Huang}
or from other experiments~\cite{Behrend,Aubert}.

Branching ratios are measured and therefore any theoretical prediction of partial width
can be transformed to the value of the \epw.
However, the theoretical predictions are spread over a relatively large range of values.
Older values e.g. 0.30-0.33~MeV~\cite{Fritzsch} and <~0.35~MeV~\cite{Deshpande}
are in line with the value of \epw\ extracted from the direct measurements~\cite{Binnie,Wurzinger},
whereas more recent theoretical results like e.g. 
0.20~MeV~\cite{Nissler} and 0.21~MeV~\cite{Beisert}
are consistent with the value obtained by the PDG group~\cite{pdg}.
 
As it was shown, issues concerning the \ep\ meson cover a broad part of modern nuclear and particle physics, however,
the value of the \ep\ total width as a tool for translating precise measured branching ratios
to partial widths is not well determined (average value from two measurements),
or is correlated with branching ratios (PDG fit value)
preventing it from an independent usage of this quantities
for the interpretation of various experiments.
Moreover,
based on the average or the fit procedure
there are two different values of the \ep\ total width available~\cite{pdg}.

There is another reason to perform a direct precise measurement of the \epw.
In spite of the fact that the \ep\ meson seems to be a well confirmed particle, it still does not fully match into the
quark model. All predictions and fits are done under the assumption that the \ep\ meson is in fact
a single state. However, the most precise signal of the \ep\ was observed in measurements with a mass resolution
of FWHM~$\approx1$~MeV/c$^2$~\cite{Binnie,Moskal4,Moskal8,Hibou}\footnote{In previous studies
of the \ep\ meson performed by \cc, DISTO and SPES3 groups
not dedicated for the total width determination the achieved mass resolutions were comparable with the experiment
performed in Rutherford Laboratory~\cite{Binnie} and amount to about 0.8~\cite{Moskal8}, 1.2~\cite{Moskal4,Moskal8},
1.5~\cite{Hibou}, 5.0~\cite{Khoukaz} and 25.0~MeV/c$^2$~\cite{Balestra}.
}
and one cannot \mbox{\emph{a priori}} exclude the possibility that some structure would be visible at higher precision.
Especially, since there was some confusion about a multiple structure of the
\ep\ signal~\cite{AguilarBenitez,Lasinski}
and there were already situations, where a better accuracy disclosed double "peaks"
where only one signal was predicted and observed with a poor resolution like
the signal of the $\omega$ meson decay into two pions~\cite{Maglich}
or a$_{1}$(1260)-a$_{2}$(1320) observed at CERN~\cite{Maglich}, it is always worth to
look at something more precisely.

As was shown in this chapter the present discrepancy of the values of the
total width of the \ep\ meson should be, at least partially, solved by a direct measurement
with a precision by an order of magnitude better than achieved so far.
The work presented in this thesis was motivated by the endeavour to achieve such a precision.

\chapter{Principle of the measurement -- simplicity is beautiful}
\label{idea}
%
A measurement of a particle's total width
can be performed via one of the following methods:
\begin{enumerate}
\item Extraction from the slope of the excitation function~\cite{Wurzinger}.
\item\label{life} Determination of the life time.
\item\label{indirect} Measurement of branching ratios~\cite{pdg}.
\item Direct measurements of mass distributions:
  \begin{enumerate}
  \item invariant mass distribution from a decay process;
  \item missing mass distribution from a production process~\cite{Binnie}.
  \end{enumerate}
\end{enumerate}
The determination of the total width from the slope of the excitation function
is model dependent due to the need of the knowledge of the influence from the
final state interactions between the ejectiles on the total cross section.
In case of the \ep\ meson a direct measurement
of the life time (decay length) is impossible at the present technological level,
because the investigated meson decays in the average after $10^{-21}$~s~\cite{pdg}.
Method~\ref{indirect} was used by the Particle Data Group and it mostly relies on the measurement of
the $\Gamma(\eta'\to\gamma\gamma)$ partial width.
A direct determination
of \epw\ from a decay process requires high precision
(at the level of $\sim1$~MeV), difficult to achieved at present. The last mentioned method,
based on the missing mass technique, was already used in the first and so far most precise direct measurement
of \epw~\cite{Binnie}.

In this thesis \epw\ is determined via the direct measurement of the \ep\ meson mass distribution.
For this purpose the \ep\ meson was produced in the \ppep\ reaction, which was investigated by determining
the four-momentum vectors of protons in the initial and final states, and the mass of the \ep\ meson
was derived using the missing mass technique according to the following equation:
\begin{eqnarray}
\label{mmeq}
m^{2}_{X}=|\mathbb{P}_{X}|^{2}=|\mathbb{P}_{beam}+\mathbb{P}_{target}-\mathbb{P}_{1}-\mathbb{P}_{2}|^2\ ,
\end{eqnarray}
where $m_{X}$ and $\mathbb{P}_{X}$ denote mass and four-momentum vector of the unregistered particle, respectively
and the $\mathbb{P}_{1}$, $\mathbb{P}_{2}$ stand for the four-momenta of the outgoing protons\footnote{The momentum of
the proton from the target can be neglected during missing mass calculation, because it is six orders
of magnitude smaller than beam momentum and two orders
of magnitude smaller than momentum spread of the beam~\cite{Dombrowski}.}.
%
The value of \epw\ will be derived by the comparison of the
experimental missing mass distribution with a set of
Monte Carlo generated distributions for several assumed values of \epw.

High precision can be achieved in the close-to-threshold region for the \ep\ meson creation
due to considerably reduced uncertainties of the missing mass determination
since at threshold the value of $\partial(mm)\slash\partial\boldsymbol{p}$ approaches zero (\emph{mm} =
missing mass, {\bf \emph{p}} = momentum of the outgoing protons)~\cite{ErykMgr}. Additionally
the signal-to-background ratio is higher close to threshold~\cite{ErykMgr}.

Since the experimental resolution for the missing mass determination
depends on the excess energy (Q) the
measurement was performed at several beam momenta in order to better control and reduce the systematic errors.

The experiment was conducted using the proton beam of the cooler synchrotron COSY and a hydrogen cluster target.
The outgoing protons were measured by means of the \cc\ detector. In order to decrease
the spread of the beam momentum the COSY beam was cooled.
Furthermore the missing mass resolution was improved by decreasing the horizontal target size and
taking advantage of the fact that due to the dispersion only a small portion of the beam
momentum distribution was interacting
with the target protons. Additionally as it will be discussed in detail in Section~\ref{cluster_target_sec}
the decrease of the interaction region improved the resolution of the momentum reconstruction of the outgoing protons.
\section{COoler SYnchrotron COSY}
\begin{figure}[!t]
 \begin{center}
    \includegraphics[width=0.60\textwidth]{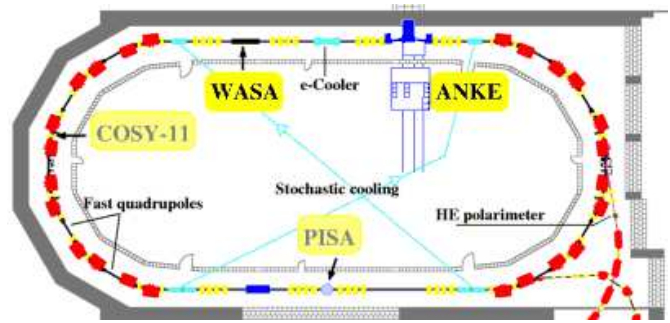}
 \end{center}
 \caption{
         Schematic view of COSY.
         Dipoles and quadrupoles are plotted as red and yellow rectangles, respectively.
         Aqua colour denotes stochastic and electron cooling devices.
         Positions of present (WASA-at-COSY and ANKE) and completed experiments (\cc\ and PISA) are shown.
         The figure is adapted from~\cite{cosywww}.
         }
 \label{COSYfig}
\end{figure}

At the COoler SYnchrotron COSY~\cite{Maier}
polarised or unpolarised proton or deuteron beams can be accelerated in the momentum range
from 600 to about 3700~MeV/c. Each kind of beam can later be used in the internal or external experiments. 
A schematic view of the accelerator part of COSY is presented in Figure~\ref{COSYfig}.
\begin{figure}[!b]
 \begin{center}
    \includegraphics[width=0.60\textwidth]{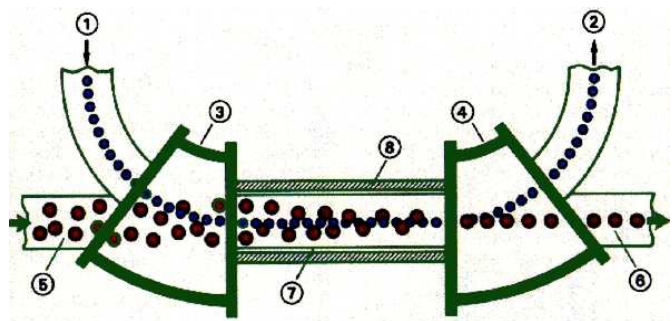}
 \end{center}
 \caption{
         The principle of the electron cooling. Bigger purple dots represent protons, while smaller blue ones - electrons.
         1. Insertion of electrons into the storage ring. 2. Extraction of the electrons.
         3. and 4. Connection of the two ion pipes via toroids. 5. Uncooled proton beam.
         6. Cooled proton beam. 7. Beam pipe. 8. Solenoid.
         The picture is adapted from~\cite{cosywww}.
         }
 \label{electronfig}
\end{figure}
The \cc\ detector was
set up\footnote{The discussed measurement was done during the last \cc\ beam time in September and October 2006.
The detector was dismounted in April 2008.} at a bending section of the synchrotron.
The COSY synchrotron is equipped with two kinds of beam cooling systems which allow for a reduction
of the momentum and of the geometrical spread of the beam.

\begin{figure}[!h]
 \begin{center}
    \includegraphics[width=0.49\textwidth]{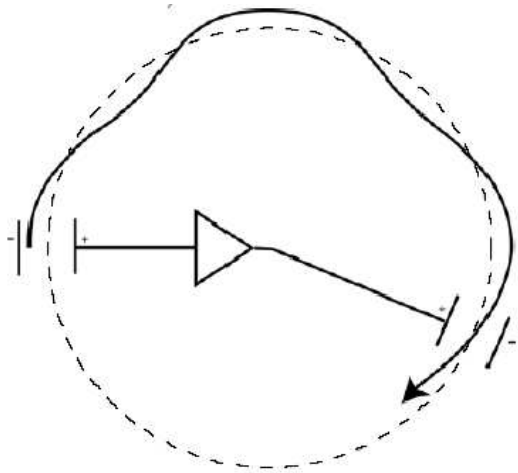}
 \end{center}
 \caption{
         The concept of the stochastic cooling. The dashed line denotes the central beam orbit,
         while the solid
         arrow represents the trajectory of some beam particles.
         The figure is adapted from~\cite{Marriner}.
         }
 \label{stochfig}
\end{figure}

The principle of electron cooling is presented in Figure~\ref{electronfig}.
The velocity of the electrons is made equal to the average velocity of the protons,
but the velocity spread of electrons is much smaller compared to the protons.
The electrons are inserted into the storage ring for a short distance where
protons undergo Coulomb scattering in the electron \emph{gas} and lose or gain energy,
which is transferred from the protons to the co-streaming electrons, or vice versa,
until some thermal equilibrium is attained~\cite{Prasuhn,Stockhorst}.
\newpage
Figure~\ref{stochfig} presents the basic concept of stochastic cooling.
It works in two steps. First, a measurement of the deviation from the nominal position of a part of
the beam is performed, then
this information is sent (by using a shorter way across the ring than the beam takes itself)
to the opposite part of the ring where the position of the measured
beam slice is corrected by electromagnetic deflection with a kicker unit~\cite{Maier}.
Accidental mixing of the particles inside the beam causes that in each cycle different groups
of the particles are corrected. The final effect occurs as a reduction of the momentum spread of the beam and as a
decrease of the size of the beam~\cite{VanDerMeer,Prasuhn,Stockhorst,Stockhorst2}.
COSY is equipped with vertical and longitudinal cooling elements which allow for the reduction of the 
emittance and decrease the momentum spread of the beam.

The above-mentioned properties of the COSY synchrotron
ensure good quality of the beam (small momentum and geometrical spread)
essential for precise measurements.
\section{Cluster target}
\label{cluster_target_sec}
\begin{figure}[!b]
 \parbox{0.40\textwidth}{
  \begin{center}
    \includegraphics[width=0.44\textwidth]{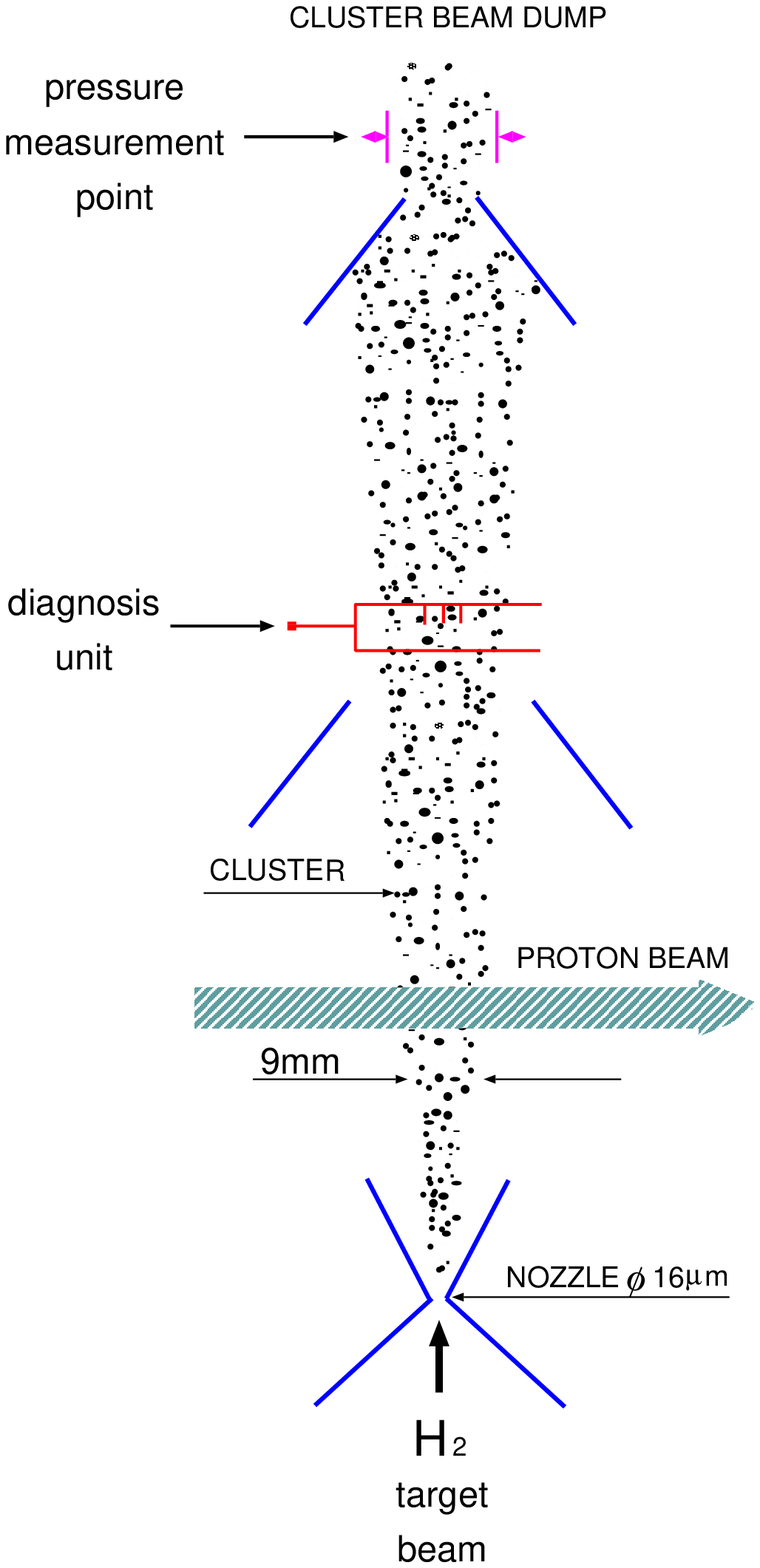}
  \end{center}
 }
 \parbox{0.60\textwidth}{
  \begin{center}
   \includegraphics[height=0.30\textwidth]{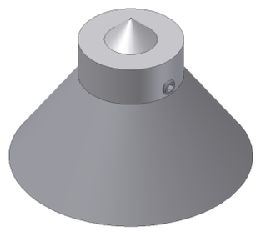}
   \includegraphics[height=0.30\textwidth,width=0.45\textwidth]{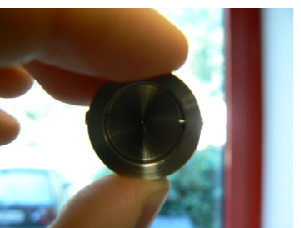}
   \includegraphics[height=0.30\textwidth,width=0.45\textwidth]{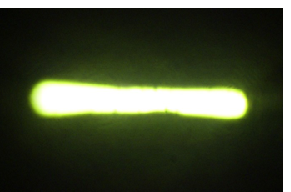}
  \end{center}
 }
 \caption{{\bf Left:} Schematic view of the cluster target setup used in the \cc\ detector setup.
                      A collimator with a $0.7\times0.07$~mm slit-shaped opening was used
                      additionally to the $\diameter=$16~$\mu$m nozzle resulting in a target width of
                      about 1~mm.
                      A description of the usage of the wire device diagnosis unit is given in Section~\ref{wire}.
                      The picture is adapted from~\cite{Moskal}.
          {\bf Right Top:} Collimator used during the \epw\ measurement. The figure is adapted from~\cite{Taeschner}.
          {\bf Right Middle:} Photo of the collimator
          from the upper part of the picture. The opening hardly is visible as a "white dot"
                        in the centre of the collimator.
          {\bf Right Bottom:} Photo of the slit in the new collimator taken with a transmitted light microscope.
                        The photo is adapted from~\cite{Taeschner}.
         }
 \label{targetfig}
\end{figure}

A cluster jet target~\cite{Dombrowski} was used in all \cc\ experiments.
The schematic view of the target setup is presented in the left part of Figure~\ref{targetfig}.
Purified hydrogen gas passes through a nozzle with an aperture diameter of $\sim$16 $\mu$m and starts
to condensate and forms nanoparticles called clusters. In order to separate the remaining gas
from the clusters, the differential pumping stages with skimmers and collimators are used.
The clusters have a divergence defined by the set of collimators
and cross the COSY beam defining the reaction region
and finally enter the beam dump.
The size of the reaction region influences
the \epw\ measurement
in two ways.
Firstly, the target setup is positioned in a bending section of the COSY ring
in a dispersive region.
It causes particles with different momenta to pass the target area at different horizontal positions.
Therefore, the size of the target stream in a dispersion
region defines the effective spread of the beam momentum,
if only the geometrical size of the stream is smaller than the beam.
Secondly, the point of the \ppep\ reaction is known only within the precision of the size of
the reaction region which is defined
as a cross section of the COSY beam and the target stream. Since the measurement of the momenta of the outgoing
protons is based on the reconstruction of their trajectories (determined from the detectors) to the centre of the reaction
region, the size of the reaction region has an influence on the accuracy of the momentum
reconstruction. Those circumstances induced us to modify the collimator of the target setup.
A slit shaped opening for the collimator was used instead of a circular opening,
in order to provide a smaller effective spread of the beam and a better reconstruction of the momenta of the
outgoing protons. The used collimator had a size of about
0.7~mm by 0.07~mm instead of a diameter of 0.7~mm~\cite{Taeschner}.
This modification ensures a decrease of the horizontal size of the target stream
in the reaction region down to \mbox{$\sim$1~mm} in the direction perpendicular to the beam line
and \mbox{$\sim$9~mm} in the direction along the beam\footnote{In the previous experiments
the target was crossing the beam as a stream with diameter of 9~mm.}.
The size of the target stream along the COSY beam axis was not reduced since the resolution of the
momentum reconstruction is not sensitive to the spread in this direction.

In order to determine the new size and position of the target stream a special diagnosis unit was designed and used.
A detailed description of the method used for the determination of the target properties
constitutes the subject of Section~\ref{wire}.
\section{\cc\ detector setup}
\begin{figure}[!h]
 \begin{center}
    \includegraphics[width=1.0\textwidth]{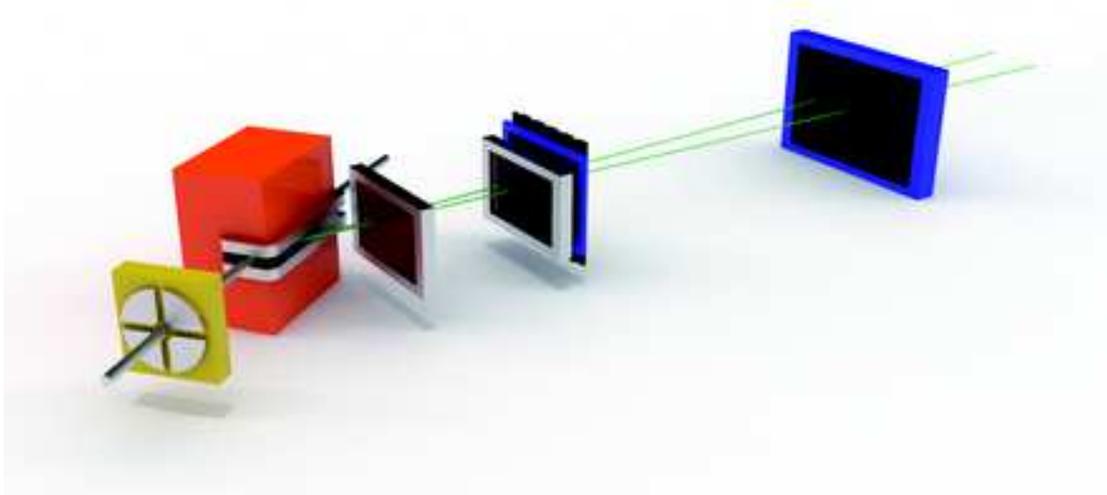}
 \end{center}
 \caption{
          \cc\ detector setup. From left to right: quadrupole (yellow) and dipole (orange) magnets of COSY,
          two drift chambers (silver) and two scintillator
          detectors (blue-black). (Picture courtesy of Barbara Wybieralska)
         }
 \label{c113d}
\end{figure}

\begin{figure}[!t]
 \begin{center}
    \includegraphics[width=0.80\textwidth]{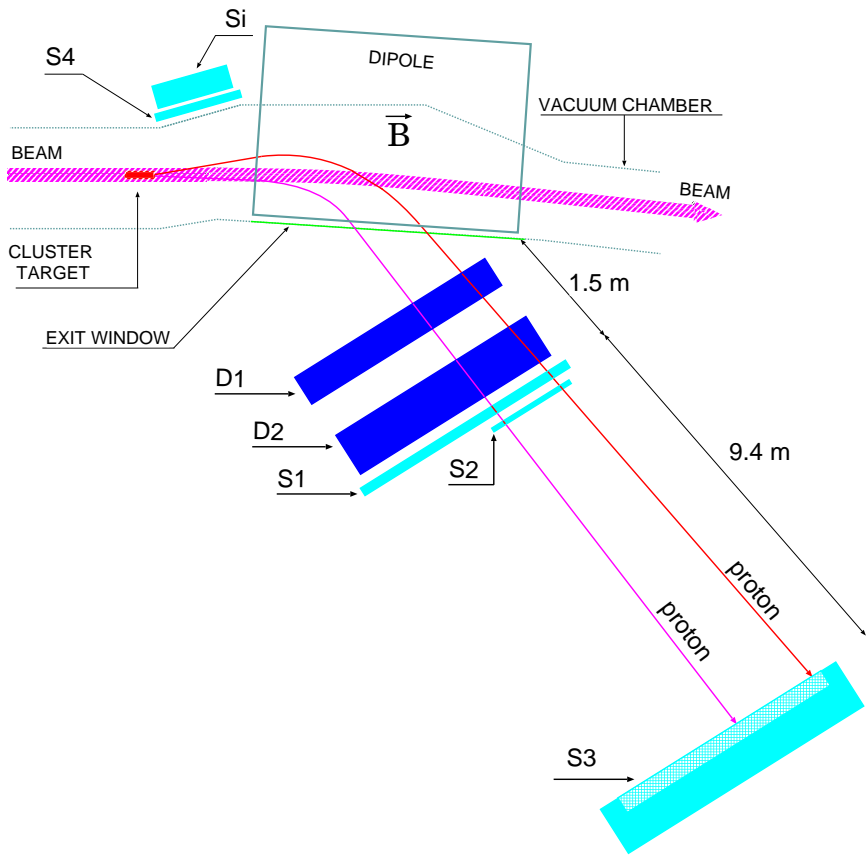}
 \end{center}
 \caption{
         Schematic view of the \cc\ detector setup (top view).
         Additionally, in comparison to picture~\ref{c113d}, 
         detectors S2, S4 and Si are shown.
         The picture is adapted from~\cite{Moskal}.
         }
 \label{c11}
\end{figure}

The \cc\ detector setup
was designed as a magnetic spectrometer used for close-to-threshold studies of the production of light mesons.
It was described in details in many previous publications
e.g.~\cite{Brauksiepe,Moskal2,Smyrski2,Czyzykiewicz,Czyzykiewicz2,Smyrski} therefore here
it is only briefly presented.
The principle of the operation of the \cc\ system is visualised in Figure~\ref{c113d} 
which shows the most important detectors for the measurement of the \ppep\ reaction.
At the left fraction of the picture
one can see a part of the COSY ring: beam pipe, quadrupole and dipole magnets.
The target setup (not shown in the figure)
is mounted between the quadrupole and the dipole magnets.
In case when a proton from the circulating beam hits a proton from the cluster target stream and
a meson is created, both protons, as a consequence of a collision, have smaller momenta than the protons in the beam.
Therefore, the reaction protons are bent stronger in the magnetic field of the dipole.
Trajectories of the two outgoing protons are shown as green traces in Figure~\ref{c113d}.
The protons leave the dipole through a special foil made of
carbon fiber layers fixed with epoxidy glue and coated with aluminium, which has a low mean
nuclear charge to reduce
straggling in the exit window~\cite{Brauksiepe}. Next, the protons fly through 
two drift chamber stacks D1 and D2 and through scintillator detectors S1, S2 and S3 (see Figure~\ref{c11}).
The measurement of the paths of the outgoing protons by means of the drift chambers
allows for the reconstruction of the trajectories back through the known magnetic field to the assumed
centre of the reaction region.
As an output of this procedure one gets the momenta of the measured particles.
Additionally, the velocity of particles is measured by the Time-of-Flight method (ToF) by means of
the scintillator detectors S1 and S3.
The information about the time when a particle crosses each detector together with the known trajectory
allows to calculate its velocity. The independent determination of particle momentum and velocity
enables its identification via its invariant mass. Since the momentum is
reconstructed more precisely than the velocity, after the identification the energy of the particle
is derived from its known mass and momentum.
The measured four-momentum vectors of the outgoing protons and the well defined properties of the beam and target
allow to calculate the mass of an unobserved particle based on the four-momentum conservation (Eq.~\ref{mmeq}).

Figure~\ref{c11} shows a schematic top view of the \cc\ detector setup. In addition to Figure~\ref{c113d}
the vacuum chamber inside the dipole, the scintillator detectors S2 and S4, as well as the part of the silicon
pad monitor detector Si are presented.
S1 and S2 consist both of 16 separate vertically oriented scintillator modules with 10~cm width for
S1 and 1.3~cm width in case of S2. The light from the scintillators is read out by photomultipliers
at the lower and upper edge of each module. The higher granularity of S2 is helpful for triggering of events
when two trajectories within one event are very close and cross
the same module of S1. In this case they are separated with S2 as long as they are not crossing the same module
of the S2.
The positioning of the S2 detector was adjusted
based on Monte Carlo simulations, prior to performing the experiment~\cite{ErykMgr}.

Detector S3 (scintillator wall) consists of one block of a $220\times100\times5$~cm$^3$ scintillator.
A light signal generated by energy loss of a charged particle inside the scintillator is read out by a matrix
of 217 photomultipliers. The centre of gravity of the signal amplitudes from individual
photomultipliers is calculated in order to determine the hit position of a particle.

Detectors S4 and Si were used for the measurement of elastically scattered protons. One of the protons
is tagged in the scintillator detector S4 and registered in the silicon
pad detector Si consisting of 144 silicon pads with the dimensions $22\times4.5\times0.28$~mm$^3$.
The pads are arranged in three layers.

For the purpose of the experiment discussed in this dissertation, additionally, to the decrease of
the horizontal target size,
the accuracy of the momentum reconstruction of the outgoing protons was also improved
at the detector level. The high voltage of the drift chambers was increased (from 1600 to 1800~V) to achieve
a better spatial resolution of the track reconstruction. This was never done
before, since such a high precision was never necessary and the new settings
of the high voltage were slightly above the standard structural safety operational level for the \cc\ drift chambers.
The applied change of high voltage caused
an improvement of the spatial resolution of the drift chambers from $\sim$250 to $\sim$100~$\mu$m.

\chapter{First steps on the way to the total width}
\label{first}
The measurement for determining the total width of the \ep\ meson
conducted by the \cc\ collaboration took place in September
and October 2006. During 23 days of data taking\footnote{This period includes 4 days break due to
a cyclotron failure.} about 360~GB of raw data were collected
from proton-proton collisions for five different beam energies.
This chapter describes the selection of events corresponding to the \ppep\ reaction
and the determination of the experimental conditions.
\section{Preselection of data}
Due to the high interaction rate
and limited data transfer
a selective hardware trigger was applied during the experiment.
The triggering of the data acquisition was based on a selection of signals from 
scintillator detectors S1, S2, S3 and S4.
The identification of the \ppep\ reaction requires the measurement of the
two outgoing protons.
Therefore the \ppep\ event candidate was stored if signals from two positively charged outgoing particles were present,
which required fulfilment of one of the following conditions:
\begin{itemize}
\item
signals from at least two modules of the S1 detector (multiplicity larger or equal to 2, $\textrm{S}1_{\mu\ge2}$).
\item
high amplitude signal from one module of the S1 detector ($\textrm{S}1_{\mu=1,\textrm{high}}$),
which corresponds to two (or more) particles
passing through one module.
\item
signals from at least two modules of the S2 detector ($\textrm{S}2_{\mu\ge2}$).
\end{itemize}
In addition to these conditions coincident signals from at least three photomultipliers (PM)
in the S3 detector were required
($\textrm{S}3_{\mu_{\textrm{PM}}\ge3}$)~\cite{Moskal}.
The complete trigger condition for a \ppep\ event candidate can be written as:
\begin{equation}
\left\{\textrm{S}1_{\mu\ge2}^{2\dots5}\lor\textrm{S}1_{\mu=1,\textrm{high}}^{3\dots5}
\lor\textrm{S}2_{\mu\ge2}^{1\dots16}\right\}\land\textrm{S}3_{\mu_{\textrm{PM}}\ge3}\ ,
\label{triggerep}
\end{equation}
where superscripts denote the range of modules taken into account for the calculations of the multiplicity $\mu$.
This range was established according to simulation of the \ppep\ reaction~\cite{ErykMgr}.
Based on the data, the threshold for the $\textrm{S}1_{\mu=1,\textrm{high}}^{3\dots5}$
signals was adjusted such that a significant amount of one track events was
discriminated with the negligible loss of two protons events~\cite{Moskal6}.

In addition, elastically scattered \pp\ event candidates were stored for monitoring target
and beam properties described in detail in Section~\ref{ellipse}.
The trigger conditions in this case required a signal
in exactly one module in the S1 hodoscope in coincidence with
one signal in the S4 detector (see Figure~\ref{c11}).

As the first step in the \emph{off-line} analysis the stored events were grouped into two categories: \pp\ and
\ppep\ event candidates. This selection was based on the signals from the drift chambers. The first group was used for
the adjustment of the position of the drift chambers, determination of the relative
beam momenta, and monitoring of the target stream properties,
while the second group was used for the calibration of the detectors and
the determination of the total width of the \ep\ meson.
The purpose of this procedure is to reduce the amount of data significantly without the application of a CPU time
consuming reconstruction.
To receive event samples
as clean as possible without using reconstruction procedures 
as a selection criterion the number of drift chamber wires with a signal above a certain threshold was used.
The drift chambers D1 and D2 consist in total of 14 planes (6 and 8, respectively).
Therefore in an ideal case 14 signals are expected for the \pp\ reaction and 28 for \ppep\, because in the
first case only one proton passes through the chambers and in the second case two protons must be registered
(see Figure~\ref{c11}). Based on the experience gained in previous \cc\ experiments~\cite{Moskal,Moskal2,Moskal3,Moskal4,
Smyrski2} the conditions for optimising the efficiency and the time of the reconstruction were obtained
when requiring that at least 12 planes responded with signals to one passing particle.
Therefore for the \pp\ event candidate additionally to signals in the S1 and the S4 detectors
at least 12 signals in drift chambers
were required. Whereas for the \ppep\ candidates signals in the S1 (or S2) and the S3 detectors and at least 24 signals
in drift chambers were demanded\footnote{For the purpose of
the described selection the number of \emph{signals} in the drift chamber in the case of the \pp\ reaction is defined as the
number of planes with at least one "fired" wire, whereas in the case of the \ppep\ reaction the number
of \emph{signals} means
the number of "fired" wires,
with the restriction that 2 or more "fired" wires in one plane are counted as exactly 2.}.

Using the above conditions the full sample of $2.1\times10^8$ registered events was reduced to
$1.1\times10^8$ \pp\ candidates and to $1.6\times10^7$ \ppep\ candidates.
\section{Calibration of detectors}
There were only two kinds of detectors used for the identification of the \ppep\ reaction:
the drift chambers (D1, D2) and the scintillator detectors (S1, S2 and S3).
In the following section their calibration based on the collected data is presented.
\subsection{Drift chambers}
The calibration of the drift chambers proceeded in three steps. First the relative time offsets between all wires
were adjusted, next the relation between the drift time and the distance to the wire was established and finally
relative geometrical settings of the drift chambers were optimised.
\subsubsection{Relative time offsets of wires}
\begin{figure}[!t]
  \begin{center}
    \includegraphics[width=0.49\textwidth]{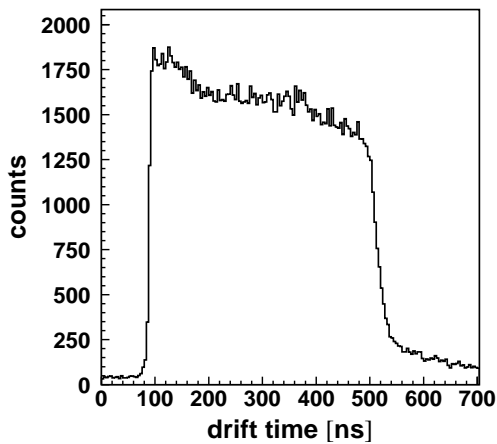}
  \end{center}
 \caption{
          Typical spectrum of the drift time for a single plane.
         }
 \label{dcwire}
\end{figure}

The measured drift time of the electrons $t_{drift}$ can be calculated from a difference
between the time signals from the drift chambers and from the S1 detector.
The arrival times of the signals from those detectors at the Time to Digital Converters (TDC)
are described by following equations:
\begin{equation}
TDC_{DC}=t_{stop}^{DC}-t_{start}^{trigger}\quad\quad
TDC_{S1}=t_{stop}^{S1}-t_{start}^{trigger}\ ,
\end{equation}
where $t_{start}^{trigger}$ denotes a common start signal for an event and $t_{stop}^{i}$ denotes
the stop signal from \emph{i}-th detector, which is a sum of the following terms:
\begin{equation}
t_{stop}^{DC}=t_{DC}^{real}+t_{drift}+C^{k}_{DC}\quad\quad
t_{stop}^{S1}=t_{DC}^{real}+\Delta t_{DC-S1}+C_{S1}\ ,
\end{equation}
where $t_{DC}^{real}$ defines the real time when the particle passes through the drift chamber,
$\Delta t_{DC-S1}$ gives the time of flight
of the particle between DC and S1, $C_{S1}$ is a constant corresponding to the time offset of the S1 detector
and $C^{k}_{DC}$ stands for the time offset of the \emph{k}-th wire of the drift chamber.
The difference between the time signals from the drift chamber and the S1 scintillator can then be written as:
\begin{equation}
TDC_{DC}-TDC_{S1}=t_{stop}^{DC}-t_{start}^{trigger}-t_{stop}^{S1}+t_{start}^{trigger}=
t_{drift}+\underbrace{C^{k}_{DC}-\Delta t_{DC-S1}-C_{S1}}_{C^{k}}\ .
\end{equation}
The $C^{k}$ offsets were adjusted based on the leading edge of the drift time spectra (see Figure~\ref{dcwire}).
The $\Delta t_{DC-S1}$ depends on the particle velocity. However, for protons from the \ppep\ reaction
(even at the largest access energy (Q~=~5~MeV) studied)
it varies only from 0.72 to 0.78 (in speed of light units)
and results in a variation of $\Delta t_{DC-S1}$ in the order of $\sim$0.3~ns,
which can be neglected in view of the 400~ns drift time for 20~mm distance (size of the one cell).
The offsets were set for each plane separately. This allows to make a single space-time calibration
for all cells in one plane.
\subsubsection{Time-space calibration}
The time-space calibration of the drift chamber
is a procedure to obtain the dependence between drift time and the distance
of the track to the sense wire. Charged particles crossing the drift chamber cause gas ionisation and
generate electron clusters moving towards the anode wires. The drift time of those electron
clusters ($t$) can be transformed to the minimum distance between the trajectory of a particle crossing the drift
chamber and the sense wire ($d$). The relation between drift time and the minimum distance ($d(t)$) has to be derived
from the experimental data and, to minimise
the influence of variations like atmospheric pressure, air humidity
and gas mixture changes~\cite{Sefzick}
on the drift velocities,
it should be determined separately for different periods of data taking. In this analysis $\sim$22-24~hours
periods were used.

\begin{figure}[!b]
  \begin{center}
    \includegraphics[width=0.49\textwidth]{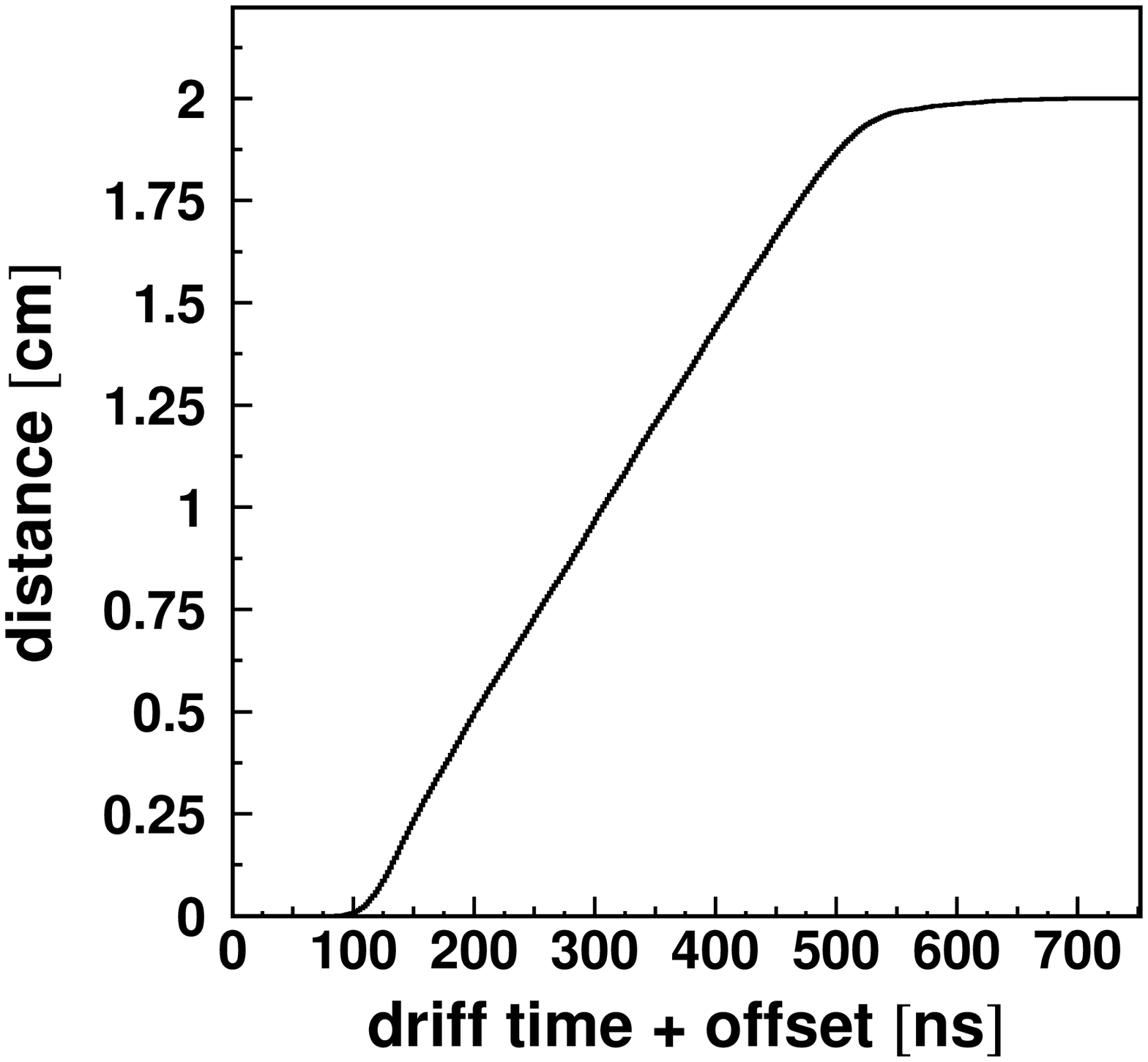}
    \includegraphics[width=0.49\textwidth]{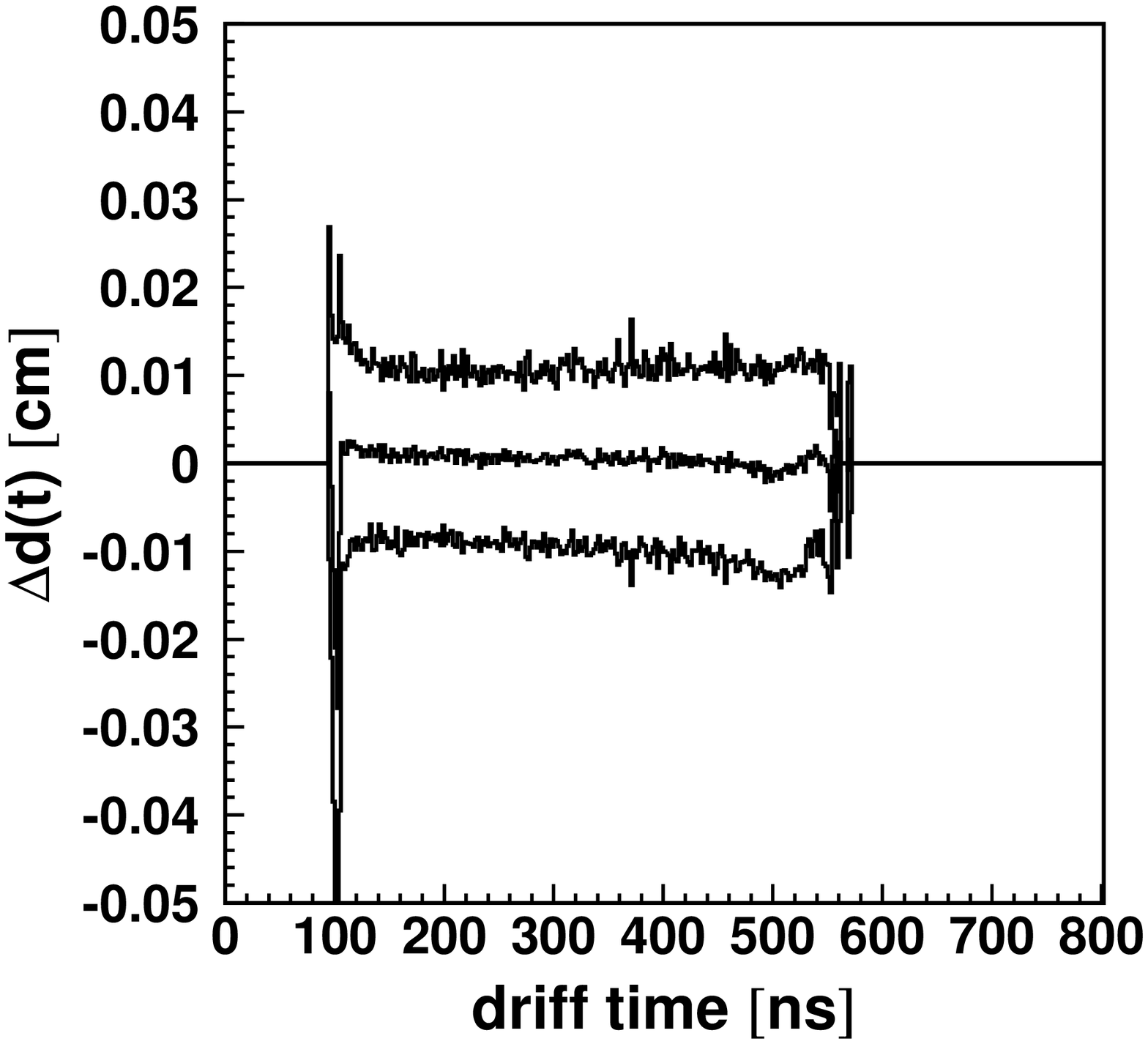}
  \end{center}
 \caption{
          {\bf Left:}~Distance from the particle trajectory to the sense wire
                      as a function of the drift time.
          {\bf Right:}~Average deviation $\Delta d(t)$ between
           the measured and the fitted distances
           of tracks from the sense wire as a function of the drift time
           as obtained after the second iteration (see text).
           The distribution around $\Delta d=0$ corresponds to the
           correction $\Delta d(t)$ and the
           lines around $\pm$0.01~cm denote one
           standard deviation of the $(d_{i}(t)-d_{i}^{fit}(t))$ distribution.
         }
 \label{dccalib}
\end{figure}

The calibration method is based on the assumption that the trajectory of a particle crossing the drift chamber
is a straight line. Starting with the approximate time-space function $d(t)$\footnote{As the approximate calibration
a function determined in the previous experiment was used.
In general one can extract the space-time relation from the shape of the drift time distribution using the
"uniform irradiation" method~\cite{Gugulski,Zolnierczuk}.}
the minimum
distance between trajectory and the sense wires were calculated.
Then, a straight line was fitted to the obtained set of points.
The minimum distance between the fitted line and the sense wire for $i$-th event is denoted as $d_{i}^{fit}(t)$.
The correction $\Delta d(t)$ of the approximate time-space function $d(t)$ has been calculated
as a function of the drift time $t$
from the
following equation:
\begin{equation}
\Delta d(t)=\frac{1}{n}\sum_{i=1}^n (d_{i}(t)-d_{i}^{fit}(t))\ ,
\end{equation}
where \emph{n} denotes the number of entries in the data sample.
Then the new calibration function was calculated as:
\begin{equation}
d^{new}(t)=d(t)-\Delta d(t)\ .
\end{equation}
The above procedure was repeated until $\Delta d(t)$ became negligible in comparison to the spatial
resolution of the chamber.
An example of the calculated time-space function for an arbitrarily chosen sense
wire in DC1 is presented in Figure~\ref{dccalib}~(left). 
In the right plot,
for an arbitrarily chosen plane of DC1,
the middle line corresponds to the average difference $\Delta d(t)$ while the upper and lower lines
denote one standard deviation of the $(d_{i}(t)-d_{i}^{fit}(t))$ distribution.
As can be inferred from the right plot of Figure~\ref{dccalib} the achieved spatial resolution
amounts to about 100~$\mu$m over the whole drift time range except for the small area very close to the sense wire.
\subsubsection{Relative positions of the drift chambers}
\label{dcpositionsection}
\begin{figure}[!b]
  \begin{center}
    \includegraphics[angle=270.,width=0.98\textwidth]{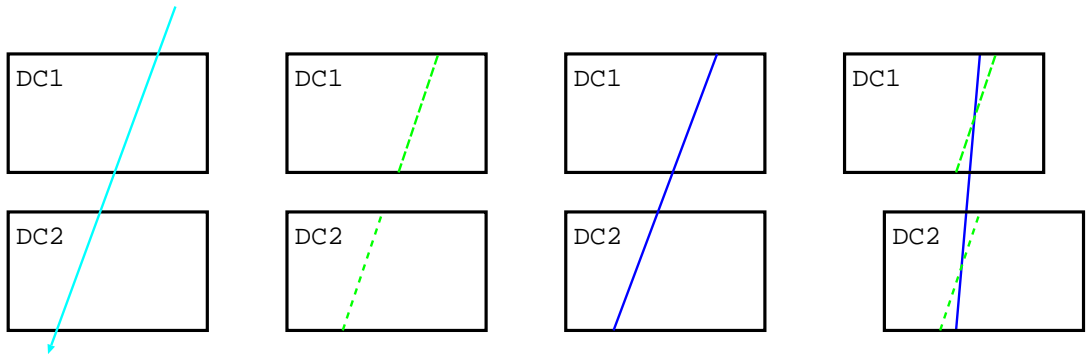}
  \end{center}
 \caption{
          The idea of the derivation of the relative position of the drift chambers.
          From left to right: (top view of the drift chambers pairs)
          a charged particle crosses two drift chambers (trajectory plotted
          as cyan arrow); positions of the trajectory (green) in each plane (derivation based on the calibration);
          straight line (blue) fitted to position information from the drift chambers, in case when
          the relative position of the drift chambers is known correctly; the same as before but in case when
          there is a discrepancy between nominal and real position of the detectors.
         }
 \label{dcshift1}
\end{figure}

The relative geometrical setting of the drift chambers was established based on the quality of the fit
of a straight line to the distances of the particle trajectory to the sense wires in both drift chambers.
The idea of the method is schematically presented in Figure~\ref{dcshift1}.
Based on the $\chi^2$ distribution of the fit the relative position of the chambers was
found to be
$\Delta x=1.4$~mm, $\Delta y=-1.8$~mm and $\Delta z=0.5$~mm (the statistical errors
are negligible).
\begin{figure}[!b]
  \begin{center}
    \includegraphics[width=0.49\textwidth]{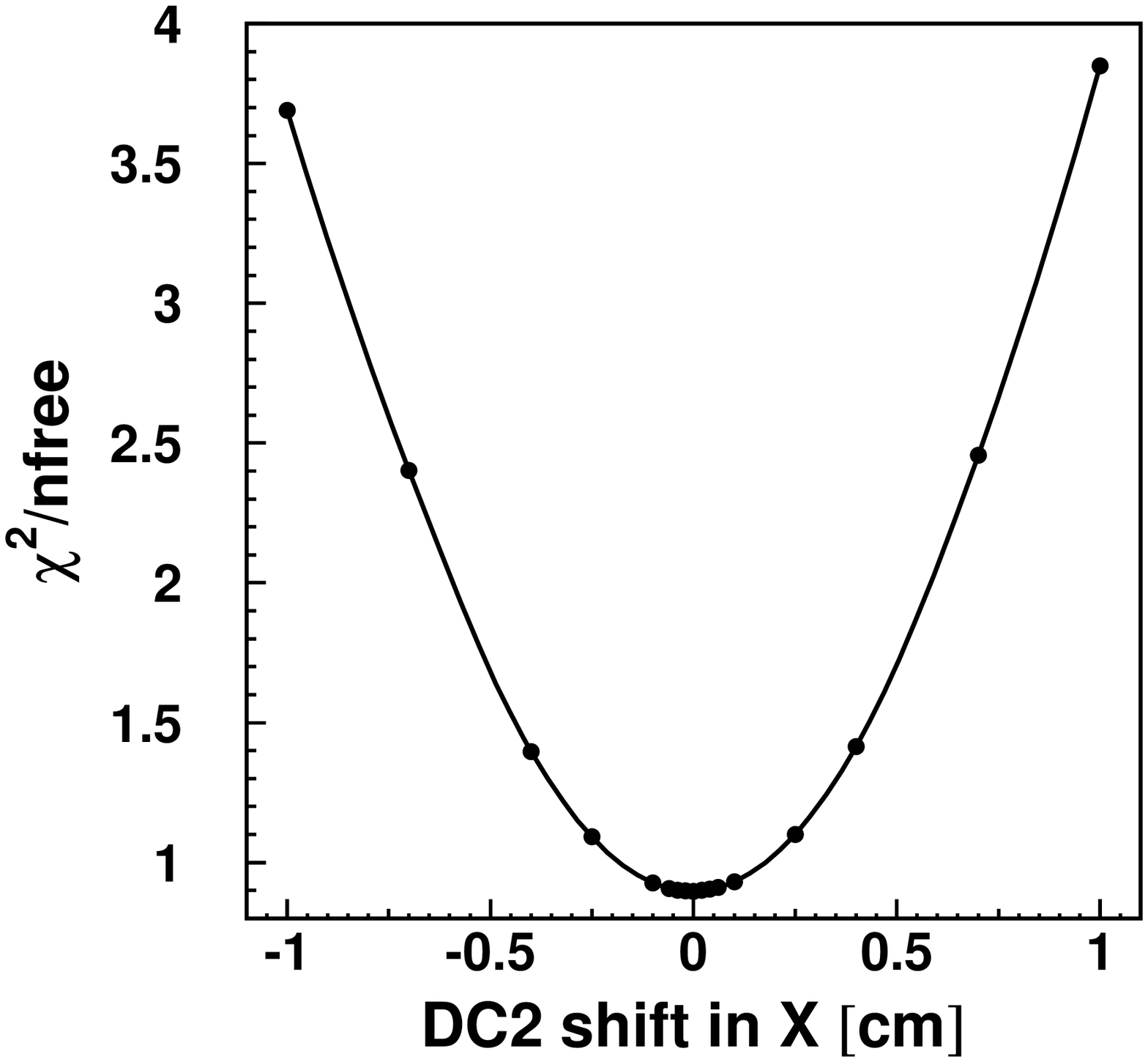}
    \includegraphics[width=0.49\textwidth]{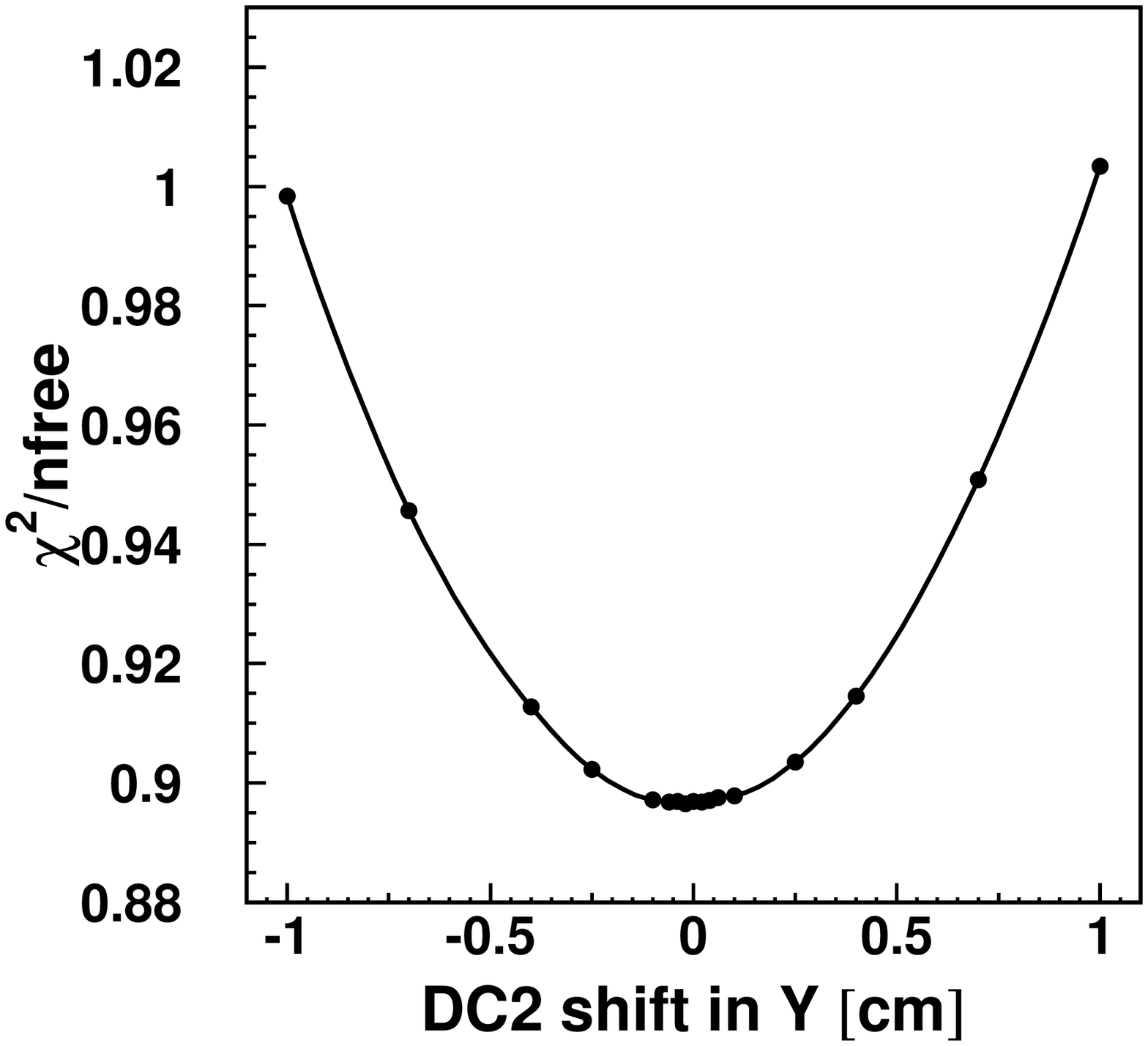}
  \end{center}
\vspace{-0.6cm}
 \caption{
           Value of reduced $\chi^2$
           for the fit of the straight line
           to the signals from both drift chambers as a function of their relative position
           in X (left) and Y (right) direction.
         }
 \label{dcshift2}
\end{figure}
Typical spectra of the $\chi^2$ values for X and Y directions are presented in Figure~\ref{dcshift2}.
The larger absolute changes of $\chi^2$
for variations of the $\Delta X$ than for the $\Delta Y$
direction correspond to the better spatial resolution
of the drift chambers in X direction compared to the Y direction. This is
due to the construction of the planes with wires oriented both vertically and inclined by $\pm$31$^\circ$~\cite{Brauksiepe}.

After the relative adjustment of the drift chambers, their position with respect to the dipole was established
based on \pp\ events.
Details are presented in Section~\ref{ellipse}.
\subsection{Timing of scintillator detectors}
The detectors S1 and S3 used for particle identification in the time-of-flight method were calibrated
in order to adjust time offsets for particular photomultipliers (PM). S1 consist
of 16 scintillating modules read out by photomultipliers on both sides, while S3 is a scintillator
wall read out by a matrix of 217 photomultipliers.
The time-of-flight is defined as the difference between times of crossing the S1 and S3 detectors
$(ToF=t_{S3}-t_{S1})$.
For the calibration of the scintillator counters
we compare the time-of-flight
obtained from signals registered in the S1 and S3
detectors and the time-of-flight calculated from the reconstructed momentum of the particle.

The experimentally available TDC values depend on the time when a particle crosses the detectors $(t_{S1},\ t_{S2})$
plus the propagation time of the created light and electrical signals.
In general, it may by expressed as:
\begin{eqnarray}
TDC_{S1}(PM) & = & t_{S1}+t(y)+t_{S1}^{walk}(PM)+t_{S1}^{offset}(PM)-t_{trigger}\nonumber\\
TDC_{S3}(PM) & = & t_{S3}+t(l)+t_{S3}^{walk}(PM)+t_{S3}^{offset}(PM)-t_{trigger},
\label{s1s3tdc}
\end{eqnarray}
where $t_{trigger}$ denotes the time of the trigger signal,
$t(y)$ denotes the time of light propagation for the distance between the cross point in the S1 module and
the scintillator edge and $t(l)$ stands for the time  of light propagation for the
distance between the hit position in S3 and the photomultiplier.
Due to the usage of leading edge discriminators
a \emph{time walk effect} is present,
i.e. a variation of the registered TDC time $t^{walk}(PM)$ as a function of the signal amplitude.
The correction of this effect can be done by applying the formula
$t^{walk}(PM)\approx constant\ \times (ADC)^{-\frac{1}{2}}$, where
\emph{ADC} denotes the signal charge value~\cite{Tanimori}.
Since the $t_{trigger}$ values are the same in both equations~\ref{s1s3tdc} for computation of \emph{ToF}
only time offset values $t^{offset}(PM)$ are unknown. However, they can be obtained by a comparison of the 
\emph{ToF} value based on the signals from scintillators $ToF_{S1-S3}$ and the time-of-flight value calculated
from the reconstructed particle momentum $ToF_{mom}=l/\beta$, where \emph{l} is the path length between
the S1 and the S3 detectors obtained
from the trajectory reconstructed in the drift chambers, and $\beta$ is the particle velocity calculated
from the reconstructed momentum with the known mass,
with the identification of the particle based on the invariant mass distribution resulting with time
offsets determined in former experiments.
Having approximate values of $t_{S1}^{offset}(PM)$ the time offsets for the photomultipliers in the S3 detector can
be determined. Then, using the determined values of $t_{S3}^{offset}(PM)$ the new set of $t_{S1}^{offset}(PM)$
can be calculated. After a few iterations the offsets for both detectors were obtained.\footnote{In case of the S1 detector,
further on in the analysis of the \ppep\ reaction the average of times from upper
and lower photomultipliers were used.
}.

\begin{figure}[!h]
  \begin{center}
    \includegraphics[width=0.49\textwidth]{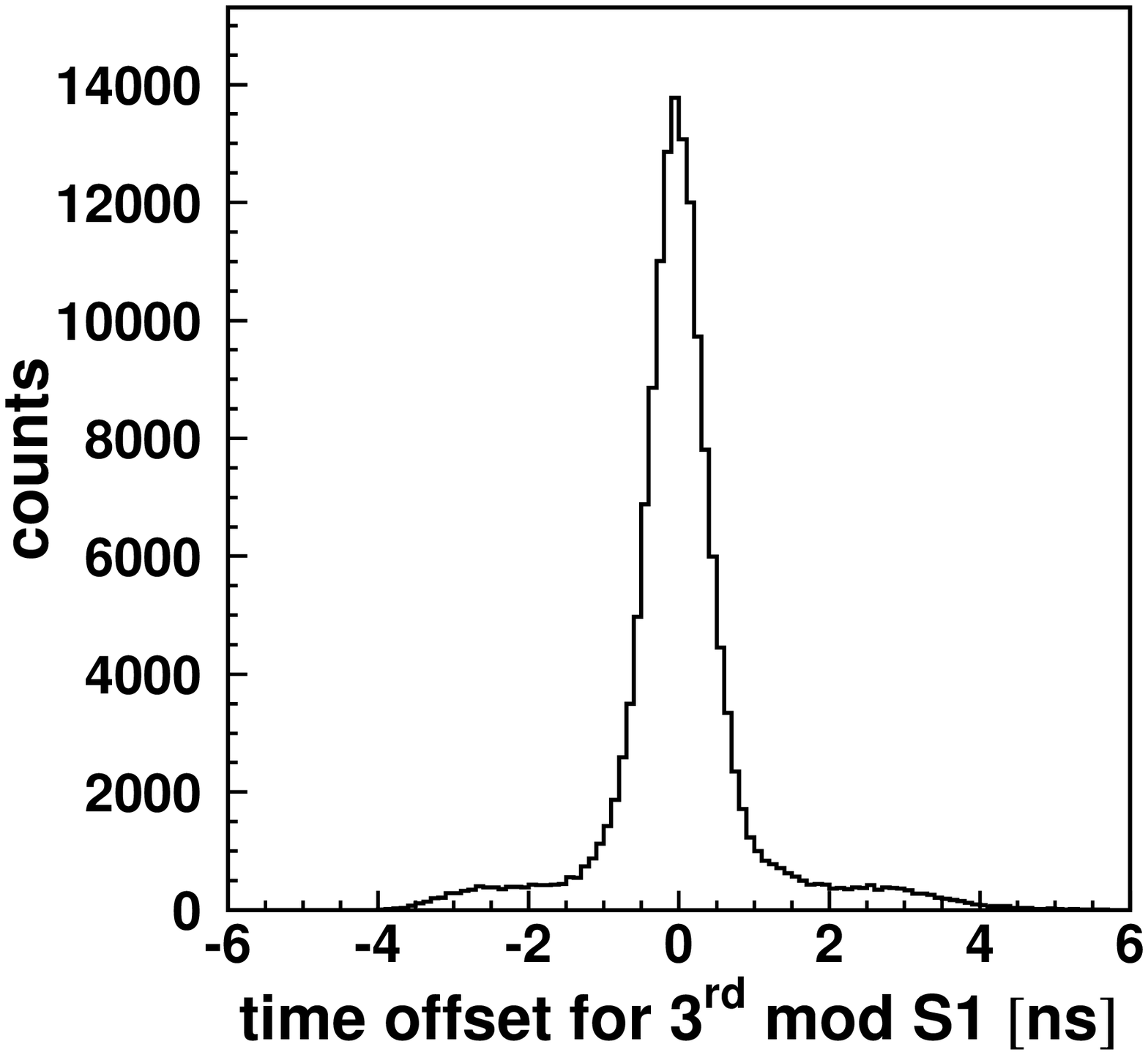}
    \includegraphics[width=0.49\textwidth]{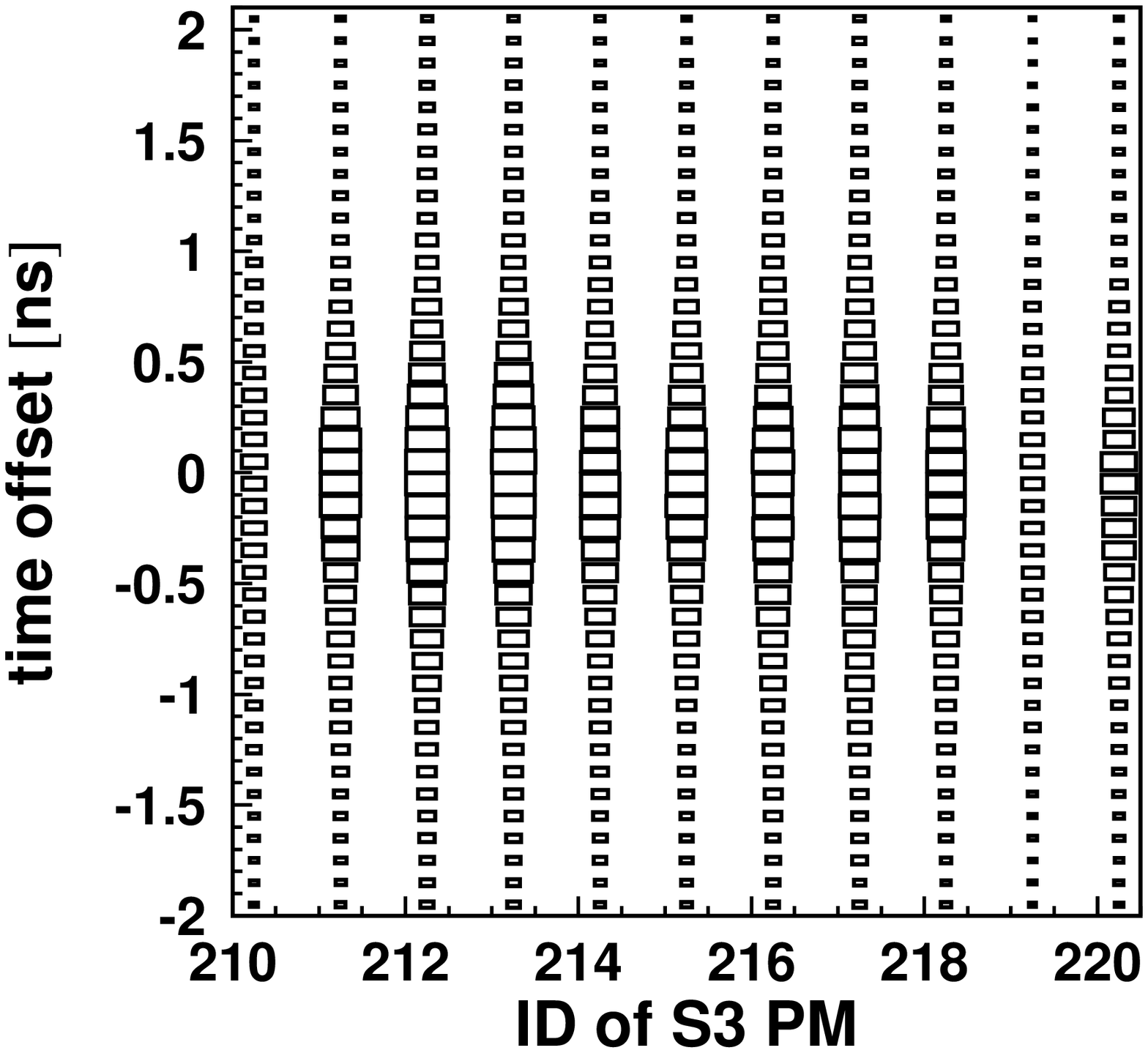}
  \end{center}
 \caption{
          Distributions of the difference determined from
          the time-of-flight measured between the S1 and the S3 detectors
          and the time-of-flight calculated from the momentum
          reconstructed based on the curvature of the trajectory in the magnetic field.
          As an example
          spectra for the 3rd S1 module and an exemplary range of photomultipliers (PM) of the S3 detector are shown.
          The counting rate of PM 210 and 219 is smaller since these photomultipliers
          are positioned at the edges of the detector.
         }
 \label{s13calib}
\end{figure}

As an example the plots in Figure~\ref{s13calib}
present results of the calibration for  arbitrarily chosen photomultipliers (PM)
of the S1 and S3 detectors.
\section{Properties of the cluster target stream}
\label{targetsection}
Since the momentum determination of the outgoing particles is based
on the track reconstruction to the centre of the reaction region
(for details see Section~\ref{protonsidentyfication}), the size and position of the target stream
influence the experimental momentum reconstruction significantly, and the accuracy of their determination
will reflect itself in the determination of systematic uncertainty of the resolution of the missing mass spectra.
Therefore, the properties of the cluster target were monitored via two independent methods:
using a dedicated diagnosis unit and inspecting a kinematic of \pp\ events.
\subsection{Diagnosis unit -- wire device}
\label{wire}
\begin{figure}[!h]
  \begin{center}
    \includegraphics[width=0.70\textwidth]{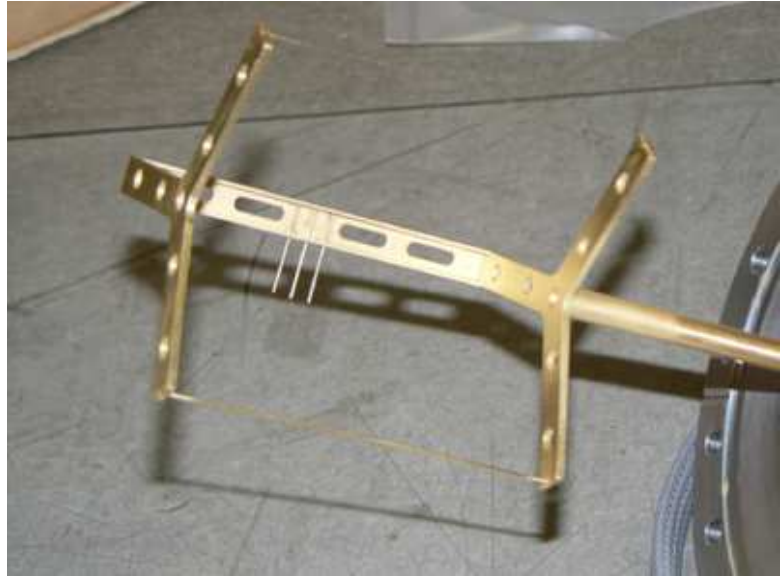}
  \end{center}
 \caption{
          Photography of the diagnosis unit. For the description see text.
         }
 \label{wiredevice}
\end{figure}

The diagnosis unit was developed for the measurement of the position and size of the target stream.
Figure~\ref{wiredevice} presents a photo of the tool.
As shown schematically in the left part of Figure~\ref{targetfig} it was installed above the reaction region downstream
the target beam\footnote{As shown
in the left part of Figure~\ref{targetfig} the target stream moves from the bottom to the top.},
allowing for the monitoring of the size and position of the target concurrently
to the measurements of the \ppep\ reaction.

The monitoring of the target properties above the beam line permits to interpolate the 
target position and size to the reaction region taking into account
the distance between collimator and reaction region ($\sim$59~cm), and the distance between
reaction region and diagnosis unit ($\sim$71~cm).

The diagnosis unit (shown in Figure~\ref{wiredevice}) consists of three \emph{arms}:
two wires with diameters of 1~mm and 0.1~mm (hardly visible in the photo)
and a broad \emph{arm}, the part with holes and three short perpendicular wires\footnote{The usage of three
perpendicular wires allowed for the determination of the target inclination.}.

\begin{figure}[!t]
\vspace{-0.5cm}
  \begin{center}
    \includegraphics[width=0.49\textwidth]{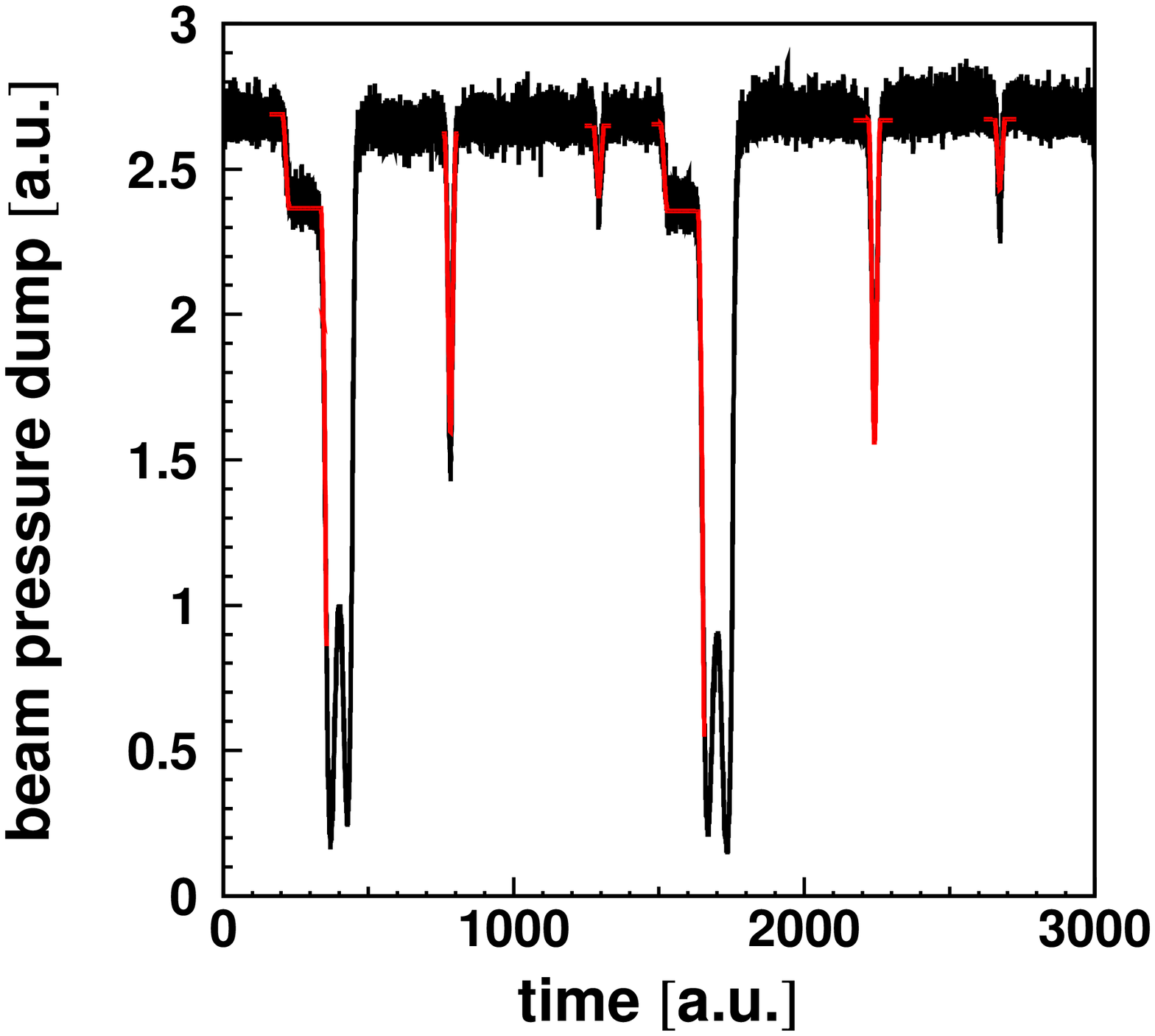}
    \includegraphics[width=0.49\textwidth]{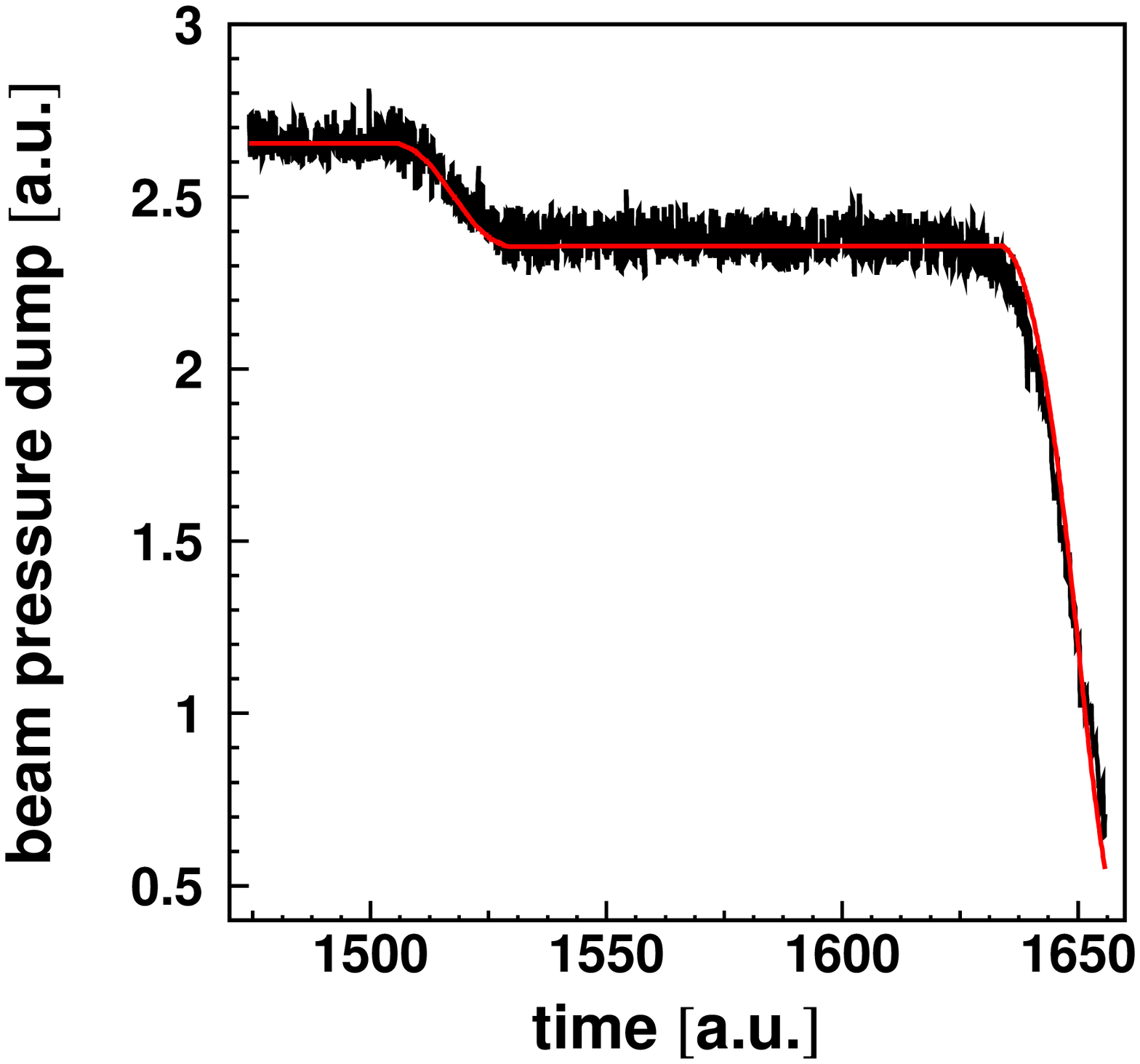}
    \includegraphics[width=0.49\textwidth]{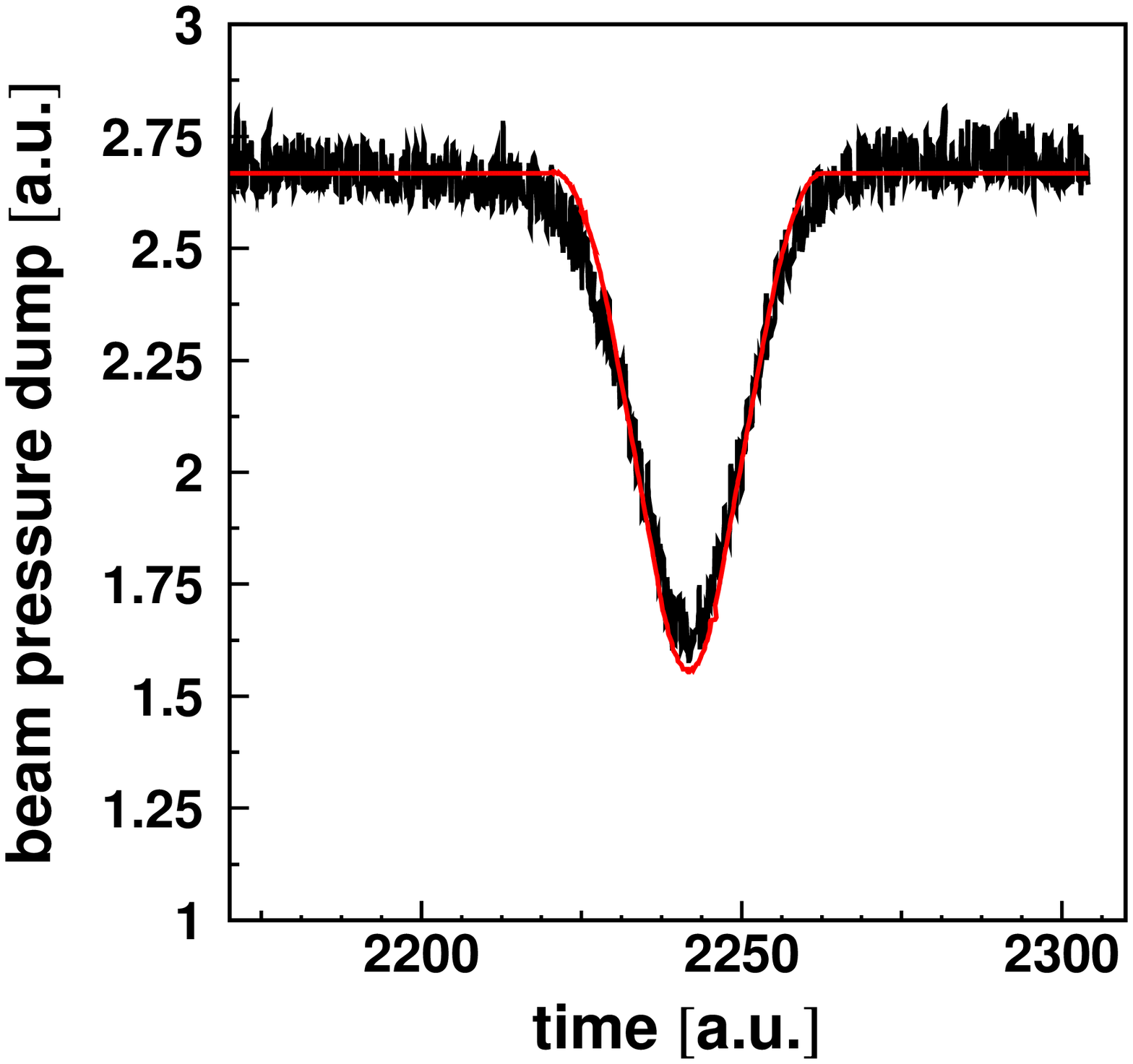}
    \includegraphics[width=0.49\textwidth]{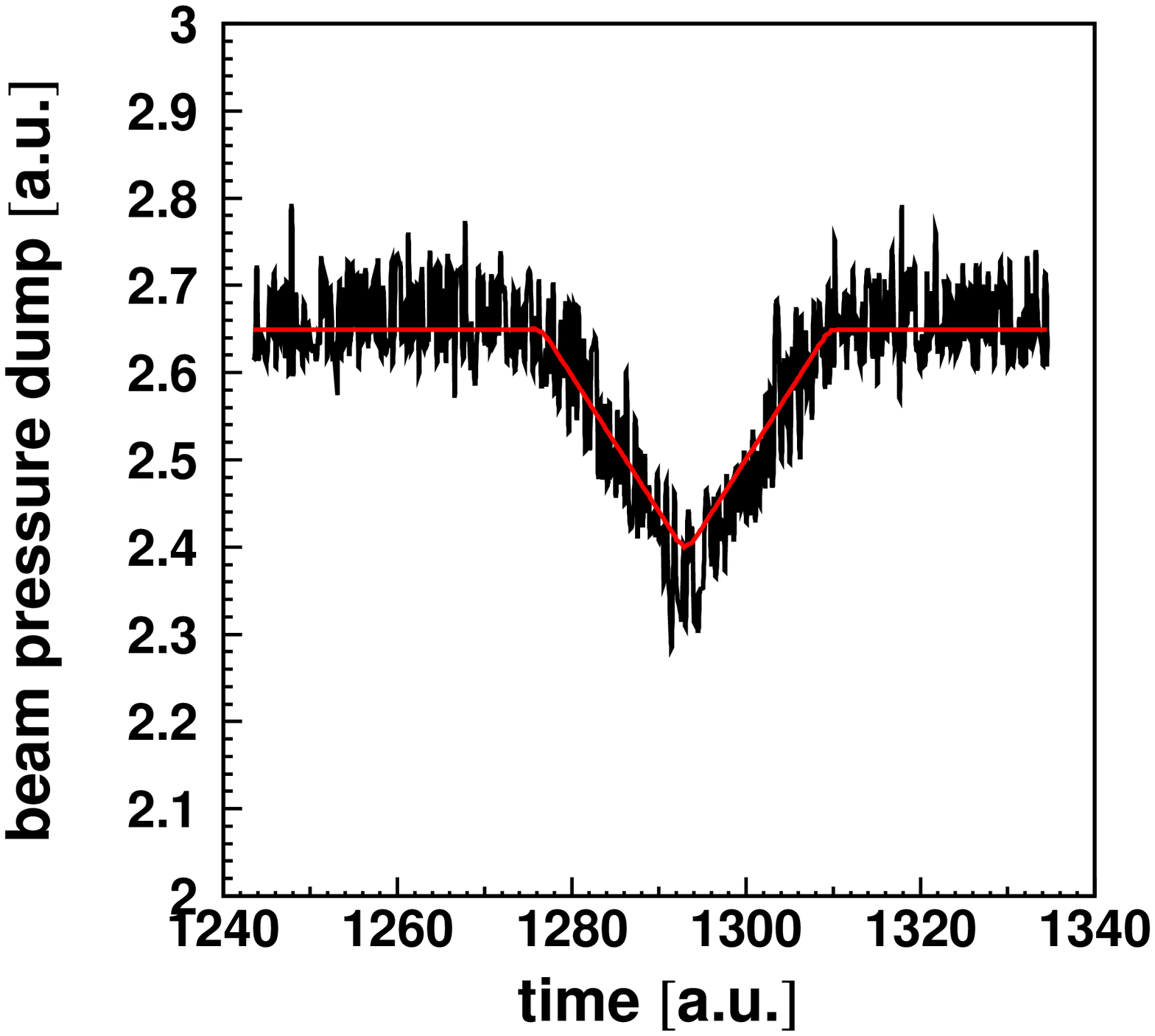}
  \end{center}
\vspace{-0.9cm}
 \caption{
          Example of changes of the target stream pressure as a function of
          the rotation time of the diagnosis unit (black). 
          Red lines correspond to the result of the simulation.
          {\bf Upper left:} Full cycle of the rotation.
          {\bf Upper right:} Close-up of the minimum due to the \emph{broad} arm and three short wires.
          {\bf Lower:} Close-ups of the minima due to the thick ({\bf lower left}) and
          thin ({\bf lower right}) wire passage.
          The width of the \emph{plateau} at the top of the pictures corresponds to the pressure fluctuations.
         }
 \label{pressures}
\vspace{-0.2cm}
\end{figure}

During the measurement the diagnosis unit rotates with constant angular velocity around the axis perpendicular to
the target stream. The \emph{arms} cross the target stream one by one, which cause changes
of the pressure in the stage above the diagnosis unit (see the left part of Figure~\ref{targetfig}).
The measured pressure values are presented in the upper left part of the Figure~\ref{pressures}
as a function of time.
The sixfold structure visible in the plot corresponds to the different \emph{arms} of the diagnosis unit
crossing the target stream (each \emph{arm} crosses the stream twice during a full rotation, once at the top and
a second time
at the bottom).
The rotation was realised by a step motor and the full rotation cycle took 2400
steps.
The first structure in the upper left part of the Figure~\ref{pressures} corresponds to the passage of the
broad \emph{arm} with three perpendicular wires
(the small \emph{step} at the leading edge) and the part with holes
(the double-well structure).
The next two sharp minima correspond to the crossing of the thick and thin wire, respectively.
The amplitudes of the minima differ slightly depending whether the \emph{arm} crosses the stream closer to or further from
the pressure measurement region.
The remaining plots in Figure~\ref{pressures} contain close-ups of structures \mbox{from the upper left part}.

\begin{figure}[!h]
  \begin{center}
    \includegraphics[width=0.37\textwidth]{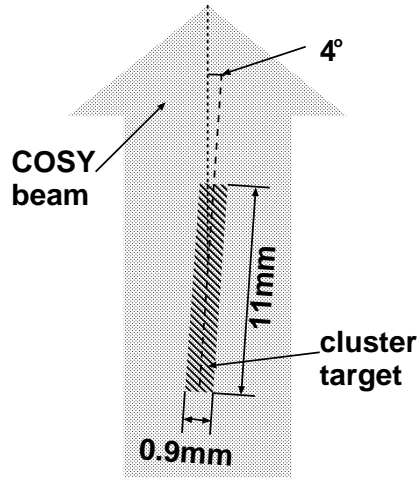}
  \end{center}
 \caption{
         Size and relative position of beam and target stream determined from the measurement
         based on the diagnosis unit. (In this thesis a Z coordinate is defined along the COSY beam line.)
         }
 \label{targetshow}
\end{figure}

The decrease of the measured pressure is proportional to the area of the wire blocking at a given moment the stream
of the target. Therefore, knowing the size of the particular parts of the diagnosis unit and velocity of the rotation
one can simulate the relative changes of the pressure as a function of time, under the assumption
of the parameters describing the size, inclination (angle) and position of the target stream.
The comparison of the results of simulations with the measured variations of the pressure allows to establish the
parameters of the target based on the minimialisation of the $\chi^2$.
The red lines in the plots in Figure~\ref{pressures} denote result of the simulation
corresponding to the parameters for which the $\chi^2$ is at a minimum value.
The determined properties of the target stream in the reaction region are:
\begin{eqnarray}
width & = & (0.089\pm 0.005)\ cm\nonumber\\
length & = & (1.053\pm 0.005)\ cm\nonumber\\
X-position & = & (0.27\pm 0.05)\ cm\\
Z-position & = & (0.02\pm 0.05)\ cm\nonumber\\
angle & = & (4.03\pm 0.01)\ deg,\nonumber
\end{eqnarray}
where the position is calculated in the nominal target system reference frame and the angle is defined
with respect to the beam direction.
The quoted uncertainties of X and Z positions include the inaccuracy of the determination of the position
of the diagnosis unit in the reference frame of the target.
Size and relative position of beam and target stream are shown in Figure~\ref{targetshow}.

There were no changes of the target stream size, angle and X-position during the entire experimental period.
However, there were changes of the Z position in the order of 1~mm.
A quantitative discussion of this variations is presented in Section~\ref{densitysection}.

It is worth to stress that the target stream width, length and angle correspond to an effective target width
of 1.06~mm.
\subsection{Kinematic ellipse from \texorpdfstring{\pp}{pp-->pp} events}
\label{ellipse}
The second and independent method used for the determination of the target stream properties is
based on the measurement
of the momentum distribution of elastically scattered \pp\ events.

\begin{figure}[!h]
  \begin{center}
    \includegraphics[width=0.48\textwidth]{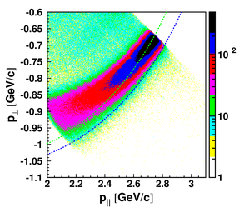}
    \includegraphics[width=0.48\textwidth]{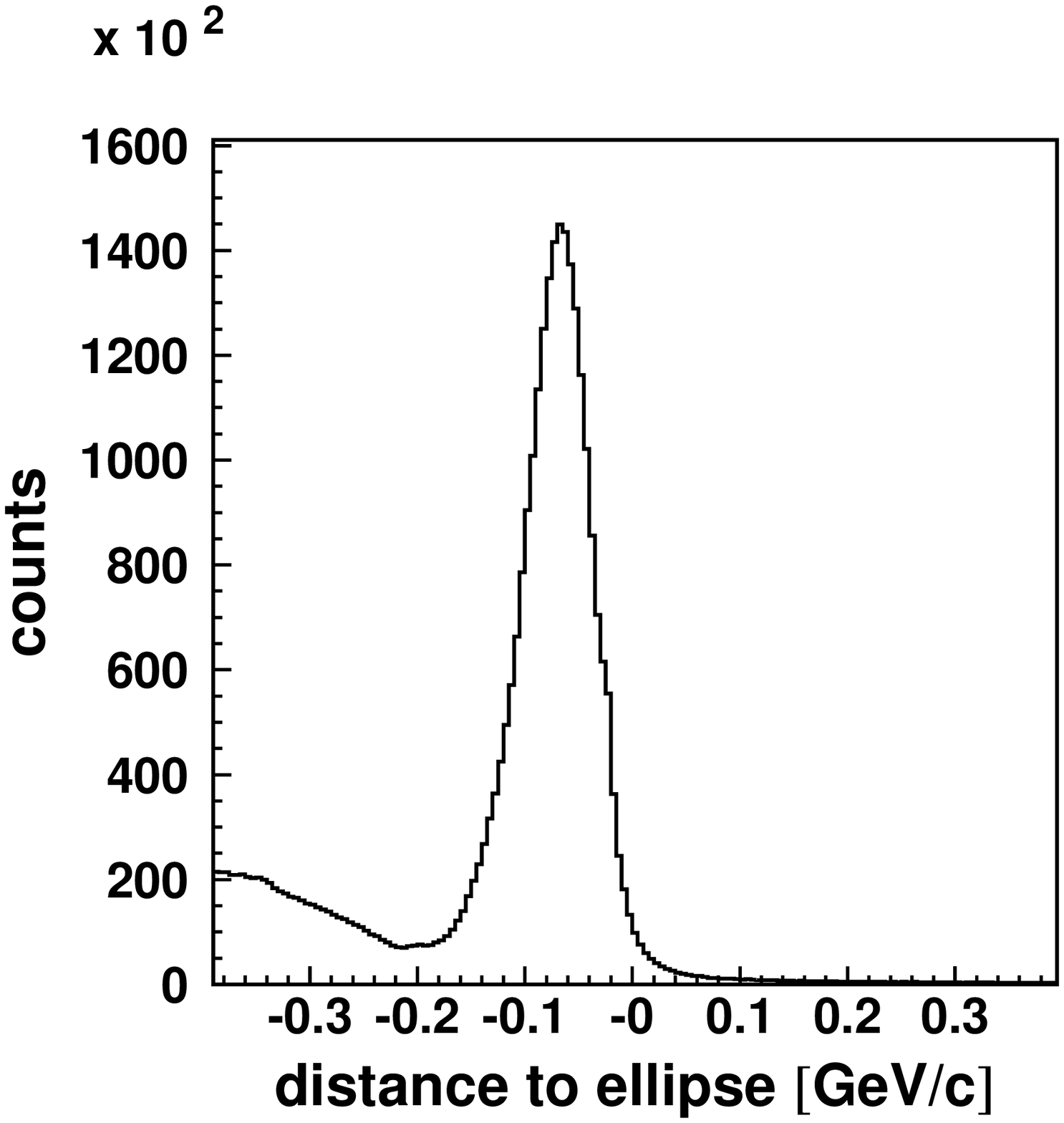}
    \includegraphics[width=0.48\textwidth]{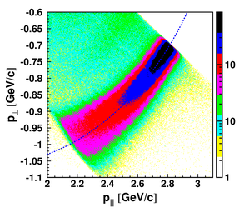}
    \includegraphics[width=0.48\textwidth]{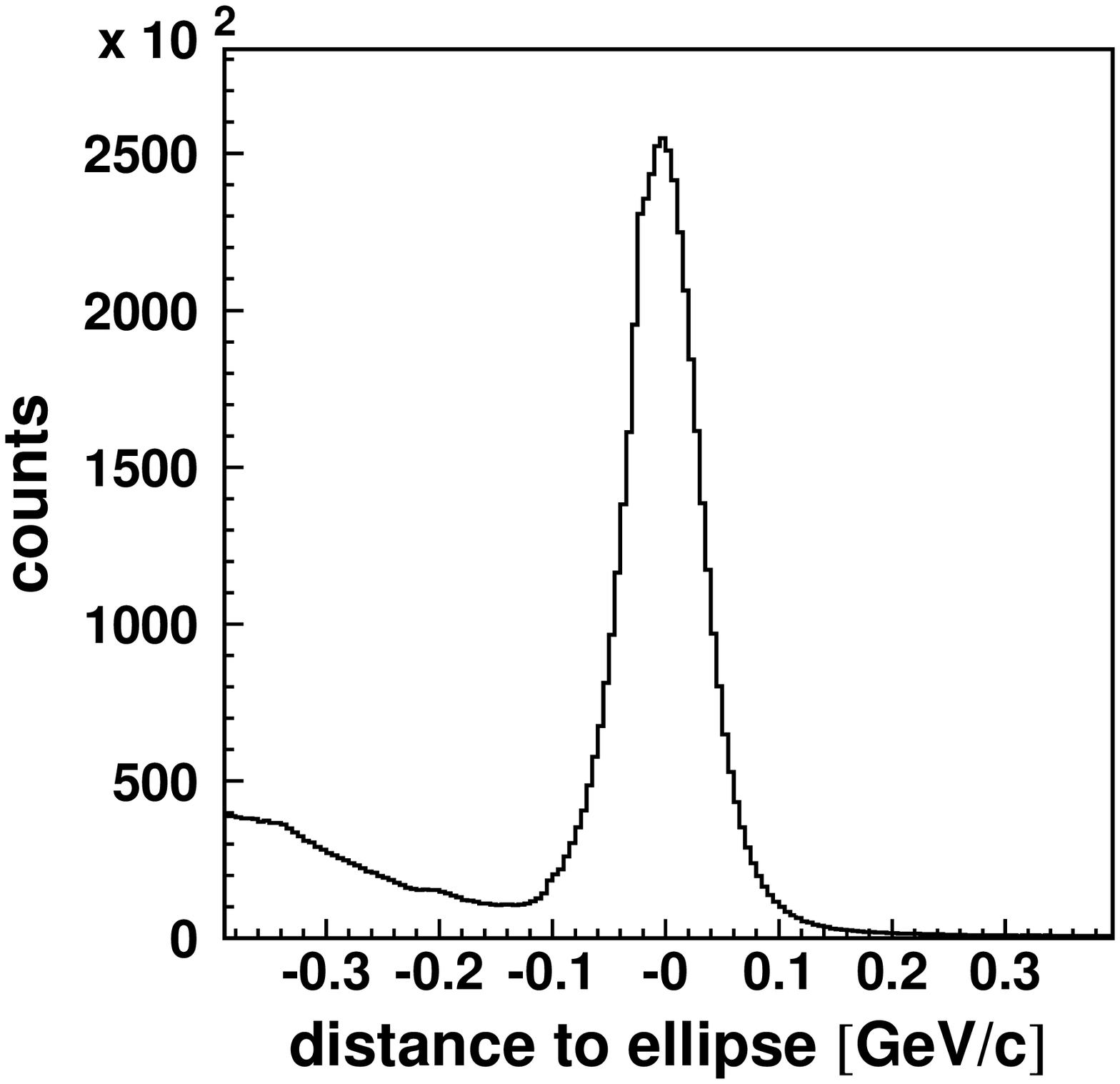}
  \end{center}
\vspace{-1.0cm}
 \caption{
         {\bf Upper left:} Part of the experimental kinematic ellipse from \pp\ events analysed for
         the nominal target stream position (x~=~0, z~=~0).
         The smooth change of the amplitude of the density distribution reflects the strong angular dependence
         of the \pp\ cross section.
         Theoretical ellipses for beam momentum values
         3211~MeV/c (nominal one) and 3111~MeV/c are plotted as blue and green lines, respectively.
         {\bf Upper right:} Projection of the distribution from upper left plot along the theoretical blue ellipse.
         {\bf Lower left:} The same data as for upper left plot but analysed for
         a target stream position of x~=~2.35~mm, z~=~0.
         {\bf Lower right:} Projection of the distribution from lower left plot along the theoretical blue ellipse.
         }
 \label{ellipses}
\end{figure}

Elastically scattered protons form an ellipsoid in momentum space in the LAB system. The projection
of the momentum components (p$_\perp$~=~perpendicular, p$_\parallel$~=~parallel to the beam direction)
constitutes an ellipse.
The acceptance of the \cc\ detector allows for the measurement of the lower right part of it
(see left part of Figure~\ref{ellipses}).
\newpage\noindent
The momentum reconstruction of positively charged particles in the \cc\ detector is based on the determination of
their trajectories by means of the drift chambers and the back-reconstruction through the known magnetic field
to the reaction region (see Figure~\ref{c11}). Since the exact reaction point is known only with an accuracy
determined by the size of the
reaction region, defined as the overlap of the beam and the target stream, back-reconstruction is performed
to the centre of this region.
This causes the spread of the points around the calculated kinematic ellipse.
Naturally the spread depends on the size of the reaction region, whereas the average shift of the points from the expected
ellipse reflect a wrong assumption of the position of the centre of the reaction region.
In principle the average shift may also be due to a wrong assumption of the absolute value of the beam momentum.
However, as proven already in the previous analysis~\cite{Moskal5} it may by safely neglected
taking into account the accuracy of the absolute beam momentum determination of 3~MeV/c~\cite{Prasuhn2}.
For the illustration of the effect, the nominal beam momentum was decreased by 100~MeV/c (see Figure~\ref{ellipses}).
One can estimate that an inaccuracy of 3~MeV/c would cause a negligible effect.
The blue line denotes the expected ellipse for the nominal beam momentum of 3211~MeV/c,
the green line for 3111~MeV/c.
The projection of the experimental points along the expected ellipse
(blue line) is presented in the upper right plot in Figure~\ref{ellipses}. 

Moreover, the momentum reconstruction is very sensitive to the assumption of the centre of the interaction region.
The ellipse presented in the upper left plot in Figure~\ref{ellipses}
was derived under the assumption that the target stream is at the nominal position
(x~=~0, z~=~0, the y-position is well defined by the plane of the circulating beam).

The ellipse in the lower left plot in Figure~\ref{ellipses}
was derived from the same data as for the ellipse from upper plot, however,
the analysis was performed under the assumption
of the target centre position: x~=~2.35~mm, z~=~0.
The blue theoretical ellipse follows the shape of the data, much better than for the (wrong) x~=~0 position, which
is also visible in the projection in the lower right plot in Figure~\ref{ellipses}.

The value of the reconstructed momentum depends also on the assumed relative settings of the drift chambers, dipole magnet
and the target. Therefore the momentum distributions of the \pp\ events are also sensitive to the
drift chamber position relative to the dipole.
However, wrong assumption about the position of the drift chambers
or about the position of the target modify those distributions
in a different ways and therefore these positions can
be established independently of each other.

\begin{figure}[!p]
  \begin{center}
    \includegraphics[width=0.49\textwidth]{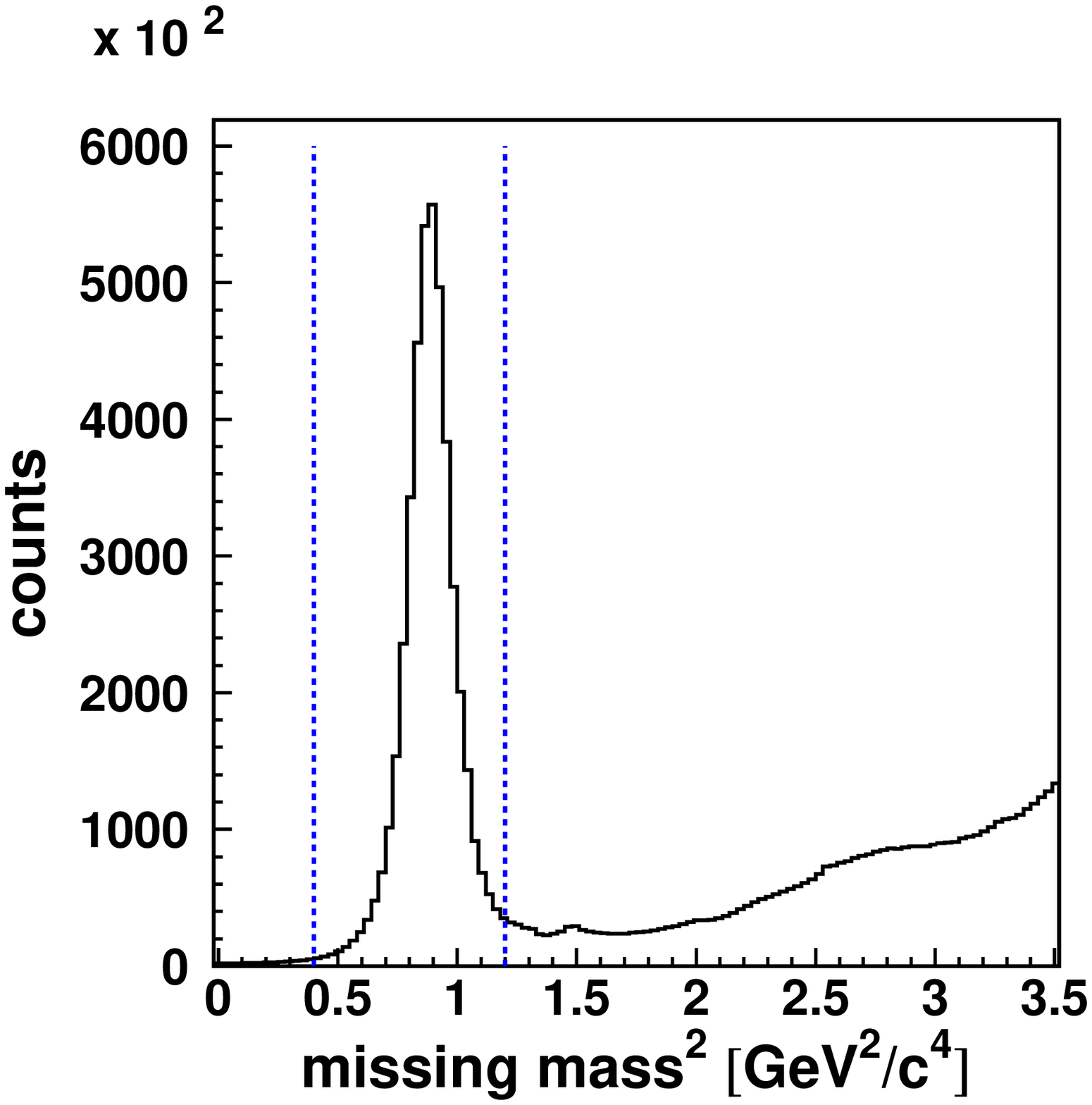}
    \includegraphics[width=0.49\textwidth]{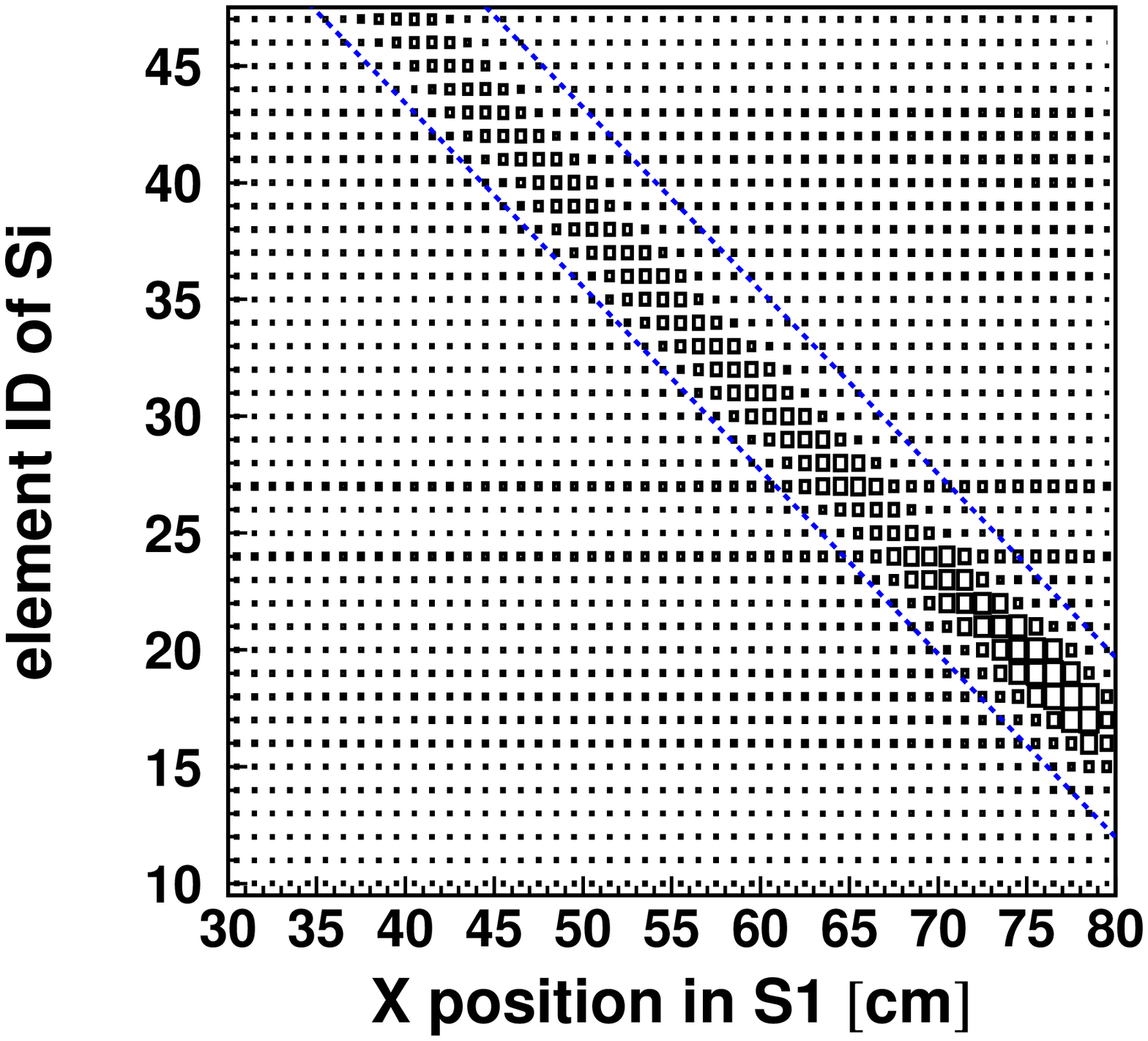}
  \end{center}
 \caption{
          {\bf Left:} Square of the missing mass to the $pp\to pX$ reaction.
          A clear signal from the protons is visible. The blue dashed lines correspond to the applied cut.
          {\bf Right:} Correlation between position of the registered particle in the S1 detector and
          the element number of the Si detector (see Figure~\ref{elasprincip}).
          The blue dashed lines correspond to the cut range.
          The relative intensity increases in channels 24 and 27 of the Si detector are due to higher noise levels
          in these detector elements.
          }
 \label{elascuts}
\end{figure}

\begin{figure}[!h]
  \begin{center}
    \includegraphics[width=0.49\textwidth]{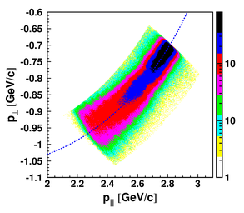}
    \includegraphics[width=0.49\textwidth]{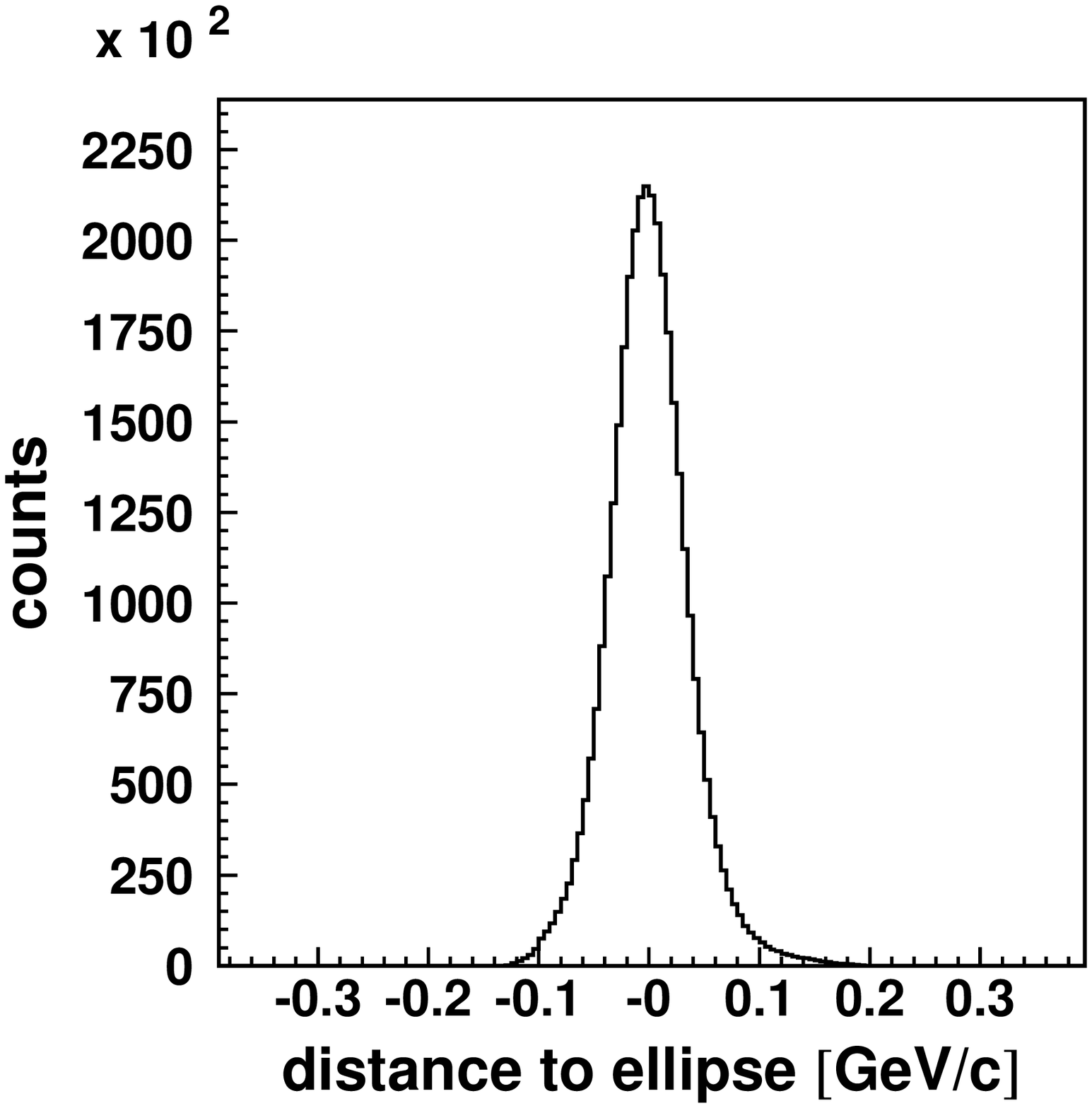}
  \end{center}
 \caption{
         {\bf Left:} Experimental kinematic ellipse from \pp\ events analysed for
         the corrected position of the target and drift chambers after application of the
         cuts on the squared invariant mass and angles correlation spectra.
         The theoretical ellipse for the nominal value of the beam momentum 
         is plotted as a blue line.
         {\bf Right:} Projection of the distribution from left plot along the theoretical ellipse.
         }
 \label{finalellip}
\end{figure}

To reduce the background contribution from the multibody production reactions
two cuts were applied. First the squared
missing mass to the $pp\to pX$ reaction 
was calculated
and then the range of the squared missing mass from 0.4 to 1.2~GeV$^2$/c$^4$ was chosen for proton selection
(see left plot in Figure~\ref{elascuts}).
\begin{figure}[!p]
  \begin{center}
    \includegraphics[width=0.78\textwidth]{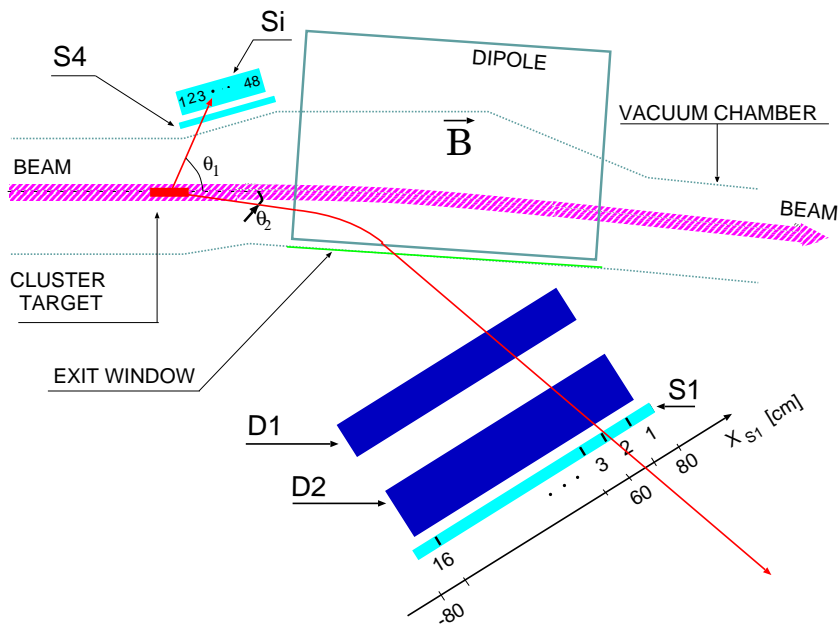}
  \end{center}
 \caption{
         Close-up of the part of the \cc\ detector used for the registration of elastically scattered events.
         The picture is adapted from~\cite{Moskal}.
         }
 \label{elasprincip}
\end{figure}
In addition, the two-body kinematics of elastic proton-proton scattering allows to combine
the scattering angles
$\Theta_1$ and $\Theta_2$ of the recoiled and forward flying protons.
In the \cc\ apparatus the scattering angles correspond to the pad number of the silicon monitor detector
(Si) and the position in the S1 detector (see Figure~\ref{elasprincip}). The correlation
is visible in the right plot in Figure~\ref{elascuts}. The cut was applied as indicated by the blue dotted lines.
The applied cuts are tight, however, the absolute number of \pp\ events is not substantial for the \epw\ analysis.
The kinematic ellipse and its projection along the theoretical
curve with adjusted position of the target after applying the mentioned cuts
is presented in Figure~\ref{finalellip}.
A negligible amount of background remained.

\begin{figure}[!t]
  \vspace{-0.5cm}
  \begin{center}
    \includegraphics[width=0.49\textwidth]{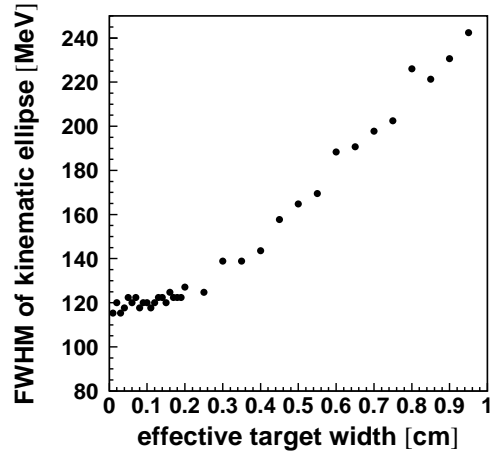}
  \end{center}
  \vspace{-1.0cm}
 \caption{
         Simulated dependence of the distribution width (FWHM) of the projection along the
         kinematic ellipse on the effective target width.
         }
 \label{fwhmVStarsize}
\end{figure}

The simultaneous comparison of the theoretical ellipses with the experimental ones derived for five different
beam momenta allows for the determination of position and effective width of the target
and position of the drift chambers.
The effective target width has an influence on the spread of points around the kinematic ellipse. However,
in practise, based on the elastically scattered events one can determine the target width only if it is greater
than $\sim0.2$~cm (as it is shown in Figure~\ref{fwhmVStarsize}).
Below 0.2~cm other effects dominate the contribution to the spread of the experimental points.
The obtained results using the described method are:
\begin{eqnarray}
effective\ target\ width & < & 0.2\ cm,\nonumber\\
target\ X-position & = & (0.235\pm 0.001)\ cm,\nonumber\\
drift\ chamber\ 1\ absolute\ position & = & (0.62\pm 0.01)\ cm,\\
drift\ chamber\ 2\ absolute\ position & = & (0.76\pm 0.01)\ cm,\nonumber\\
drift\ chamber\ angle & = & (0.045\pm 0.005)\ deg.\nonumber
\end{eqnarray}
The obtained target position and effective size
are in the good agreement with the values derived from the measurement with the diagnosis tool.
\subsection{Density distribution of the cluster target stream}
\label{densitysection}
\begin{figure}[!p]
  \begin{center}
    \includegraphics[width=0.49\textwidth]{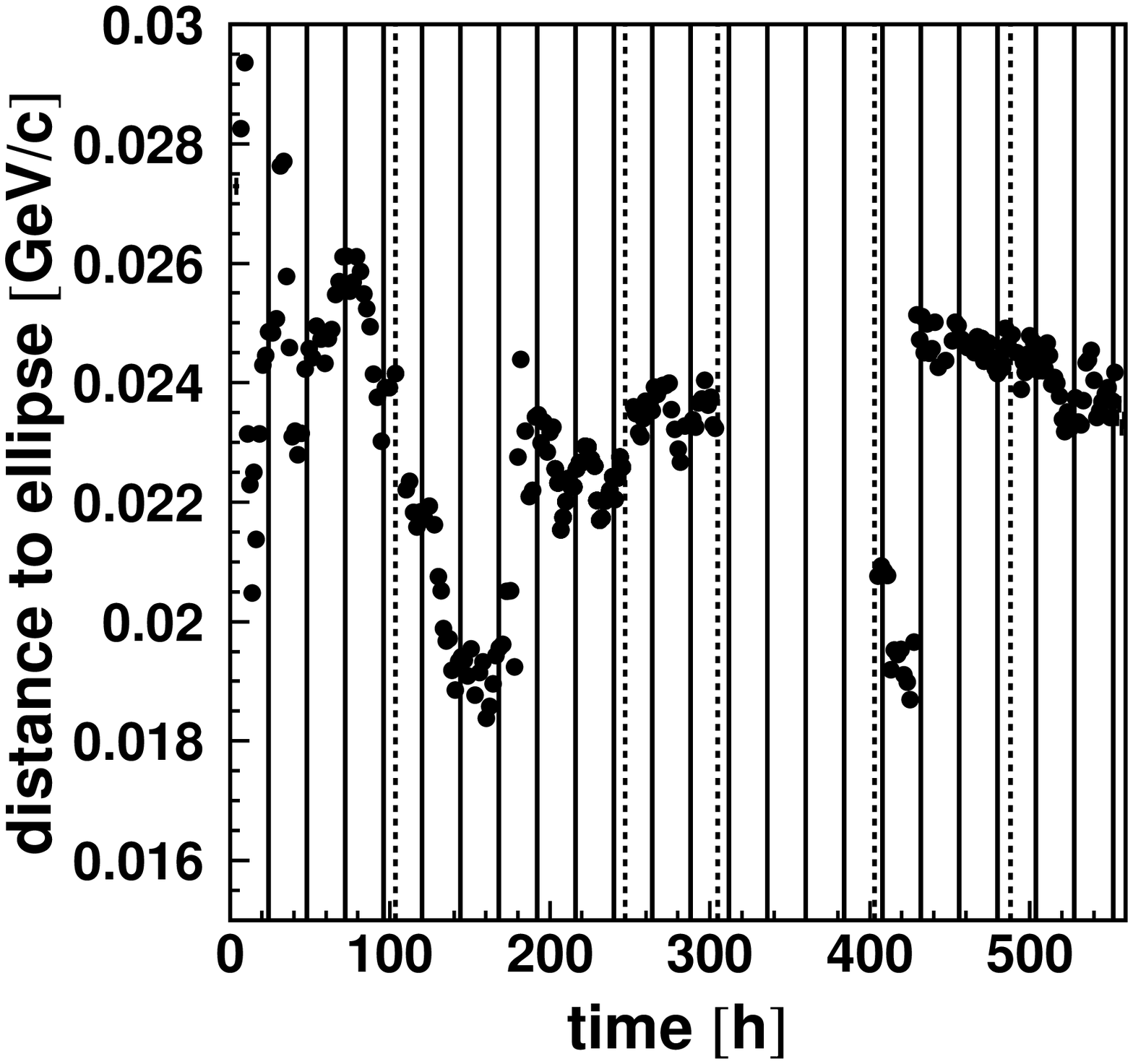}
    \includegraphics[width=0.49\textwidth]{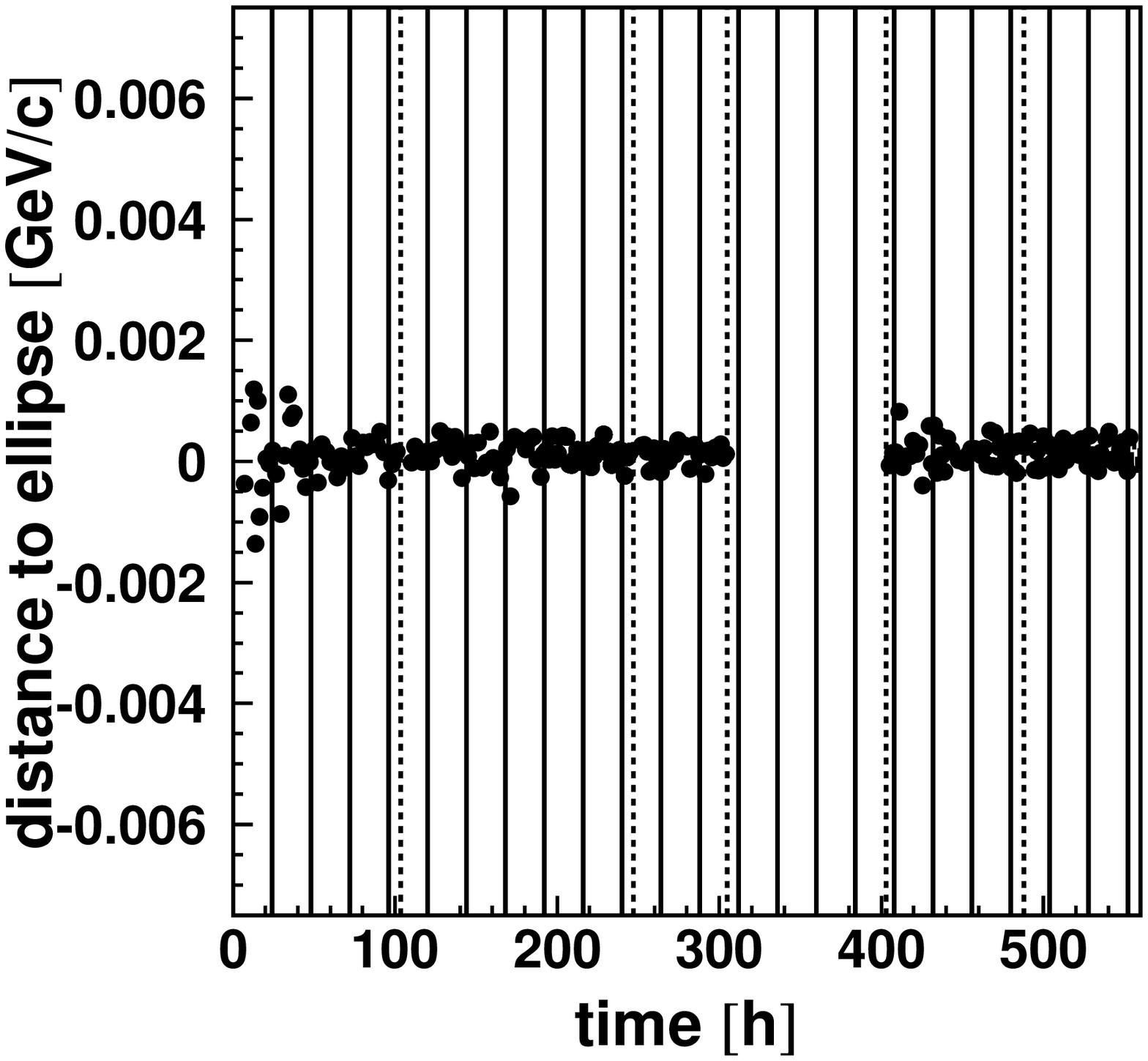}
  \end{center}
 \caption{
          {\bf Left:} Distance between the theoretical ellipse
          and the experimental distribution for the elastic kinematics plotted as a function of time.
          Points denote results averaged over $\sim$2 hours. Solid lines denote succeeding days of the
          measurement, while the dashed lines separate periods with different beam momenta
          (3218, 3211, 3214, 3213 and 3224~MeV/c).
          The values are not around 0 since at this stage the analysis was performed
          without corrections for the target and drift chamber positions as described
          in the previous section.
          The four days gap after $\sim$300 hour of measurement time is due to a cyclotron
          down time.
          {\bf Right:} The same data as presented on the left plot but analysed
          after correction of the changes of the centre of the density inside the target stream
          (see text for the details). Both plots have the same scale on the vertical axis.
         }
 \label{wahania}
\end{figure}

\begin{figure}[!p]
  \begin{center}
    \includegraphics[width=0.49\textwidth]{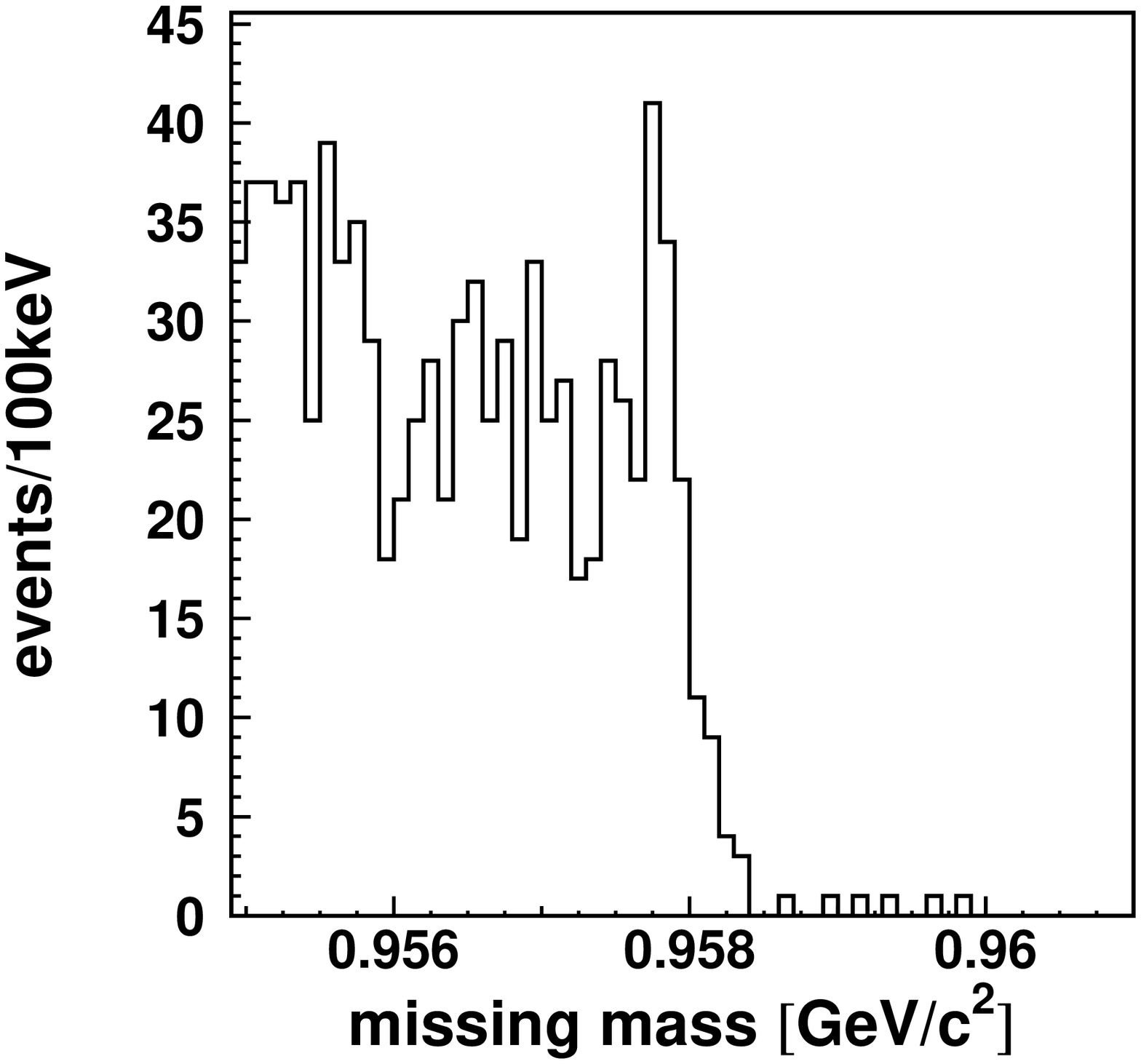}
    \includegraphics[width=0.49\textwidth]{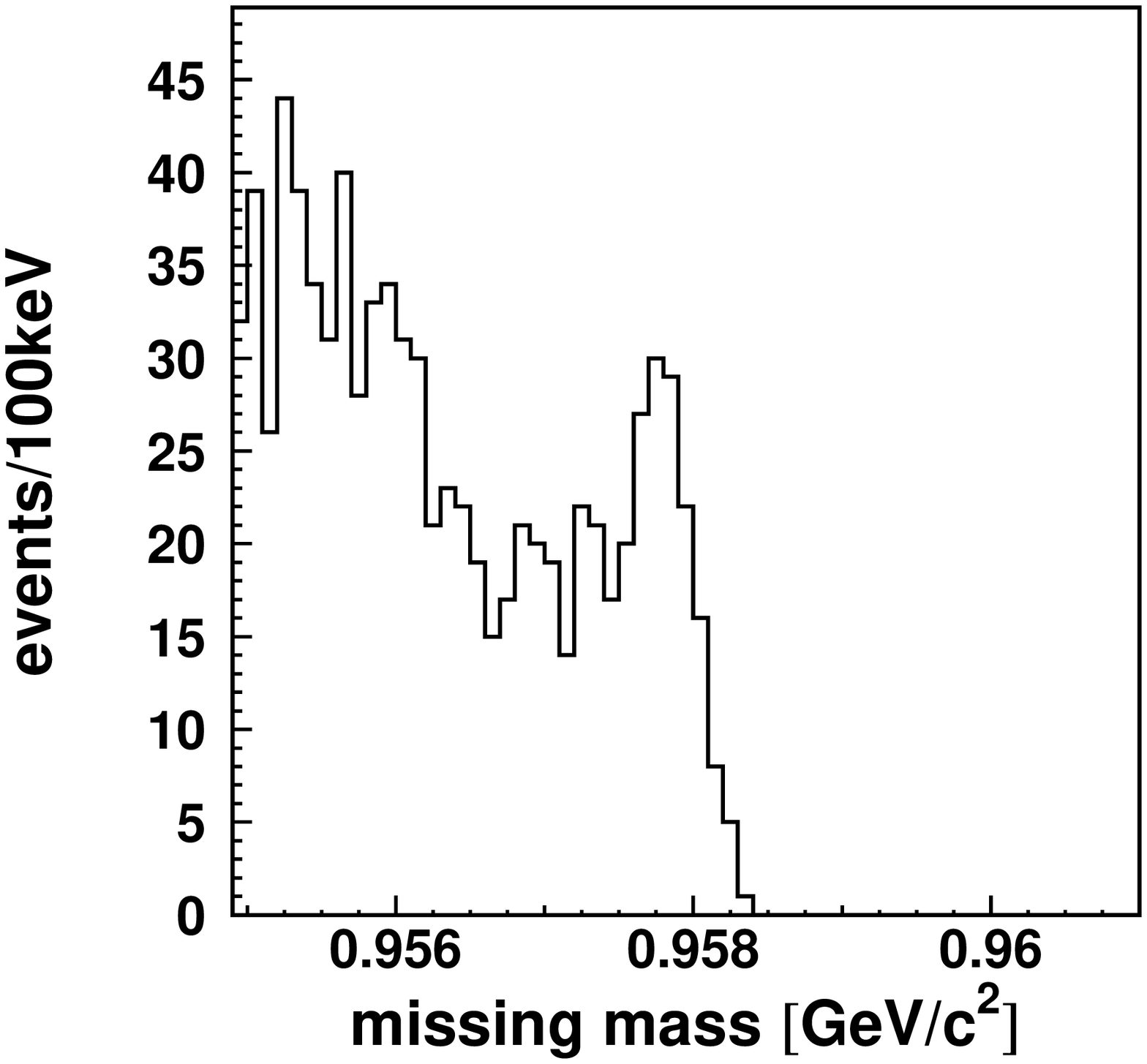}
  \end{center}
 \caption{
          Missing mass spectra extracted from data collected for 3211~MeV/c beam momentum.
          {\bf Left:} Result from $\sim$100~h to $\sim$175~h of the experiment.
          {\bf Right:} Result from $\sim$175~h to $\sim$250~h of the experiment.
         }
 \label{3211half}
\end{figure}

\begin{figure}[!h]
  \begin{center}
    \includegraphics[width=0.49\textwidth]{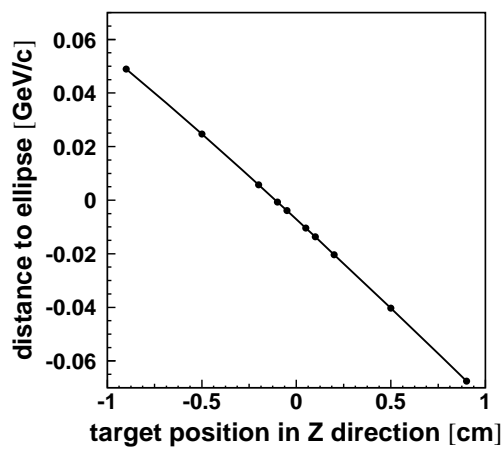}
  \end{center}
 \caption{
          Average deviation of the experimental points distribution from the expected kinematic ellipse
          as a function of target position in the direction parallel to the COSY beam.
         }
 \label{beamshiftz}
\end{figure}

Figure~\ref{wahania} shows the
changes of the average distance between the theoretical
ellipse and the experimental distribution as a function of time.
These variations correspond to changes of the centre of the density distribution of the target and hence
influence the resolution of the missing mass.
As an example of the effect,
the missing mass spectra for the \ppx\ reaction obtained for the first and second half of the measurement 
at 3211~MeV/c momentum are presented in Figure~\ref{3211half}, where the first part of the measurement (from $\sim$100~h
to $\sim$175~h)
corresponds to the large variation of the kinematic ellipse position, while the second part (from $\sim$175~h
to $\sim$250~h) to the small variations.
The presented spectra differ and the \ep\ signal is better visible in the data collected in the period with smaller
variations
(a detailed description of the missing mass technique will be presented in section~\ref{mmsection}).
Such fluctuations could be explained by variations of the beam momentum due to the changes of the
beam optics\footnote{The variations of the beam optics could be caused by e.g. the variations of the dipole
currents, or small deformation of the dipole shape caused by temperature changes.}
or by small fluctuations of the density distribution of the target stream.
However, the beam optics variation is excluded by the results of the monitoring of the stability of the
COSY beam (as described in the next section).
The observed variations of the kinematic ellipse position can be
 plausibly
explained by density
changes inside the target stream in beam direction.
In this direction the target length is about 1~cm
(see section~\ref{wire}) and expected fluctuations of the density
can cause the changes of the centre of the target stream distribution in the order of 1~mm.
Such variations along the z-axis were observed by means of the diagnosis unit as mentioned in
Section~\ref{wire}. Therefore,
in the following analysis we assumed that the observed deviations are due to the target density
variations and corrected them by continuous changes of the nominal value of the centre of the target
along the z direction
(see Figure~\ref{beamshiftz}).
The corrections were interpolated between points calculated for each $\sim$2 hours interval of beam time.
The average distance to the expected kinematic ellipse after the correction for
target density fluctuation is presented in the right panel in Figure~\ref{wahania}.
The first two days of the measurement were used for different test of the
detection system and the optimisation of the beam optics
which cause somewhat larger fluctuations during this period.
Therefore the data collected during those two days were not used for
the final \epw\ determination.
\section{Monitoring of the stability of the proton beam}
\label{schottkysection}
Although the frequency of the circulating beam is monitored routinely several times per minute, during the described
experiment a measurements of additional parameters were performed in order to provide a better control
of the stability of the proton beam.
\subsection{Synchrotron parameters}
\begin{figure}[!h]
  \begin{center}
    \includegraphics[width=0.49\textwidth]{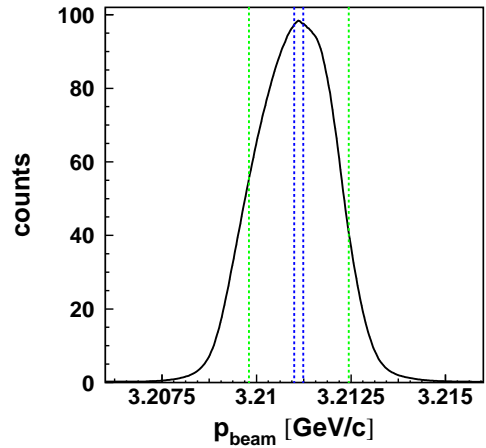}
  \end{center}
 \caption{
          Example of the momentum spectrum for the measurement with the nominal beam momentum of 3211~MeV/c.
          The dashed lines limit the effective spread of the beam due to the dispersion relation
          for a target width of 1~mm (blue) and 1~cm (green).
         }
 \label{schottky}
\end{figure}

The standard technique for monitoring the beam momentum at the COSY accelerator is the measurement of the
frequency distribution of the circulating beam.
Based on the equation~\cite{Prasuhn,Prasuhn2,Schepers}:
\begin{equation}
\frac{f-f_0}{f_0}=\eta_{beam}\frac{p-p_0}{p_0},
\end{equation}
where \emph{f} denotes
the frequency and \emph{p} denotes the beam momentum,
($f_0$ and $p_0$ correspond to their nominal values, respectively)
one can transform the frequency into the beam momentum. 
(The determination of the real value of the beam momentum is described in section~\ref{absbeammom}.)
The $\eta_{beam}$ parameter depends on the settings of the accelerator and for the described measurement
was equal to $-0.10~\pm~0.01$~\cite{Prasuhn2}.
As an example a spectrum transformed to the momentum coordinate for the lowest beam energy used in the experiment
is presented
in Figure~\ref{schottky}.
The beam momentum distribution is smooth and its spread is equal to 2.5~MeV/c (FWHM).
However, due to the position of the \cc\ target system in a bending section of the COSY ring
in a dispersive region, the effective spread of the beam (the momentum range \emph{seen} by target)
is smaller.
The dispersion relation is:
\begin{equation}
\Delta x=D\frac{\Delta p}{p_0},
\end{equation}
where $\Delta x$ and $\Delta p$ denote the difference between the real and nominal values of
particle orbit and momentum, respectively.
The dispersion in the \cc\ target system was set to $D=14.15~m$,
which (taking into account the 1.06~mm effective target width) results in an
effective beam spread of $\pm0.06$~MeV/c.
The relevant momentum range is marked by blue lines in Figure~\ref{schottky}.

After variations of the position of the kinematic ellipse have been observed during an \emph{on-line} analysis,
measurements of additional (to the frequency spectrum) parameters were implemented.
To check the stability of the beam optics the current through the dipole magnets was controlled,
however, no fluctuations on the level of $10^{-5}$ during beam cycles were found.
Furthermore the temperatures of the incoming and outgoing water used for the magnets cooling were monitored\footnote{The
variation of the temperature could cause changes of the dipole dimensions and hence changes of
the magnetic field \emph{shape}.} --
also without any correlations with the observed behaviour of the ellipse position.
\subsection{Atmospheric conditions}
In addition to the control of the hardware status, the atmospheric conditions were monitored
inside and outside the COSY-tunnel.
Thermograph and hygrograph were installed in the COSY tunnel close to the \cc\ detector
and information about air temperature, pressure and humidity outside the building were delivered
by the meteorology station
(courtesy of Dr. Axel Knaps).
No correlations to the observed behaviour of the kinematic ellipse was found.

The stability of the parameters described in this and the previous sections raised our confidence that the fluctuations
of the kinematic ellipse position was due to the density fluctuation inside the target stream.

\chapter{Identification of the \texorpdfstring{\ppep}{pp-->pp eta prime} reaction}
\label{identyfication}
\vspace{-0.6cm}
The identification of protons via the invariant mass method
and the determination of missing masses of unregistered particles allow to select \ppep\ events.
\vspace{-0.6cm}
\section{Identification of the outgoing protons}
\label{protonsidentyfication}
\begin{figure}[!b]
 \vspace{-0.4cm}
  \begin{center}
    \includegraphics[width=0.49\textwidth]{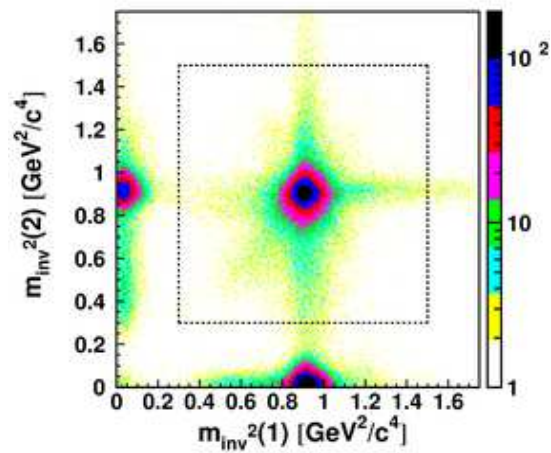}
  \end{center}
\vspace{-0.6cm}
 \caption{
         Distribution of the invariant mass of two registered particles. The superimposed black dashed square
         represents the applied cut. Outside the selected area $\pi^{+} p$ and $p\pi^{+}$
         events are visible. Note that the number of entries is given in the logarithmic scale.
         }
 \label{invmassplot}
\end{figure}

The invariant mass of registered
particles was calculated from the equation:
\begin{equation}
m_{\textrm{inv}}^2=\frac{p^2(1-\beta^2)}{\beta^2},
\end{equation}
where momentum \emph{p} and velocity $\beta$ of the particle were determined by means of drift chambers
(track reconstruction through the known magnetic field) and the S1-S3 hodoscope
(ToF), respectively. The correlations of the two invariant masses for events with two reconstructed tracks are presented
in Figure~\ref{invmassplot}. Appropriately chosen cuts allow to select events corresponding only to two registered protons.
\section{Determination of the relative beam momenta}
\label{relatbeammom}
Due to the inaccuracy of the orbit length~\cite{Prasuhn2},
the COSY crew can set the absolute momentum of the beam (nominal value)
only with an accuracy of 3~MeV/c.
Since this was not sufficient for our experiment,
we decided to use collected \ppep\ and \pp\ events for the determination of the absolute beam momenta.
Based on the position of the \ep\ signals in the missing mass spectra the absolute beam momenta can be derived.
For this procedure the mass of the \ep\ meson has to be determined as will be discussed later.
Since, on the other hand,
the correct signal position can be obtained only for a background-free missing mass spectrum,
the background has to be subtracted first. The method used for the background subtraction (see the next section)
requires information about the relative beam momenta, which can be determined by the comparison
of kinematic ellipses from \pp\ events (see Section~\ref{ellipse}).
Although the position of the kinematic ellipse depends stronger on the target position than on the beam momentum
(see Section~\ref{ellipse})
it can be used for calculations of the relative beam momenta,
since the fluctuations during the measurement were corrected before.
As a result we obtained the following momenta relative to the lowest measured one:
\begin{eqnarray}
\Delta p_{1-2} & = &\ \ 1.85\pm 0.01\ \textrm{MeV/c}\nonumber\\
\Delta p_{1-3} & = &\ \ 2.82\pm 0.01\ \textrm{MeV/c}\nonumber\\
\Delta p_{1-4} & = &\ \ 6.53\pm 0.01\ \textrm{MeV/c}\nonumber\\
\Delta p_{1-5} & = & 12.66\pm 0.01\ \textrm{MeV/c}\nonumber
\end{eqnarray}
\section{Missing mass spectra and background subtraction}
\label{mmsection}
The determined missing mass spectra of the five measured energies were used
(i)~to determine the background,
(ii)~to evaluate the absolute beam momenta and
(iii)~to calculate the width of the \ep\ meson
(see next chapter).
\subsection{Experimental background from different energies}
The missing mass spectrum of the multipion background can be
established either by Monte Carlo simulations
(see e.g.~\cite{Czyzykiewicz}) or by the usage of the experimental background collected
for another energy~\cite{Moskal7}. The second method was used in the analysis and is described in this
dissertation, since it allows to avoid additional approximations and assumptions.

\begin{figure}[!b]
  \begin{center}
    \includegraphics[width=0.80\textwidth]{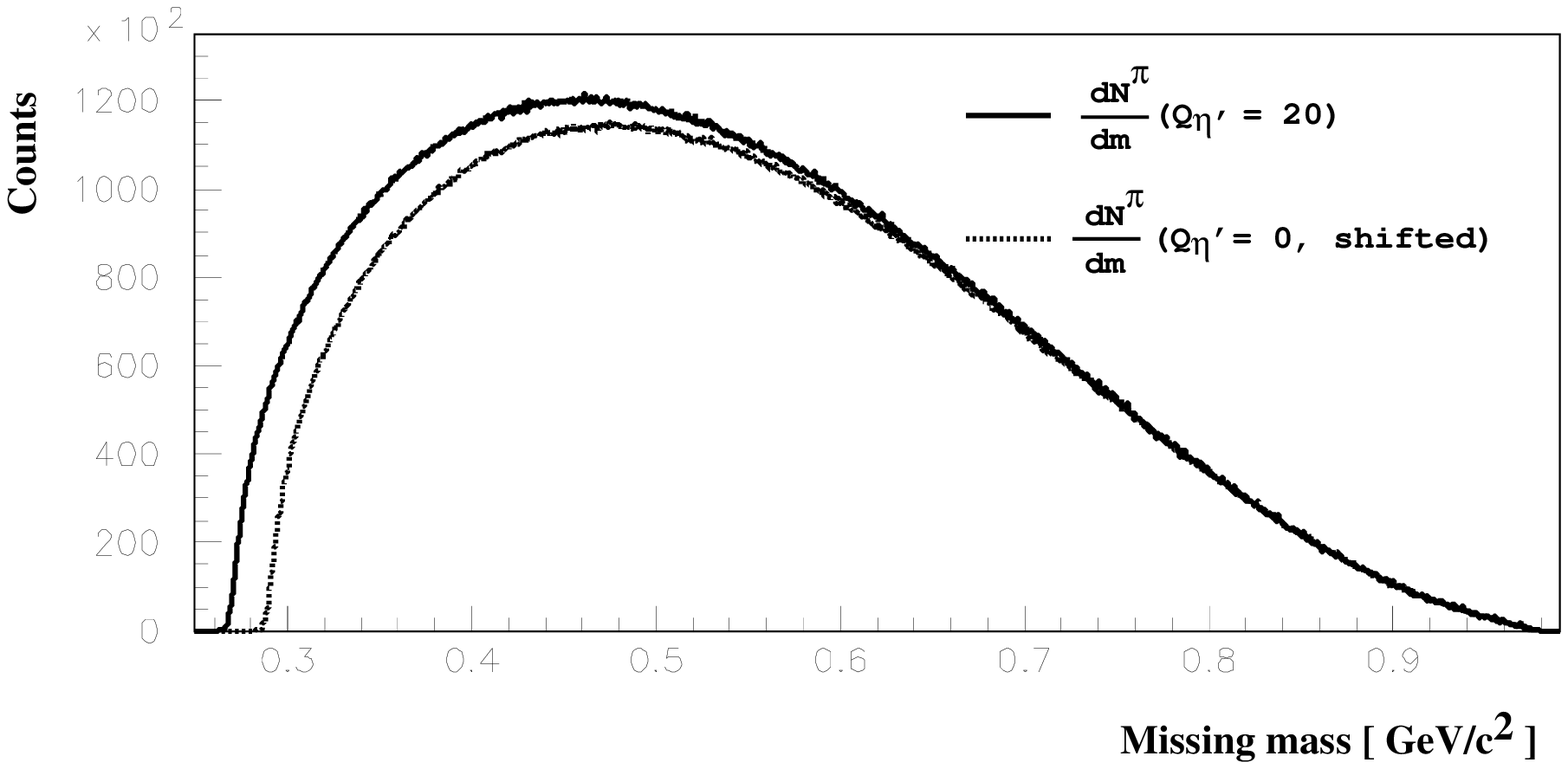}
    \includegraphics[width=0.80\textwidth]{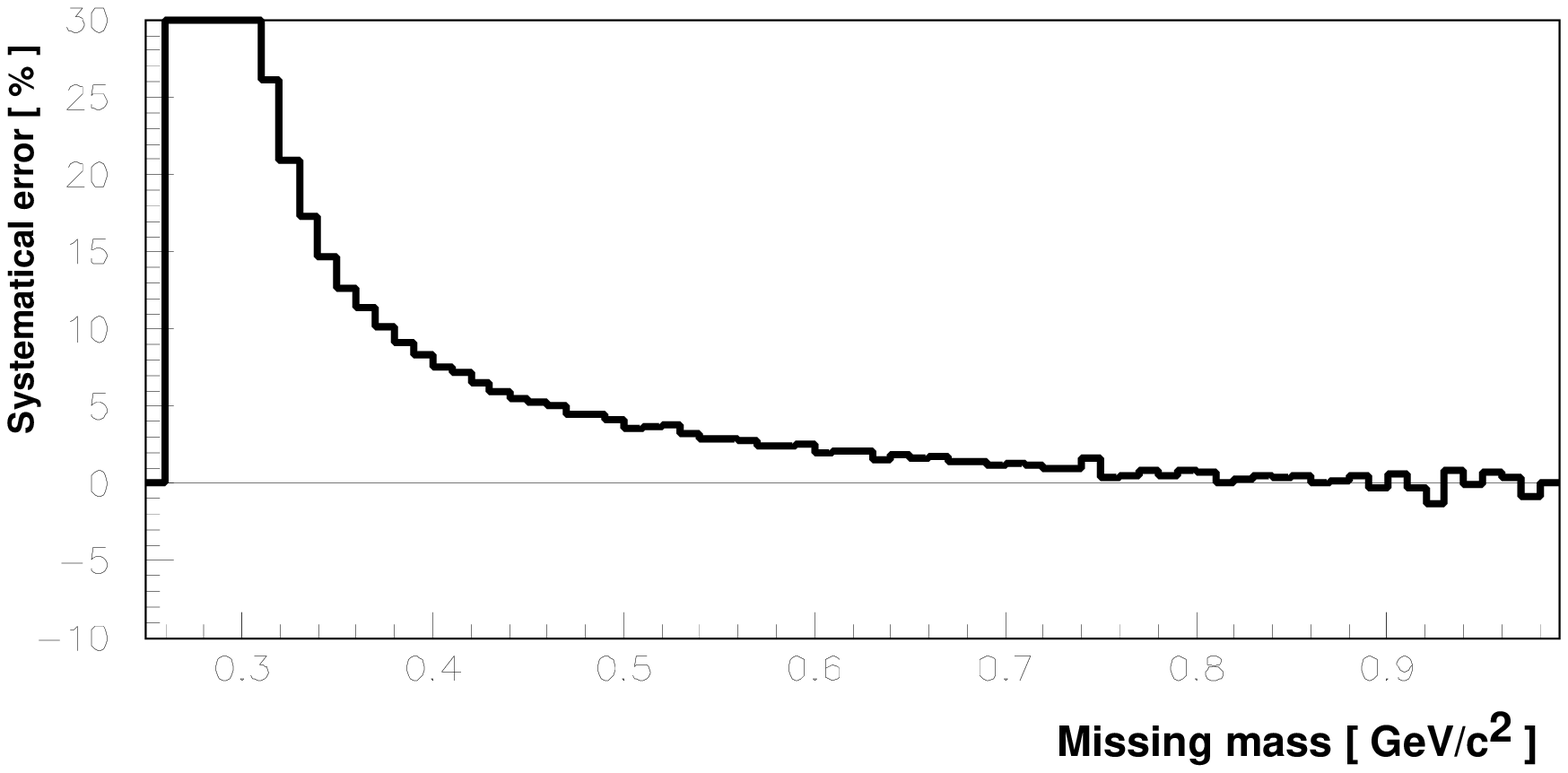}
  \end{center}
 \caption{
          {\bf Upper:}~Missing mass distribution with respect to the
           $pn$ system calculated from Monte Carlo data for the $pn\to pn \pi\pi$ process
           for a beam energy at the threshold
           (dotted line) and 20~MeV above the
           threshold for the $pn\to pn \eta^{\prime}$ reaction (solid line).
           The dotted histogram was shifted by 20~MeV.
           {\bf Lower:}~Difference between the spectra in the upper plot normalised to the solid line, which gives
           the systematical error due to the background determination.
         The plots are adapted from~\cite{Moskal7}.
         }
 \label{bcg_shift}
\end{figure}

The used background subtraction method was described in detail in article~\cite{Moskal7}.
It~is based on the observation that the shape of the multipion
mass distribution does not change when the excess energy for
the \ppep\ reaction varies by a few MeV only since the dominant background
originates from $2\pi$ and $3\pi$ production~\cite{Moskal4} and
the excess energy for either of these reactions is larger than 0.5~GeV.
Figure~\ref{bcg_shift} demonstrates that the change in the shape of the missing mass distributions is
in the order of 1\% over a range more than 0.2~GeV and in this experiment a range of about 0.005~GeV is important.
From the measurement below the threshold for the \ep\ production, the signal-free
background can be obtained and
 used for the close-to-threshold production process.
Also the measurements sufficiently high above the threshold, with an excess energy larger than the resolution of the missing
mass determination,
provides a signal-free background for close-to-threshold production data.

\begin{figure}[!h]
  \begin{center}
    \includegraphics[width=0.98\textwidth]{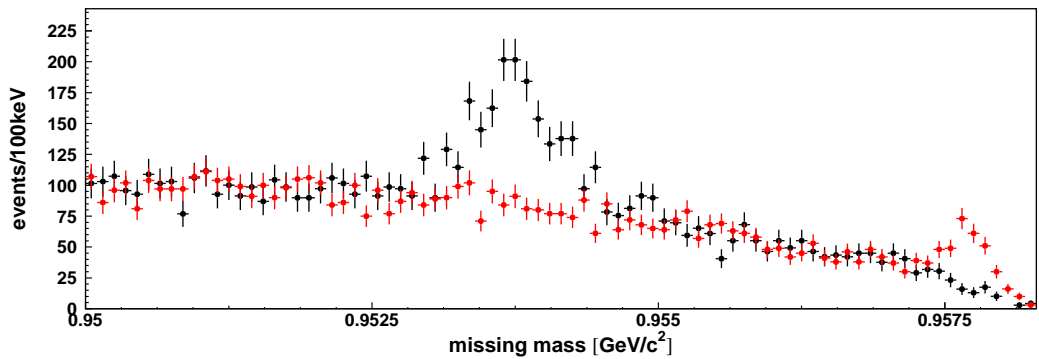}
  \end{center}
 \caption{
          Missing mass spectra for the \ppx\ reaction.
          Red points correspond to the measurement at 3211~MeV/c beam momentum
          (the lowest one in the described experiment), while black ones represent
          the measurement at the highest beam momentum (3224~MeV/c).
          The black points were shifted by the difference between the kinematic limits
          and normalised to the red points (see text for details).
         }
 \label{smoothbcg}
\end{figure}

The excess energy in the centre of mass system (CM) is defined as:
\begin{equation}
Q=\sqrt{s}-\sum_i^Nm_i\ ,
\end{equation}
where $\sqrt{s}$ denotes the centre-of-mass total energy of the colliding protons system
and \emph{N} is the number of outgoing particles.
In order to determine the background shift between the spectra
the relative values of Q have to be known.
Since Q depends inter alia on the beam momentum from which the relative changes can be
controlled via the
position of the experimental distributions of the kinematic ellipse, relative Q values can be obtained.
Missing mass spectra for the \ppx\ reaction
obtained for the lowest (3211~MeV/c) and highest (3224~MeV/c) beam momenta in the
described experiment are presented in Figure~\ref{smoothbcg}. It is important to stress that the background
distribution is smooth in the whole range studied. The spectrum for the beam momentum of 3224~MeV/c
was shifted according to the described method and normalised to the data from lower energy.
As one can see the background shape in the signal-free region is the same with respect to the statistical errors
for both energies.
This confirms the correctness of the above described method for background determination.
\subsection{Background parametrisation with polynomial fit}
\label{bcgparsect}
\begin{figure}[!p]
\vspace{-1.0cm}
  \begin{center}
    \includegraphics[width=0.49\textwidth]{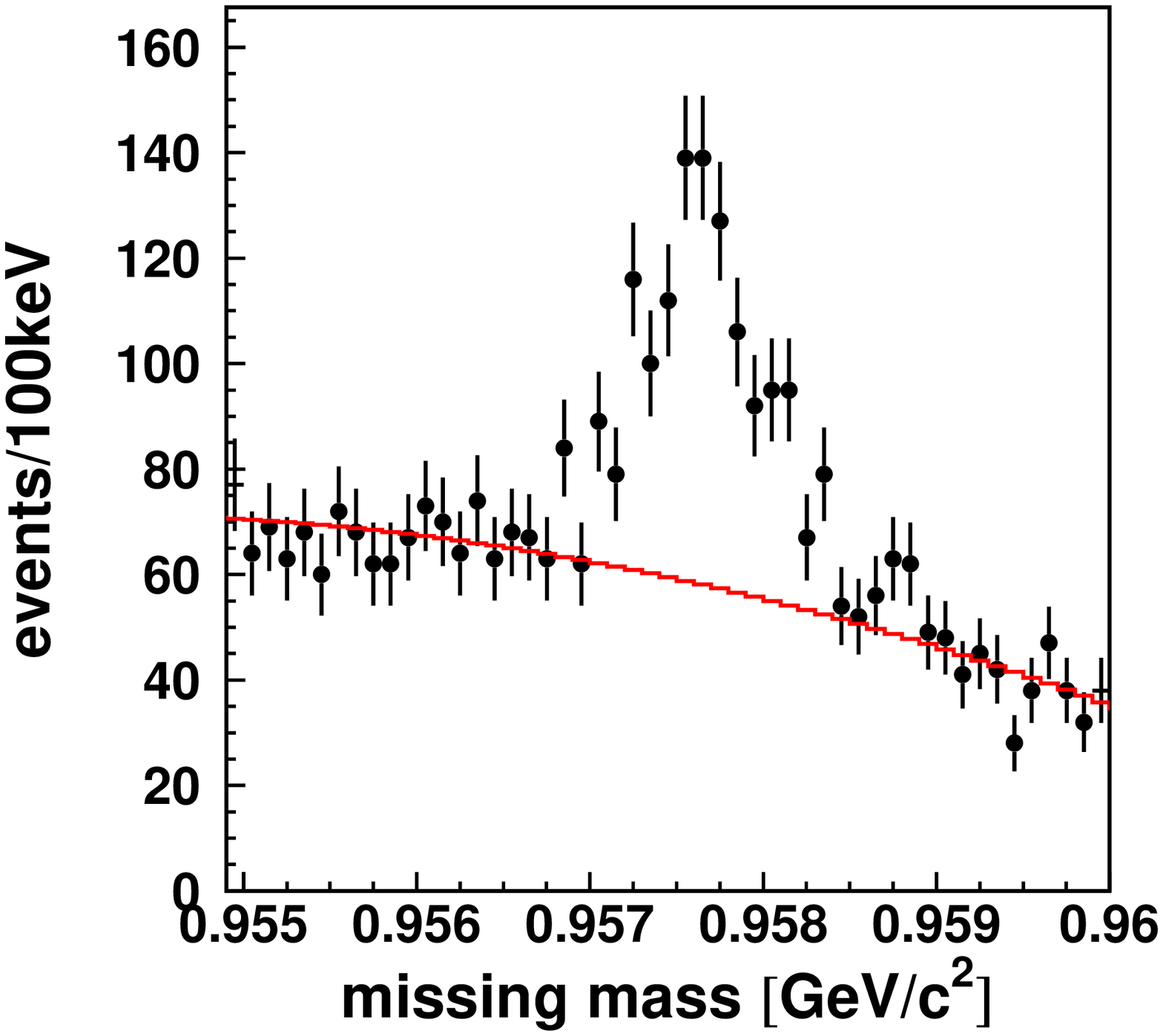}
    \includegraphics[width=0.49\textwidth]{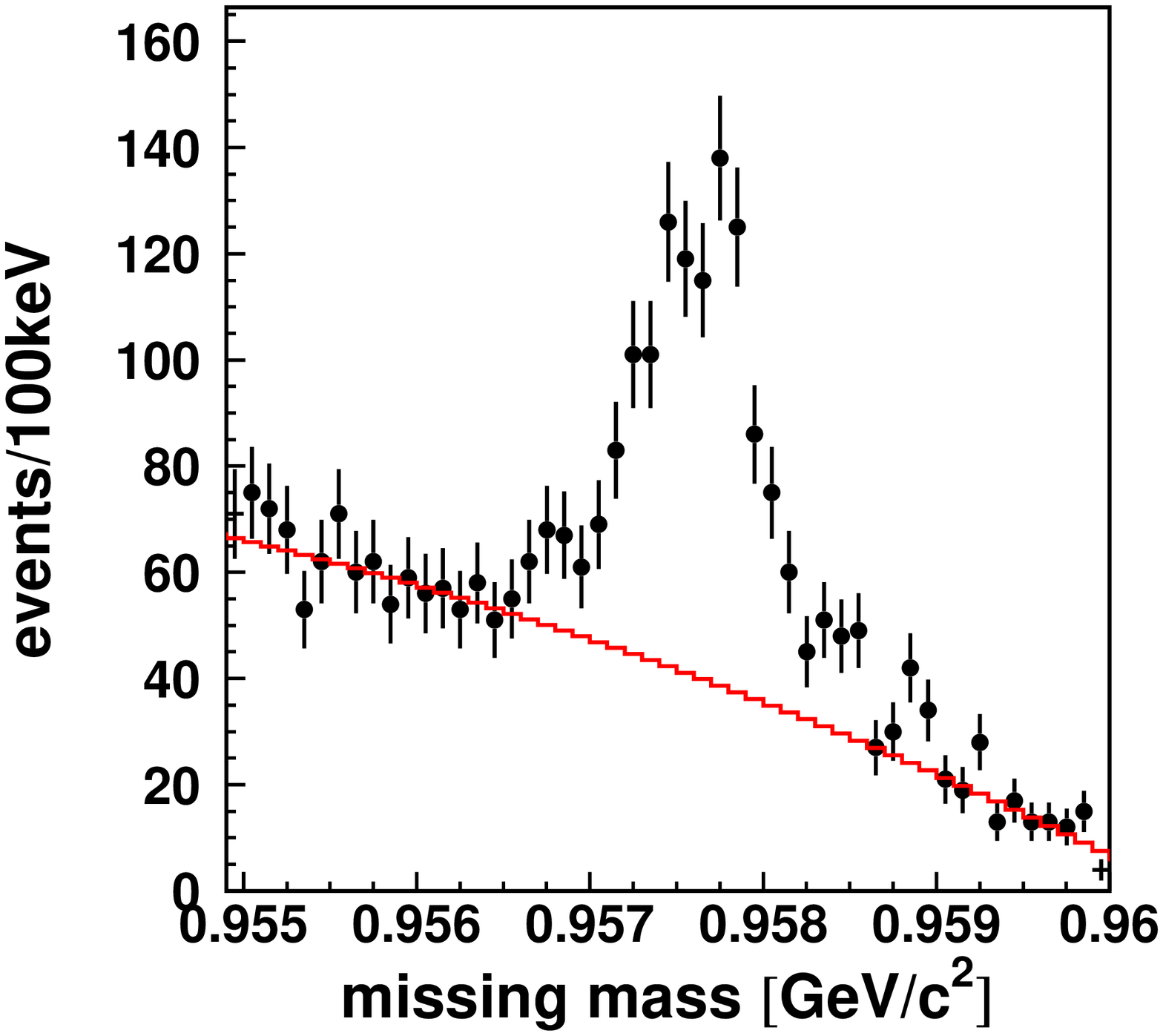}
    \includegraphics[width=0.49\textwidth]{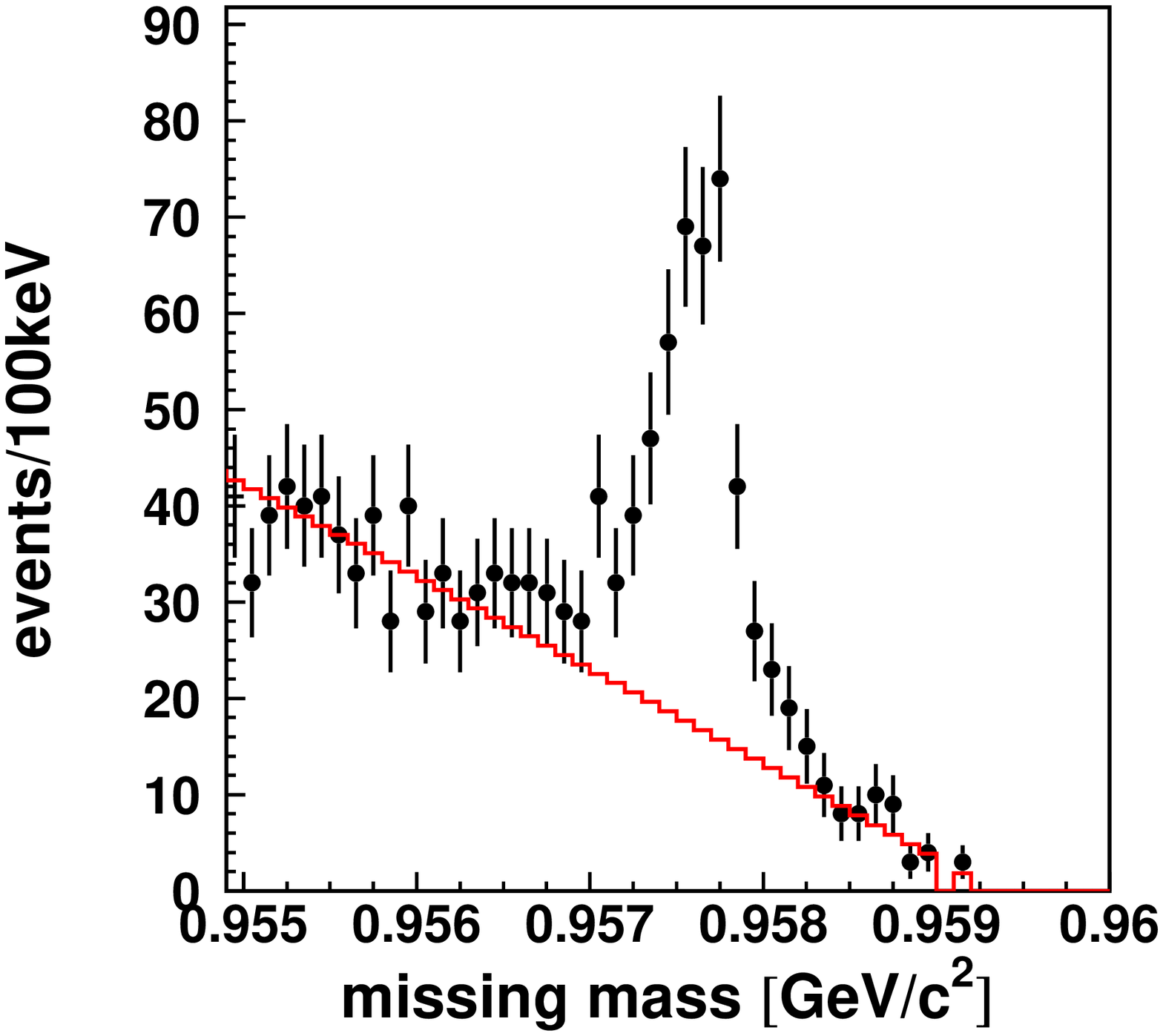}
    \includegraphics[width=0.49\textwidth]{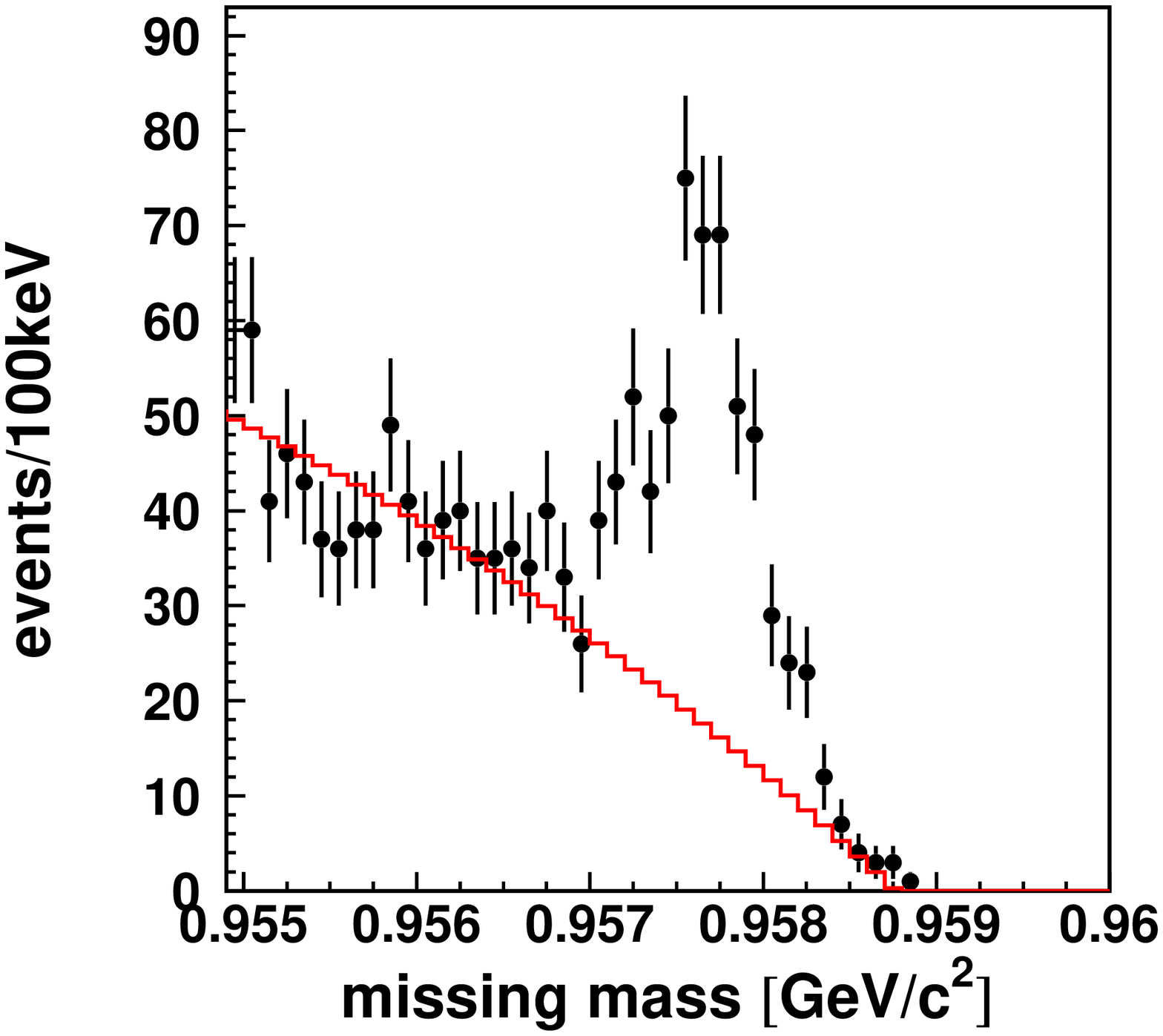}
    \includegraphics[width=0.49\textwidth]{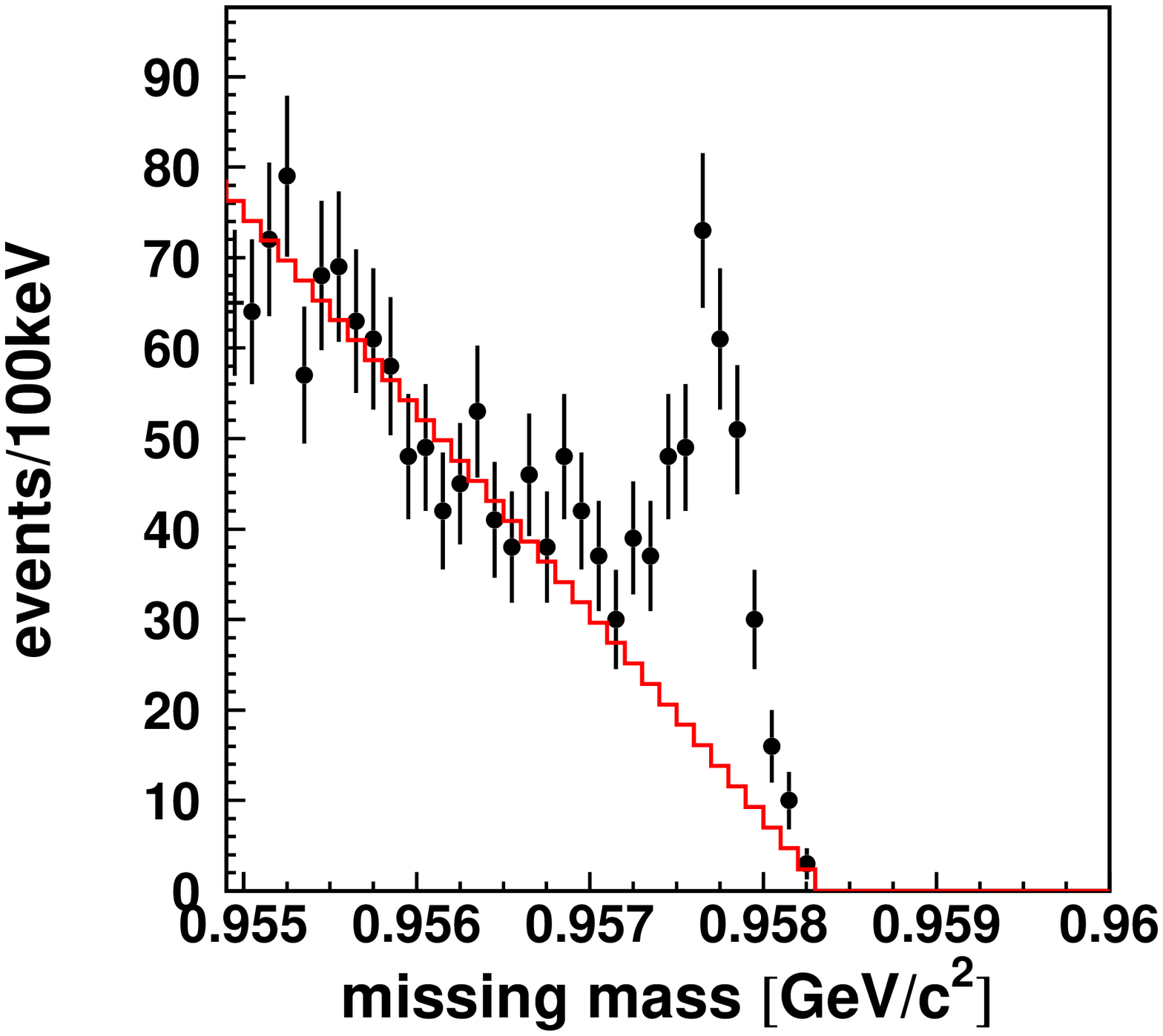}
  \end{center}
\vspace{-.5cm}
 \caption{
         Missing mass spectra for the \ppx\ reaction obtained for the beam momenta
         of 3224, 3218, 3214, 3213 and 3211~MeV/c
         (from left to right, top to bottom).
         The black points represent the experimental data, while the red curves are the shifted and normalised
         second order polynomials obtained from the fit to the background
         (for the description see text).
         }
 \label{mmbcg}
\end{figure}
To decrease the influence of the statistical fluctuation of the background and based on the
smooth change of the background in the signal region (see Figure~\ref{smoothbcg}) the background for each energy
was determined as a second order polynomial which was derived from data at a different energy and was shifted
and normalised to the actual data.
The full procedure for background determination consists of the following steps:
\begin{enumerate}
\item
fit of a second order polynomial to the missing mass spectrum in the signal-free region for a lower$\slash$higher
beam energy;
\item
shift of the obtained curve according to the calculated Q difference;
\item
normalisation of the curve to the data obtained for the actual beam energy.
\end{enumerate}
The results are presented in Figure~\ref{mmbcg}. 
The determined curves agree well with the background data within
the statistical accuracy.
\section{Absolute beam momentum determination}
\label{absbeammom}
The knowledge of the relative beam momenta
(see Section~\ref{relatbeammom}) and the background shape (see previous section)
allows to determine the absolute beam momenta based on the position of the \ep\ signal.
The procedure relies on the comparison of the position of the \ep\ signal with
the \ep\ mass\footnote{$m_{\eta'}=(957.78\pm0.06)$~MeV/c$^2$~\cite{pdg}}.
Due to the best signal-to-background ratio and the sharpest signal, for the derivation of the absolute value of Q,
the missing mass obtained for the lowest beam momentum was used,
while the other four beam momenta were adjusted with respect to the relative differences obtained
in Section~\ref{relatbeammom}.
Table~\ref{nomilanvsreal} presents the values of the nominal and real beam momenta and corresponding real Q values.
\begin{table}[!h]
\begin{center}
\begin{tabular}[c]{c|c|c}
\multicolumn{2}{c|}{beam momentum $\left[\textrm{MeV/c}\right]$} & real excess energy\\
~~nominal~~ & real & $\left[\textrm{MeV}\right]$\\
\hline \hline
3211 & 3210.7 & 0.8\\
3213 & 3212.6 & 1.4\\
3214 & 3213.5 & 1.7\\
3218 & 3217.2 & 2.8\\
3224 & 3223.4 & 4.8\\
\end{tabular}
\end{center}
\caption{
        Nominal and real beam momenta for the measurement of the \ppep\ reaction and corresponding real values
        of  excess energy.
        }
\label{nomilanvsreal}
\end{table}
For all measurements the real beam momentum is lower by about 0.5~MeV/c.
The accuracy of the real beam momentum derivation depends on the accuracy of both the knowledge
of the \ep\ mass as well as determination of the relative beam momenta.
The first component is of systematic type and 
contributes as an error of $\pm0.2$~MeV/c whereas the second one is of statistical nature and its contribution
is negligibly small.

The systematically lower value of the real beam momentum of about 0.5~MeV/c matches well with
the range of accuracy
in beam momentum setup
being typically $\pm3$~MeV/c~\cite{Prasuhn2}
and the results are in line with previous experiences at COSY where also the real beam momentum
was smaller than the nominal one~\cite{Smyrski3,Moskal6}.
%

\chapter{Determination of the total width} 
\label{determination}
The determination of the total width of the \ep\ meson was based on the simultaneous comparison of all
experimental missing mass
spectra with the Monte Carlo generated ones, where \epw\ was varied in the range from 0.14 to 0.38~MeV.
\section{Comparison of experimental data with simulations}
The \ppep\ reaction and trajectories of the outgoing protons were simulated and
detector signals were generated by the GEANT3-based program~\cite{geant3www} for the five investigated beam energies.
The program itself contains the implementation of the whole geometry of the \cc\ detector setup.
It takes into account also known physical processes like
multiple scattering and nuclear reactions in the
detector material, as well as the detector and target properties established and described
in the previous chapters, like: position and spatial resolution of the drift chambers,
size and position of the target stream and value and spread of the beam momentum.
%

\begin{figure}[!h]
  \begin{center}
    \includegraphics[width=0.49\textwidth]{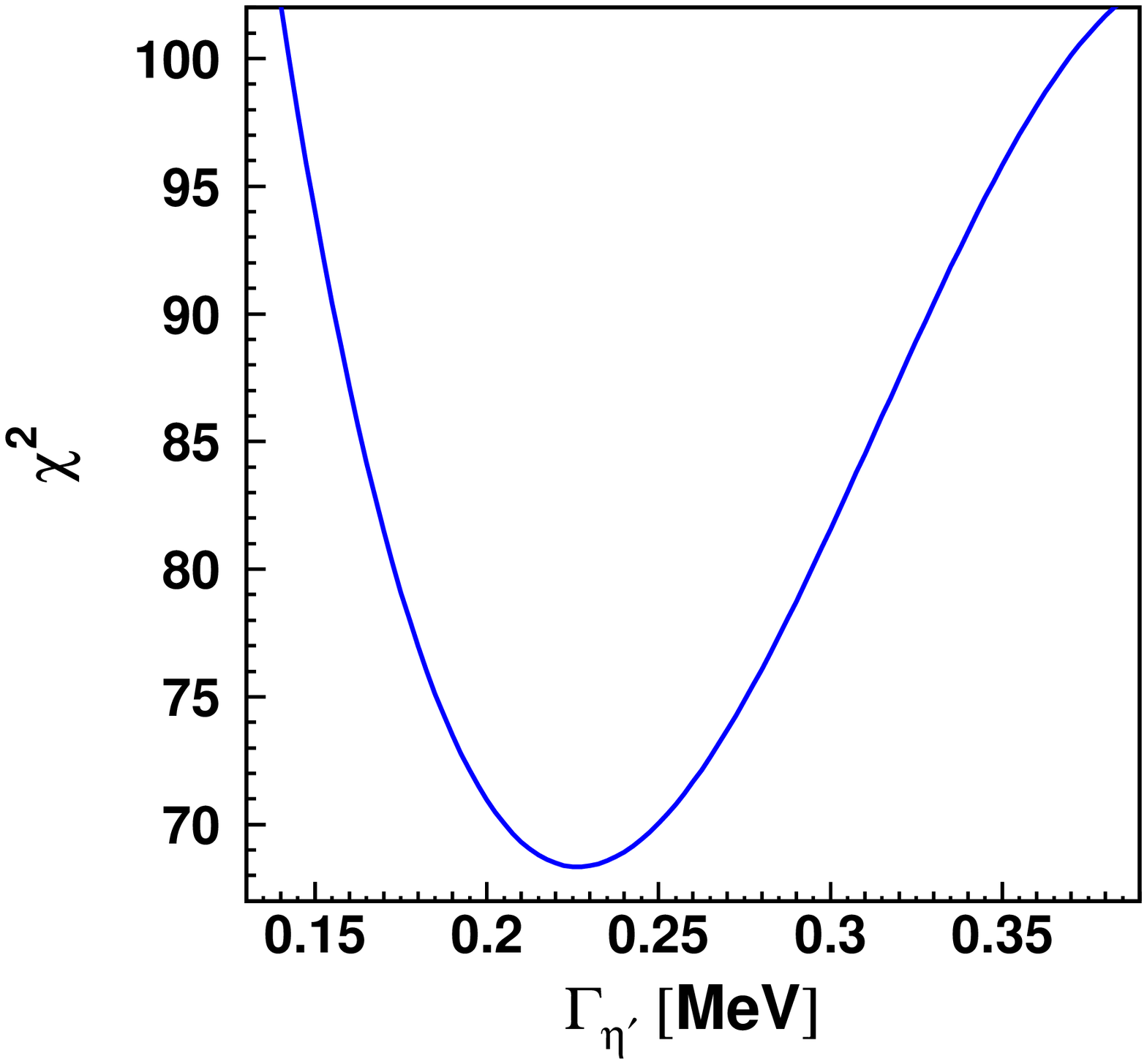}
    \includegraphics[width=0.49\textwidth]{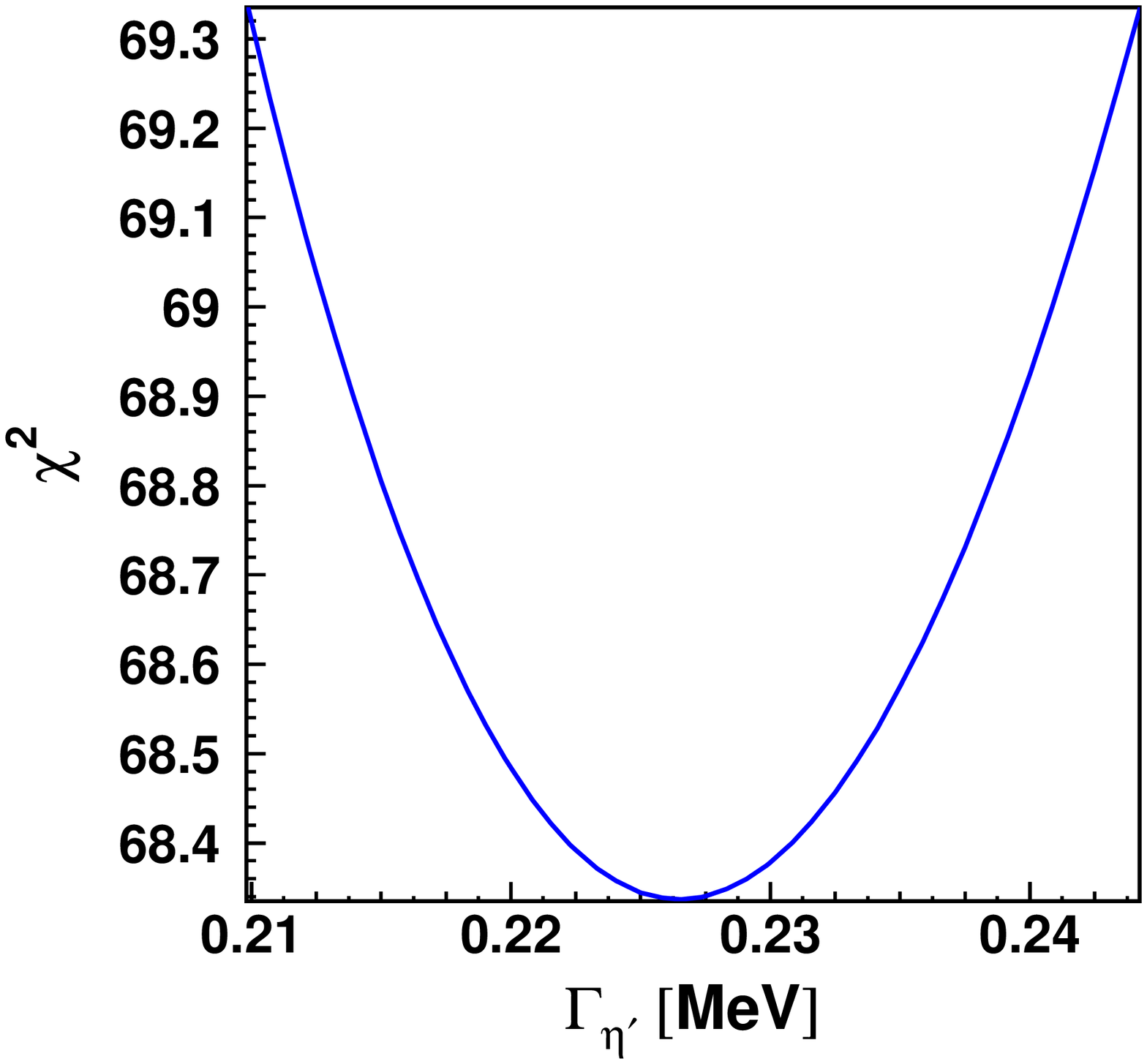}
  \end{center}
 \caption{
          {\bf Left:}~Similarity (as a value of $\chi^2$) of the missing mass spectra obtained from the measurement
          and from Monte Carlo simulations. The minimum value corresponds to $\Gamma_{\eta'}=0.226$~MeV.
          {\bf Right:}~Close-up of the left plot with the range where $\chi^2=\chi^2_{min}+1$, which corresponds
          to the value of the statistical error of the measurement.
         }
 \label{mmchi}
\end{figure}

\begin{figure}[!p]
\vspace{-1.0cm}
  \begin{center}
    \includegraphics[width=0.49\textwidth]{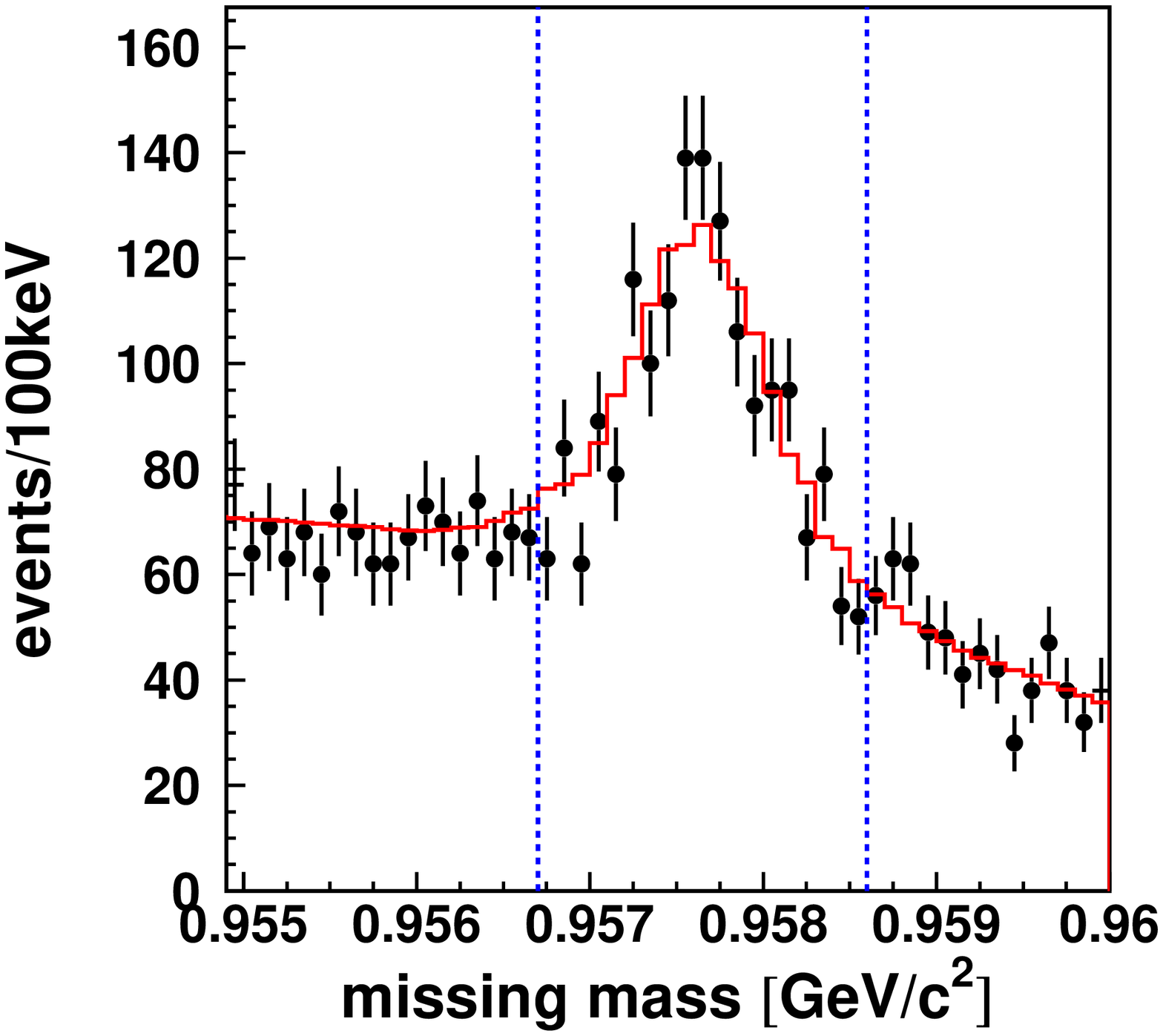}
    \includegraphics[width=0.49\textwidth]{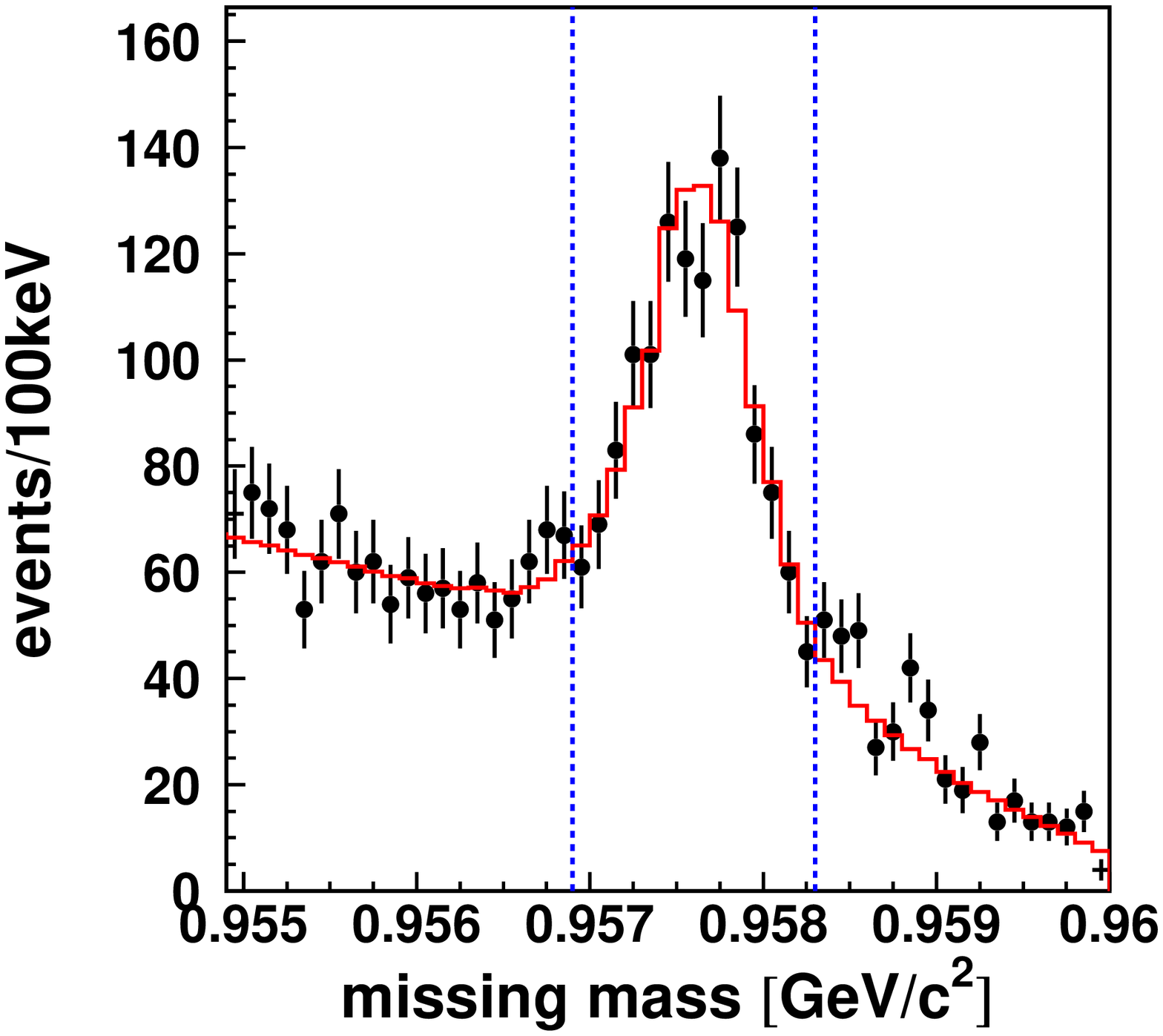}
    \includegraphics[width=0.49\textwidth]{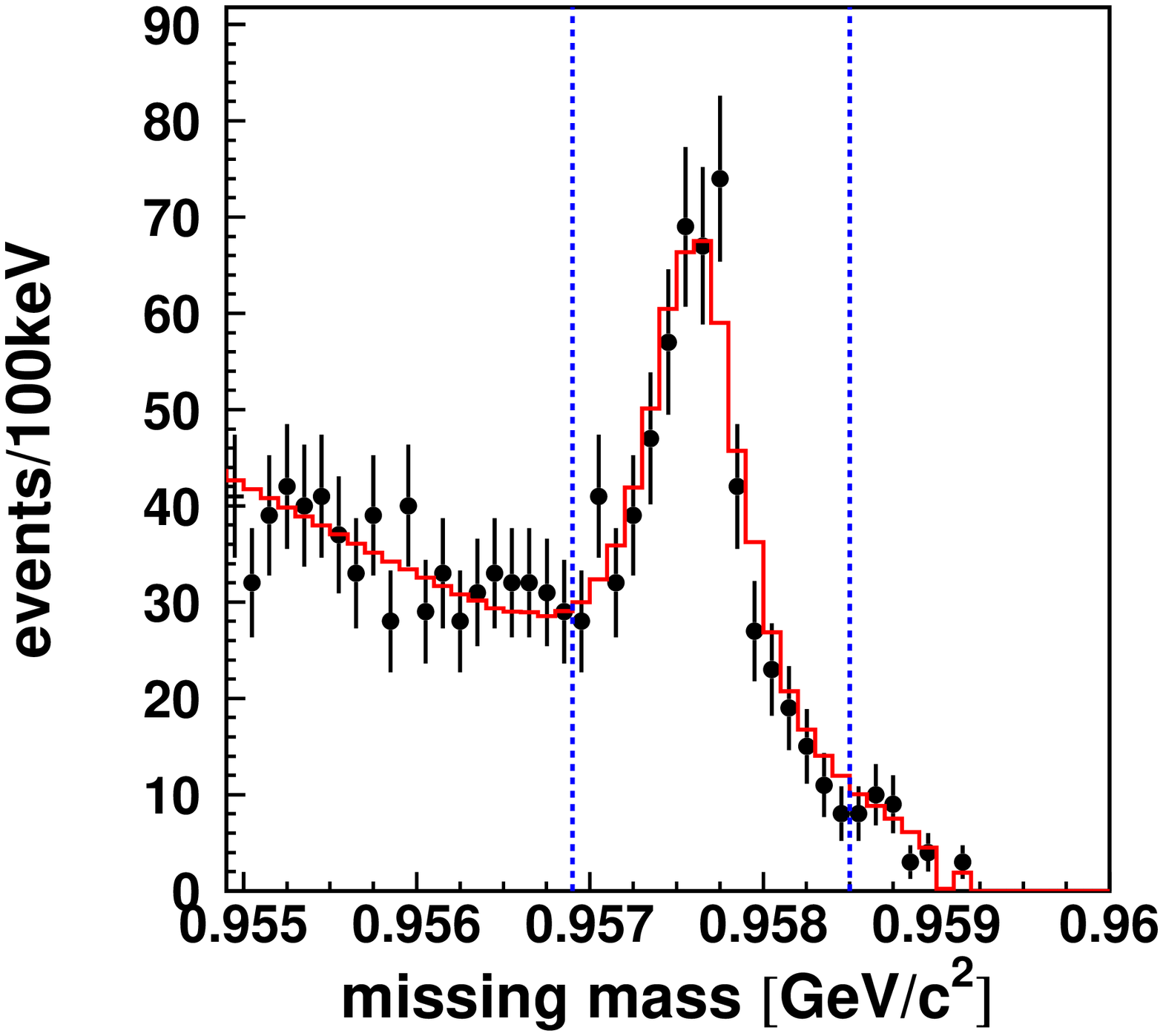}
    \includegraphics[width=0.49\textwidth]{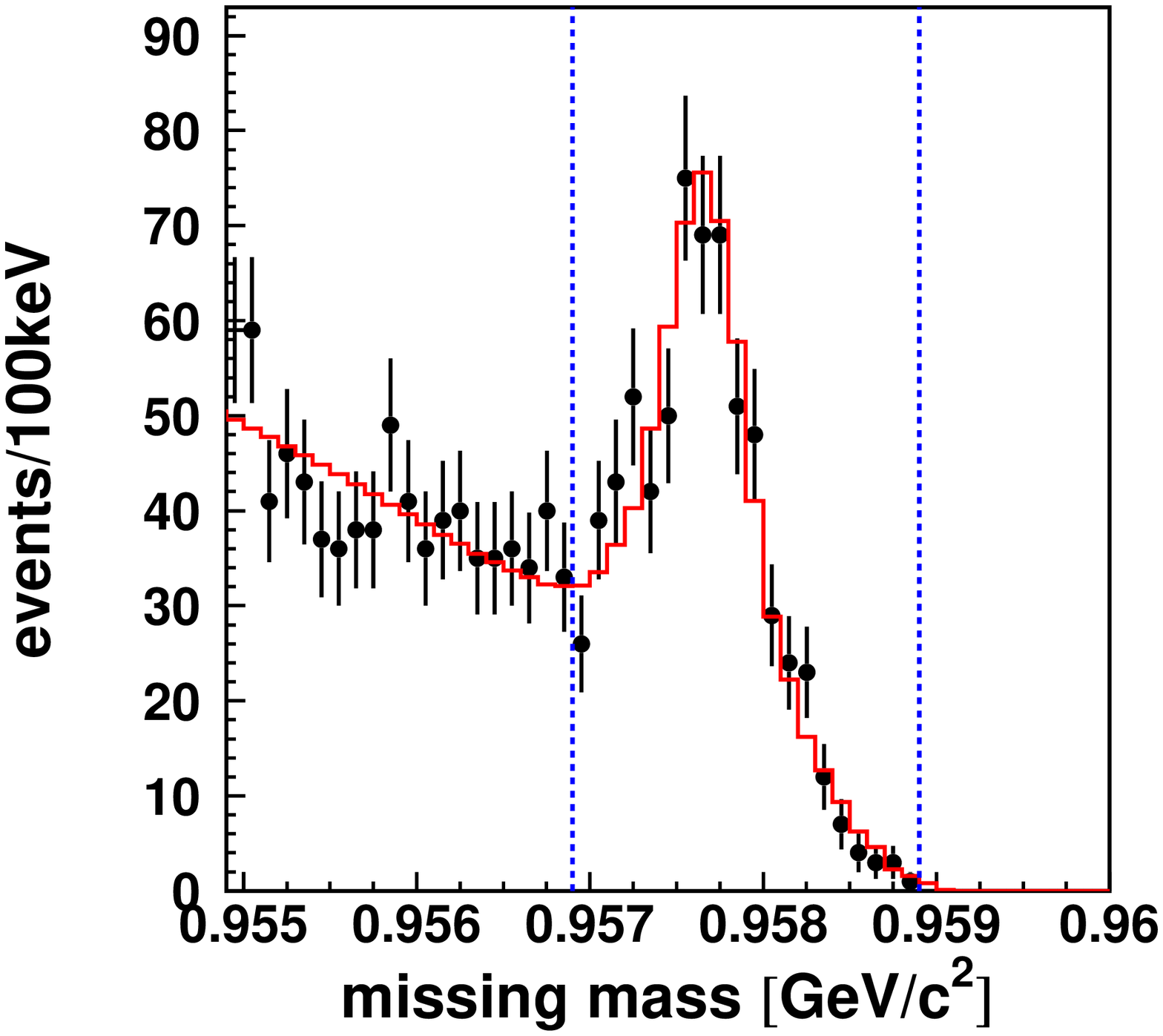}
    \includegraphics[width=0.49\textwidth]{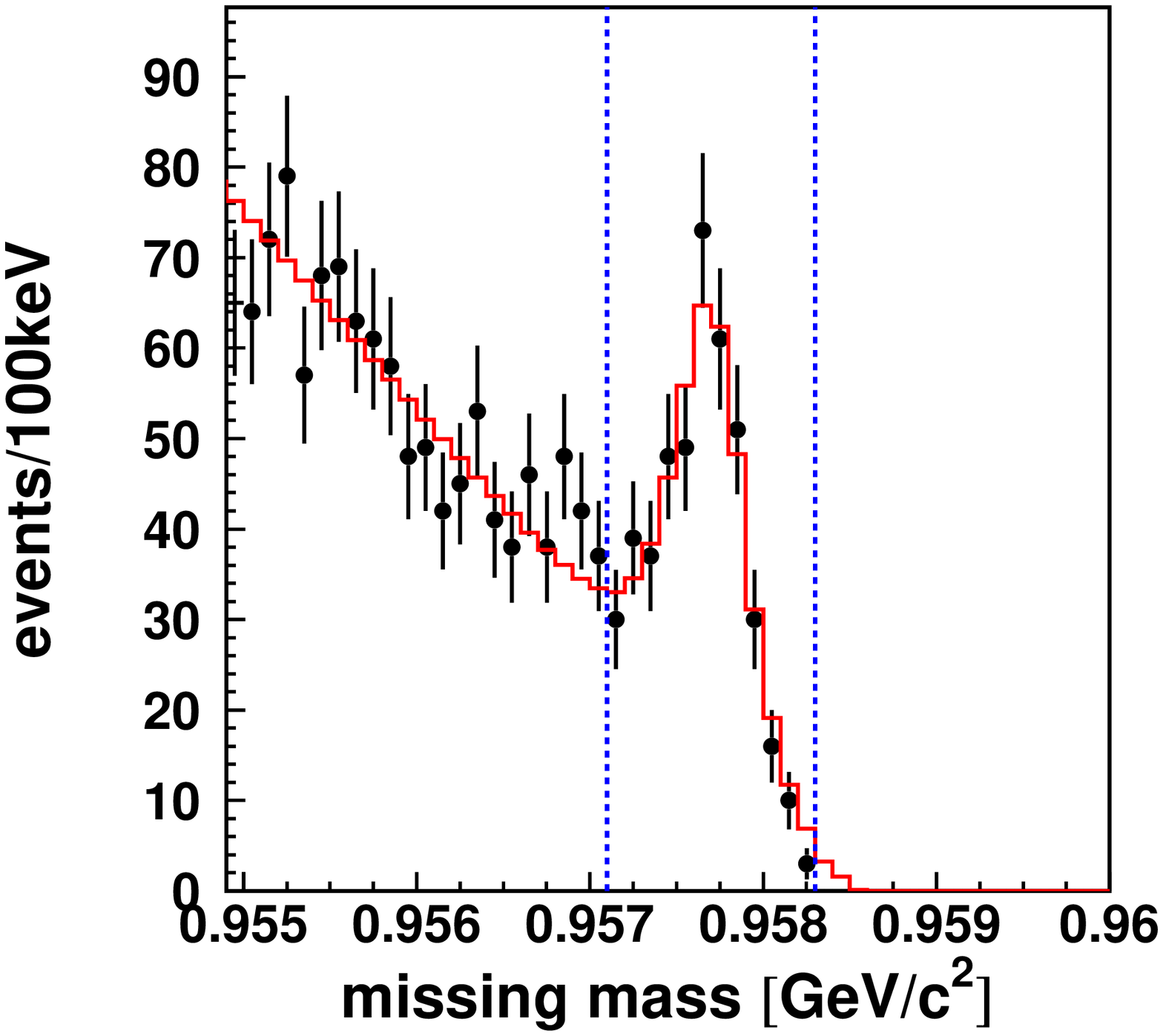}
  \end{center}
\vspace{-1.0cm}
 \caption{
         The missing mass spectra for the \ppx\ reaction for excess energies in the CM system equal to
         4.8, 2.8, 1.7, 1.4, and 0.8~MeV
         (from left to right, top to bottom).
         The \ep\ meson signal is clearly visible. The experimental data are presented
         as black points, while in each plot the red line corresponds to the sum of the Monte Carlo generated signal
         for $\Gamma_{\eta'}=0.226$~MeV and the shifted and normalised second order polynomial obtained as
         a fit to the signal-free background region for another energy.
         }
 \label{mmbcgfit}
\end{figure}

Afterwards the generated events were analysed in the same way as the experimental data and
sets of missing mass spectra for the five measured energies
were obtained
for the values of \epw\ ranging from 0.14 to 0.38~MeV.
For simulations of the mass distribution of the \ep\ meson
the Breit-Wigner formula was used. 
Finally the Monte Carlo missing mass spectra with an \ep\ signal were added to the second order polynomial
fitted to the experimental backgrounds (see Section~\ref{bcgparsect}).
The obtained spectra were compared to the experimental ones via calculating the $\chi^2$ derived from the
maximum likelihood method~\cite{Baker,Feldman}. The following formula was used for the $\chi^2$ computation:
\begin{equation}
\label{chieq}
\chi^2=2\sum^{K}_{i=1}\left[\alpha N^{\textrm{MC}}_{i}+B_i-N^{\textrm{exp}}_i+N^{\textrm{exp}}_i ln\left(
\frac{N^{\textrm{exp}}_i}{N^{\textrm{MC}}_{i}+B_i}\right)\right]\ ,
\end{equation}
where $K$ denotes the number of bins in the range where the histograms were compared, $\alpha$ is the
free parameter of the fit which describes the normalisation
factor of the Monte Carlo spectra with the \ep\ signal.
The numbers of entries in the $i$-th bin in the Monte Carlo spectra, the background and the experimental spectra
are denoted as
$N^{\textrm{MC}}_{i}$, $B_i$ and $N^{\textrm{exp}}_i$, respectively.
The dependence of the calculated $\chi^2$ -- quantifying a difference
between the experimental and Monte Carlo spectra -- on the applied \epw\ value is
presented in Figure~\ref{mmchi}.
The minimum value of $\chi^2$ corresponds to $\Gamma_{\eta'}=0.226$~MeV, which is the most likely value
of the total width of the \ep\ meson. The right plot in Figure~\ref{mmchi} is the close-up
of the left plot in the region of the minimum, where the range of the horizontal axis
corresponds to the range where $\chi^2$ differs by one with respect to its minimum value.
Since the calculated value of $\chi^2$ is not normalised to the number of degrees of freedom, the range
of \epw\ where $\chi^2=\chi^2_{min}+1$ corresponds to the statistical error of the
measurement~\cite{pdg,Binnie,Kamys},
which in case of the reported measurement is $\pm0.017$~MeV.
The experimental spectra of the missing mass superimposed with the sum of the background polynomial and the 
Monte Carlo
generated signals for $\Gamma_{\eta'}=0.226$~MeV are presented in Figure~\ref{mmbcgfit}.
The blue dashed lines mark the range where the experimental histograms were compared to
the result of the simulations.
The presented missing mass signals are the convolution of the total width of the \ep\ meson
and the experimental resolution.
\begin{figure}[!p]
\vspace{-1.0cm}
  \begin{center}
    \includegraphics[width=0.49\textwidth]{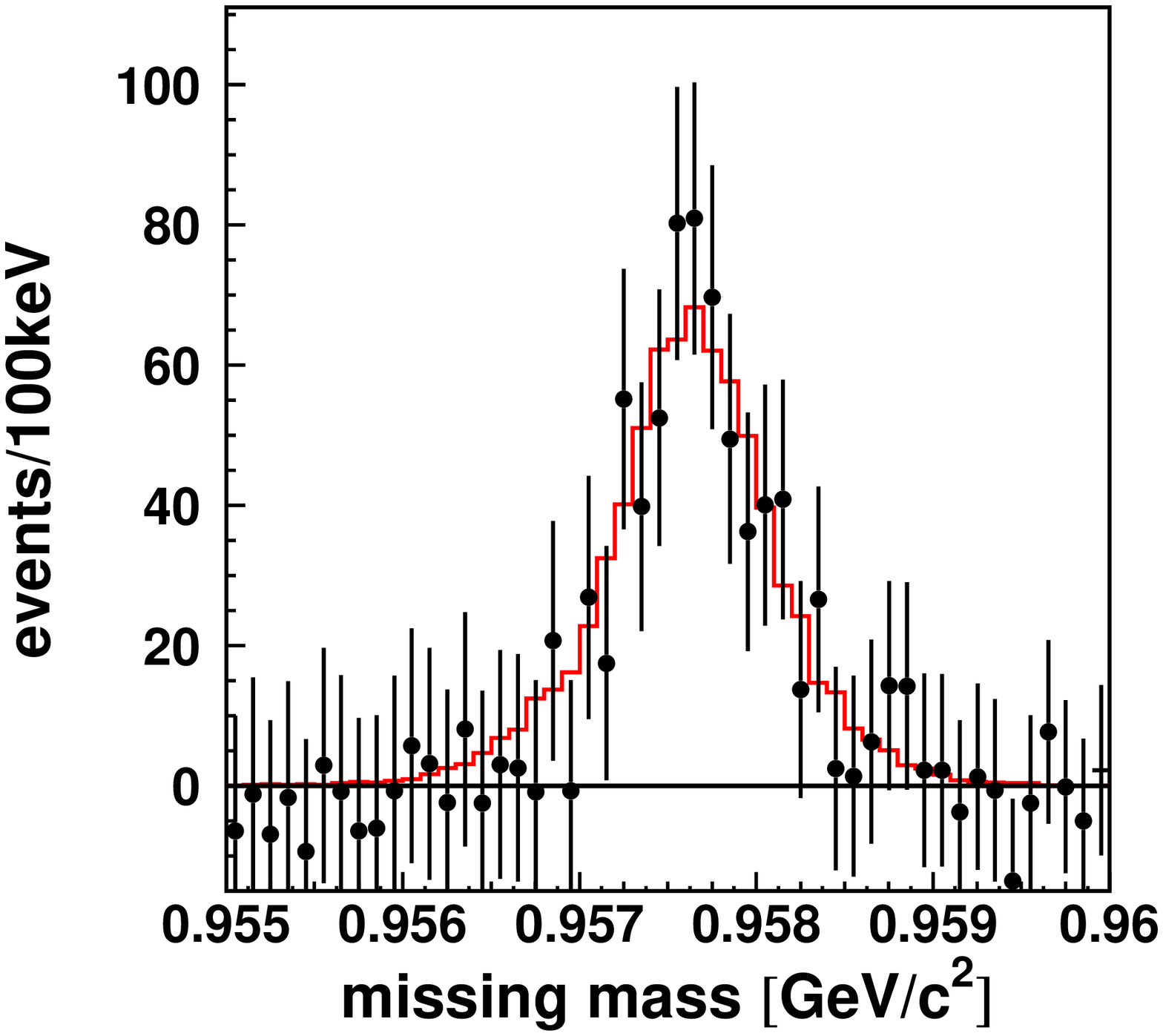}
    \includegraphics[width=0.49\textwidth]{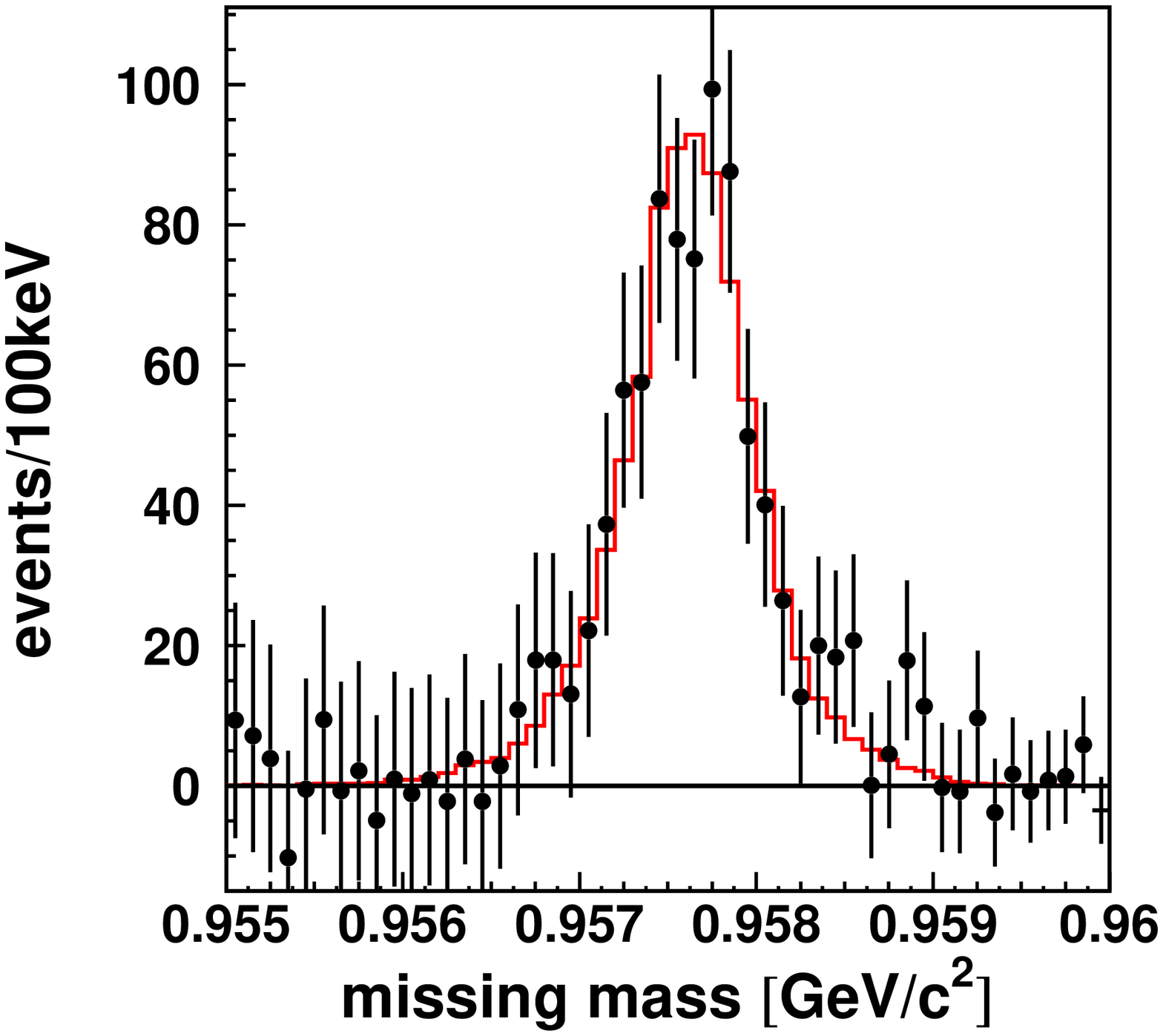}
    \includegraphics[width=0.49\textwidth]{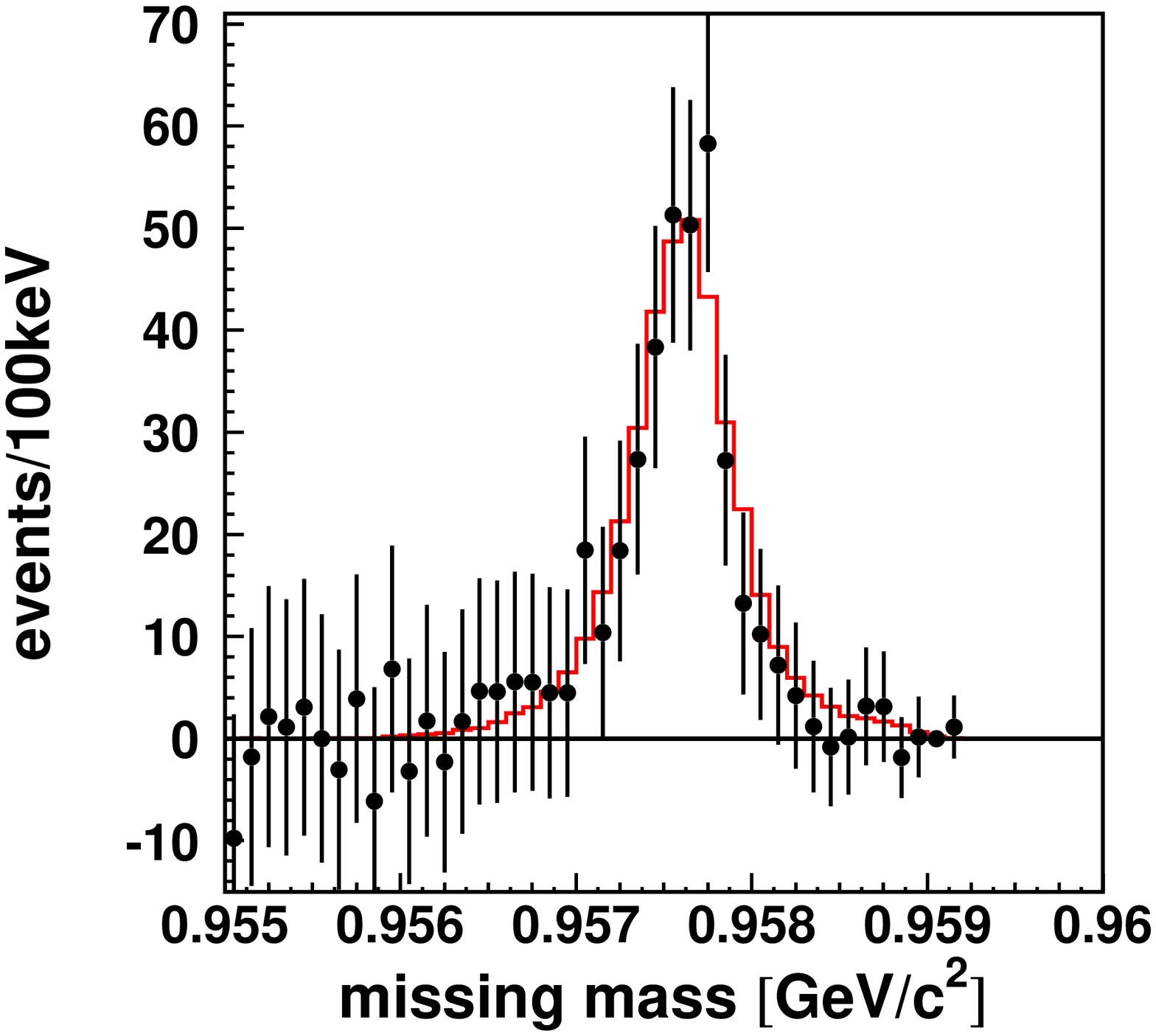}
    \includegraphics[width=0.49\textwidth]{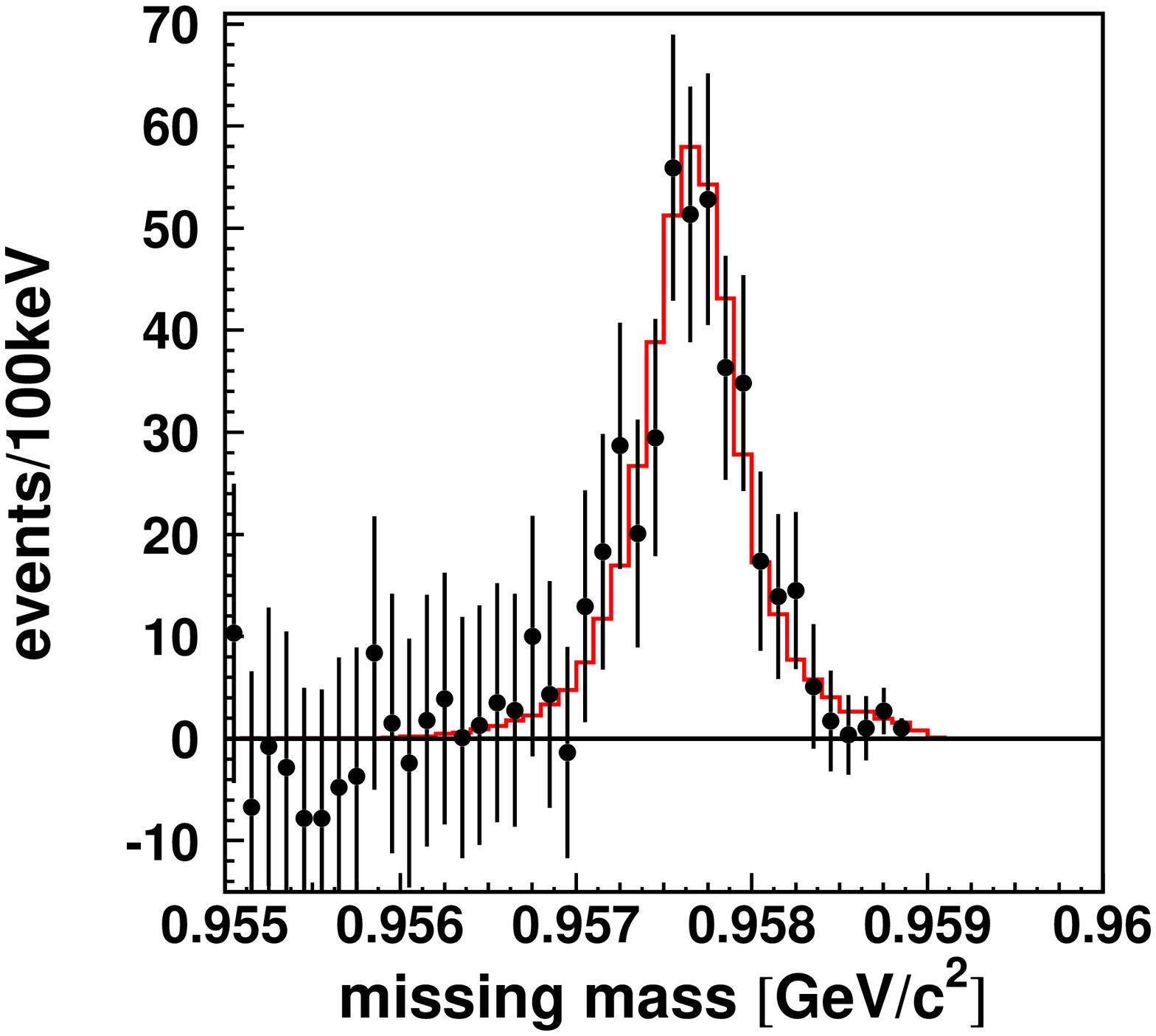}
    \includegraphics[width=0.49\textwidth]{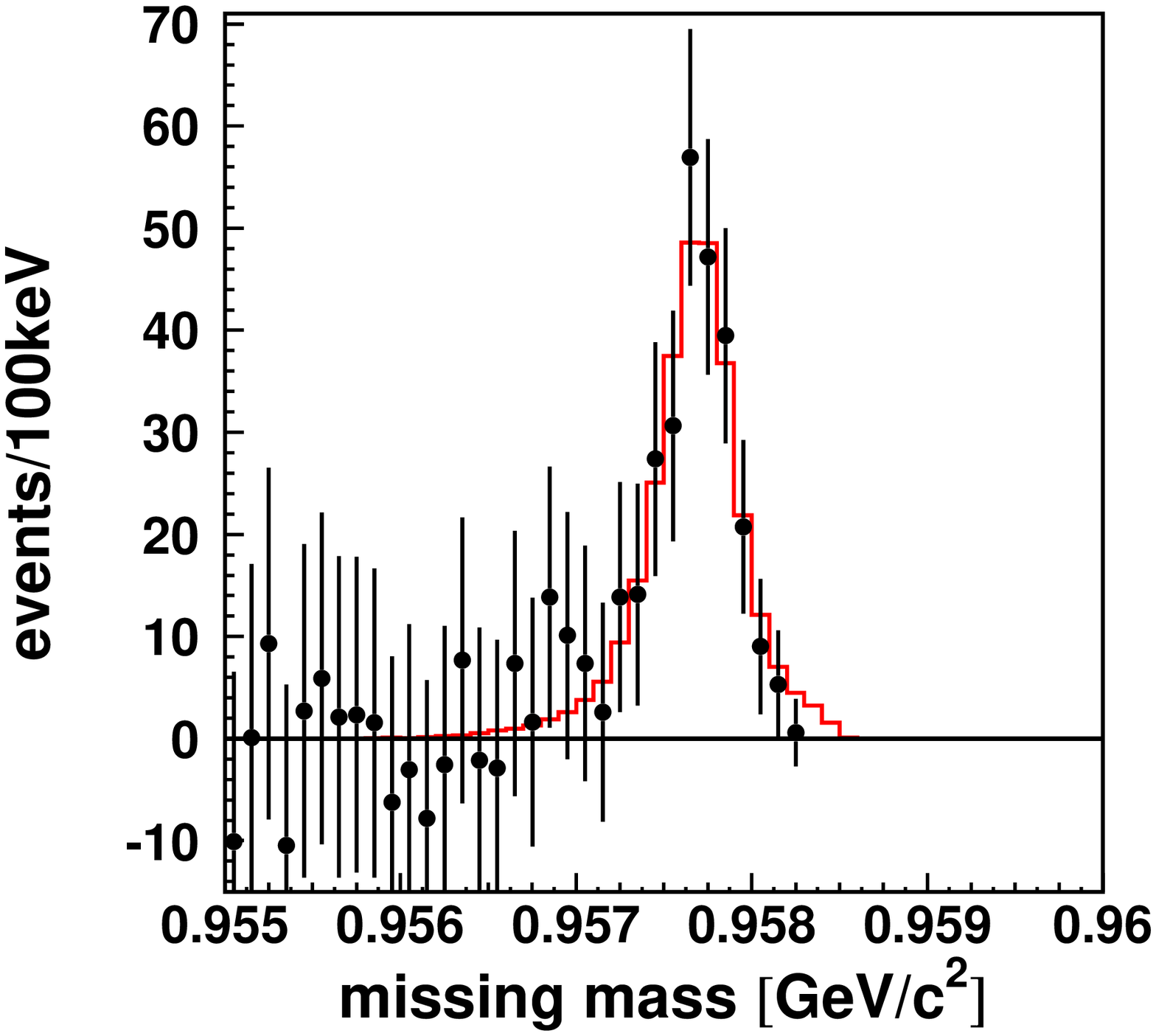}
  \end{center}
 \caption{
         Background-free missing mass spectra for the \ppx\ reaction for excess energies
         4.8, 2.8, 1.7, 1.4, and 0.8~MeV (from left to right, top to bottom) in CM system.
         The experimental data are presented
         as black points, while the red lines correspond to the Monte Carlo generated signal
         for $\Gamma_{\eta'}=0.226$~MeV.
         }
 \label{mmnobcg}
\end{figure}

The observed dependence of the width of the
missing mass signals on the excess energy (see Figure~\ref{mmbcgfit} and \ref{mmnobcg})
is due to the propagation of errors
of protons momenta involved in the
missing mass calculations~\cite{ErykMgr}.
Since the Monte Carlo program is reproducing the changes of the experimental spectra with energy very well,
this confirms the correctness of the established detector and target characteristics.

For completeness the experimental missing mass spectra for the \ppx\ reaction after background subtraction are presented
in Figure~\ref{mmnobcg}.
More than 2300 \ep\ mesons were reconstructed and
achieved experimental mass resolution for the lowest measured energy
amounts to FWHM~=~$(\textrm{FWHM}_{\textrm{missing\ mass}}^2-\Gamma_{\eta'}^2)^{1/2}=0.33\ \textrm{MeV/c}^2$.
\section{Systematic error estimation}
The accuracy of the determination of collected in Table~\ref{systtable} parameters of
the \cc\ detector and the COSY accelerator contribute to the systematic error of the derivation of the \ep\ width.
The estimated values of the influence of the accuracy
of each parameter on the final result are presented.
The contributions from the accuracy of the target position and size,
the map of the magnetic field, the position of the drift chambers
and the absolute beam momentum determination were estimated as the difference between the derived result
of the \epw\ and the \epw\ values
established by changing particular parameter by its error value.
\begin{table}[!h]
\begin{center}
\begin{tabular}[c]{r|p{2.9cm}}
parameter & contribution to the \mbox{systematic} error $\left[\textrm{MeV}\right]$\\
\hline \hline
map of the magnetic field & 0.007\\
target position & 0.006\\
background subtraction method & 0.006\\
ranges of missing mass spectra, where $\chi^2$ was calculated & 0.005\\
bins width & 0.004\\
absolute beam momentum & 0.003\\
final state interaction (FSI) between protons & 0.003\\
effective target width & 0.002\\
position and orientation of the drift chambers & 0.001\\
\end{tabular}
\end{center}
\caption{
        Summary of the parameters contributing to the systematic error of the \epw\ measurement at the \cc\ detector.
        }
\label{systtable}
\end{table}
The systematic error due to the method of the background subtraction was established as the difference
between \epw\ values determined using experimental background shapes from different
energies\footnote{For a missing mass spectrum at a given energy each of four remaining spectra could be used
for the background determination.}.
The bin width was changed from 0.1 to 0.04~MeV/c$^2$,
while the ranges of the missing mass spectra, where $\chi^2$ was calculated,
were enlarged by 0.7~MeV/c$^2$ at each side (which corresponds to seven bins in plot~\ref{mmbcgfit}).
The estimation of the influence of the final state protons-proton interaction is very conservative.
The reported value is the difference between the case where the FSI was and was not taken into account~\cite{Naisse,Moskal9}.

The total systematic error was calculated as a square root of the sum of the squared values listed
in Table~\ref{systtable} and the final result of the measurement of the total width of the \ep\ meson
conducted with the \cc\ detector is $\Gamma_{\eta'}=0.226\pm0.017(\textrm{stat.})\pm0.014(\textrm{syst.})$~MeV.

\chapter{Summary}
\label{summary}
The aim of this work was to determine the total width of the \ep\ meson with
a unique precision and independently of the other properties of this meson, like e.g.
partial widths or production cross sections.
The motivation for the measurement of the total width of the \ep\ meson as well as the
experimental method and the final result have been presented.

The value of \epw\ was established directly from the measurement of the mass distribution of the \ep\ meson.
The \ep\ meson was produced in proton-proton collisions via the \ppep\ reaction and its
mass was reconstructed
based on the information about the momentum vectors
of the protons before and after the reaction.
The experiment was conducted in the Research Centre J{\"u}lich in Germany.
The accelerated and stored protons were circulating through
the stream of the hydrogen cluster target in the ring of the cooler synchrotron COSY.
The two outgoing protons were measured by means of the \cc\ detector.
The reconstruction of a particle trajectory through the known magnetic field allows for the momentum determination, while
the ToF method provides information about the velocity.
The identification of the particle is
an outcome of the combination of those two informations, while
the \ppep\ reaction was identified via the missing mass technique.
Altogether more than 2300 \ppep\ events were reconstructed.
The comparison of the derived experimental missing mass spectra with Monte Carlo generated ones results in a $\chi^2$
dependence on the \epw\ value used in the simulation.

The statistical error of the final result was obtained directly from a $\chi^2$ vs \epw\ plot at
$\chi^2=\chi^2_{min}+1$ value~\cite{Binnie,pdg,Kamys}. 
A small
systematic error of the final result was achieved due to:
\begin{itemize}
\item
the excellent properties of the stochastically cooled proton beam;
\item
the application of a decreased size of the target stream which resulted in a
small beam momentum spread and a small geometrical size of the reaction region;
\item
monitoring of the properties of the reaction region by two independent methods:
by a specially developed diagnosis unit and by examining of the momentum
distributions of the elastically scattered protons;
\item
the close-to-threshold measurement --
where $\partial(mm)\slash\partial\boldsymbol{p}$ approaches zero.
\item
the verification of the characteristics of the synchrotron beam, target stream and detector setup
by a comparison of the results obtained for five different beam momenta.
\end{itemize}

The value of $\Gamma_{\eta'}=0.226\pm0.017(\textrm{stat.})\pm0.014(\textrm{syst.})$~MeV
determined in the analysis described in this dissertation is
three times more precise than the best classified measurement until now
($\Gamma_{\eta'}=0.28\pm0.10$~MeV)~\cite{Binnie}
and the achieved accuracy is in the same order as the value obtained by the PDG from a fit to 51
measurements of branching ratios ($\Gamma_{\eta'}=0.204\pm0.015$~MeV)~\cite{pdg}.
It is also important to note that
the achieved mass resolution amounts to FWHM~=~0.33~MeV/c$^2$
and is of the same order as the total width of the \ep\ meson itself, which excludes the possibility
of a multistructure in the \ep\ signal at this level.

The value of the \ep\ meson total width
recommended by Particle Data Group~\cite{pdg} is correlated with the partial width for the
$\eta'\to\gamma\gamma$ decay~\cite{Biagio},
while the average of two available measurements~\cite{Binnie,Wurzinger} has an 30\% error.
The result of the measurement and analysis presented in this dissertation 
has an accuracy of 13\% and agrees within the error bars with the
value provided by the PDG fit, however, the value established in this work
is independent of any of the branching ratios and
the $\Gamma(\eta'\to\gamma\gamma)$ measurements.
Therefore, it can be used as a \emph{tool} to translate branching ratios to partial widths and vice versa,
and applied for the investigations of e.g. the gluonium component in the \ep\ meson~\cite{Aloisio,Ambrosino}
and, indirectly, for studies of the quark mass difference $m_{d}-m_{u}$~\cite{Borasoy2,Zielinski,Kupsc}.
%
%

This was the last measurement conducted by the \cc\ group. It therefore could take advantage of the methods
developed
in the course of the nearly eleven years of experiments~\cite{booklet},
which, as shown in this work, resulted
in the unique mass resolution.

\cleardoublepage
\thispagestyle{plain}
\hspace{1cm}\\
\vfill
\begin{flushright}
\emph{The journey is the reward.}
\\Chinese Proverb
\end{flushright}

\thispagestyle{plain}
\cleardoublepage
\pdfbookmark[-1]{Last but not least}{}
\phantomsection
\addcontentsline{toc}{chapter}{Acknowledgments}
\pagestyle{empty}
\begin{center}
{\bf Acknowledgements}
\end{center}
\bigskip

The list of people who contributed into this work is to long to be written in a short note.
I would like to thank all of you for your indispensable help during the preparation of this thesis.
In particular those of you who are not mentioned by name -- thank you.

\smallskip
Especially I would like to express my gratitude to
my supervisor Prof. Pawe{\l} Moskal, a man who give me an opportunity for better
understanding of experimental physics, for his suggestions, advices and patience. I would like to thank you, Pawe{\l},
for possibility to work together during all of these years, although this work was rather an adventure than obligation.
All of ours discussions were great pleasure for~me. Thank~you.

\smallskip
In addition I would like to thank to:\\
--~Prof. Bogus{\l}aw Kamys for discussion and suggestions concerning this work and for support during the period of
studies;\\
--~Prof. James Ritman for his outstanding support during my stay in the Research Centre J{\"u}lich and constructive
advices and comments during presentations of the analysis progress;\\
--~Prof. Walter Oelert for the corrections of this dissertation and help since my very first visit in J{\"u}lich;\\
--~Dr Dieter Grzonka for the suggestions and help during all of the stages of the data analysis, for developing
the diagnosis unit,
which allowed us to achieve unique precision of the measurement, and also for the corrections of this dissertation;\\
--~Dr Thomas Sefzick for all his help during my work, for the corrections of this thesis and for very fast
arrangement of the hygrograph and thermograph during the beam time;\\
--~Prof. Jerzy Smyrski for the discussions concerning the properties of drift chambers and for the corrections of this
work;\\
--~Dr Magnus Wolke for the time spent on explanations of the acquisition and coding of the data at \cc\ system;\\
--~Alexander T{\"a}schner for
taking care of the \cc\ target setup during the experiment and for the tests and developing of the
new collimator with rectangular opening;\\
--~Dr Peter W{\"u}stner for always fast reactions in case of problems with the Alpha machines;\\
--~the COSY team for providing a good quality beam during the experiment;\\
--~the reviewers, Prof. Agnieszka Zalewska and Dr hab. Tomasz Kozik, for agreeing to devote
their time to reference this dissertation;\\
--~Prof. Lucjan Jarczyk for all his comments, questions and provoked discussions about the \ep\ meson and data analysis;\\
--~Dr Rafa{\l} Czy{\.z}ykiewicz for the corrections of this dissertations and many, many discussions which reach behind
the horizon, also for the chess and volleyball games;\\
--~Adrian Dybczak for the discussions and collective studies during all the time spent in Cracow;\\
--~Wojciech Krzemie{\'n} for the answers to the not so trivial questions and all the discussions on the IKP corridor;\\
--~Cezary Piskor-Ignatowicz for putting me in touch with Pawe{\l} Moskal and for very deep and serious discussions;\\
--~Dagmara Rozp\k{e}dzik and Wojciech Bruzda for the support in the typical PhD student problems;\\
--~Ma{\l}gorzata Hodana and Benedykt Jany for all the movie evenings in J{\"u}lich;\\
--~Barbara Rejdych-Iwanek, Micha{\l} Silarski, Joanna and Pawe{\l} Klaja, Jaros{\l}aw Zdebik, Marcin Zieli{\'n}ski
and Micha{\l} Janusz for a nice atmosphere of work;\\
--~Ma{\l}gorzata Fidelus, Izabela Ciepa{\l}, Borys Piskor-Ignatowicz and Rafa{\l} Sworst for the working climate
in room 03a.

\smallskip
Finally, I am indebted to my beloved Alina Ptaszek for her support, understanding and patience.

\smallskip
Last but not least I would like to say thank you to my Parents:
Dzi\k{e}kuj\k{e} Wam, Mamo i Tato, za
pomoc i wsparcie na wszystkich drogach, kt\'orymi kroczy{\l}em.\\

\pagestyle{fancy}

\backmatter
\phantomsection
\addcontentsline{toc}{chapter}{Bibliography}

\end{document}